\documentclass[aps,prb,superscriptaddress,reprint,nofootinbib]{revtex4-2}
\usepackage{bm}
\usepackage{graphicx}
\usepackage{amsmath}
\usepackage{amssymb}
\usepackage{dcolumn}
\usepackage[colorlinks=true,citecolor=blue,linkcolor=blue,urlcolor=blue]{hyperref}
\usepackage{subcaption}
\usepackage{xspace}
\usepackage{paralist}
\usepackage{multirow}
\usepackage[normalem]{ulem}

\newcommand{\fref}[1]{Fig.~\ref{#1}}
\newcommand{\tref}[1]{Table~\ref{#1}}
\newcommand{\eref}[1]{Eq.~\eqref{#1}}
\newcommand{\sref}[1]{Section~\ref{#1}}

\newcommand{\ang}[0]{\AA\xspace}
\newcommand{\hnn}[0]{hNN--Gr$_x$\xspace}
\newcommand{\act}[0]{h}

\graphicspath{{figures/}{./}}

\begin{document}

\title{Hybrid neural network potential for multilayer graphene}

\author{Mingjian Wen}
\affiliation{Department of Aerospace Engineering and Mechanics, University of
Minnesota, Minneapolis, MN 55455, USA}

\author{Ellad B. Tadmor}
\email[Author to whom correspondence should be addressed: ]{tadmor@umn.edu}
\affiliation{Department of Aerospace Engineering and Mechanics, University of
Minnesota, Minneapolis, MN 55455, USA}

\date{\today}

\begin{abstract}
Monolayer and multilayer graphene are promising materials for
applications such as electronic devices, sensors, energy generation and storage,
and medicine.
In order to perform large-scale atomistic simulations of the mechanical and
thermal behavior of graphene-based devices, accurate interatomic potentials are
required.
Here, we present a new interatomic potential for multilayer graphene structures
referred to as ``\hnn.''
This hybrid potential employs a neural network to describe short-range interactions
and a theoretically-motivated analytical term
to model long-range dispersion. The potential is trained against a large dataset of
monolayer graphene, bilayer graphene, and graphite configurations obtained
from \emph{ab initio} total-energy calculations based on density functional
theory (DFT). The potential provides accurate energy and forces for both
intralayer and interlayer interactions, correctly reproducing DFT results for
structural, energetic, and elastic properties such as the equilibrium layer
spacing, interlayer binding energy, elastic moduli, and phonon dispersions
to which it was not fit.
The potential is used to study the effect of vacancies on thermal conductivity
in monolayer graphene and interlayer friction in bilayer graphene.
The potential is available through the OpenKIM interatomic potential repository
at \url{https://openkim.org}.
\end{abstract}

\maketitle

\section{Introduction}
\label{sec:intro}

Since the discovery of graphene \cite{novoselov2004electric}, two-dimensional
(2D) materials have been shown to possess remarkable electronic, mechanical,
thermal, and optical properties, with great potential for nanotechnology
applications, such as ultrasensitive sensors and medical
devices \cite{neto2009electronic, hendry2010coherent,sevik2014assessment, lee2008measurement}.
Stacked 2D materials are even more exciting as they offer an opportunity
to create completely new materials with remarkable properties by controlling
the stacking order and orientation \cite{geim2013van, novoselov20162d}.
A striking example is the recent discovery of unconventional superconductivity
in bilayer graphene with an imposed twist of about $1.1^\circ$ \cite{graphene_sc_2018}.

Stacked 2D materials can be simulated accurately using a first-principles
density functional theory (DFT) calculation, which involves a numerical solution
to the Schr\"{o}dinger equation. However, due to hardware and algorithmic limitations,
DFT is typically limited to small molecular systems and crystalline
materials comprised of several hundreds of atoms at most. For example, the supercell
required to simulate a graphene bilayer with $1.1^\circ$ twist has too many atoms to be
simulated by first principles.\footnote{A DFT calculation of a twisted bilayer
employs a commensurate supercell. By increasing the size of the supercell it is
possible to approach arbitrarily close to any twist angle, but the supercell can be
quite large. For example, commensurate supercells for $1.084^\circ$ and $1.103^\circ$
(close to $1.1^\circ$) include 11,164 and 42,204 atoms, respectively, which is far beyond
DFT capabilities.}
In contrast, empirical interatomic potentials are computationally far less costly
and can therefore be used  via molecular simulations to compute static and dynamic properties that are inaccessible
to first-principles calculations \cite{mishin1999interatomic, wen2015interpolation, wen2017potfit}.

Development of an interatomic potential for stacked 2D materials is challenging
due to very different nature of the intralayer and interlayer bonding, and the
different energy scales associated with these interactions. Multilayer graphene
exhibits strong $sp^2$ covalent bonds within a layer and weak dispersion and
orbital repulsion interactions between layers. The cohesive energy of monolayer
graphene, characterizing intralayer bonding, is $8.06~\text{eV/atom}$, whereas the
interlayer binding energy of bilayer graphene is only $0.02263~\text{eV/atom}$.
Although weak, it is the interlayer interactions that define the function of
many nanodevices such as nanobearings, nanomotors, and
nanoresonators \cite{kolmogorov2005registry}, and also drive incommensurate
to commensurate structural transitions \cite{zhang:tadmor:2017, zhang:tadmor:2018}, which
lead to novel transport properties \cite{graphene_sc_2018,yoo:engelke:2019}.

There have been several efforts to develop an interatomic potential for carbon
systems. Early efforts include the bond-order Tersoff
\cite{tersoff1988empirical,tersoff1989modeling} and
REBO \cite{brenner1990physical,brenner2002rebo} potentials,
which modulate the strength of bonds based on their atomic environments.
These potentials provide a reasonable description for strong covalent bonds,
but do not account for dispersion interactions and thus are inherently
short-ranged in nature. To address this limitation,
the AIREBO \cite{stuart2000airebo} potential adds a 6--12 Lennard--Jones \cite{lennardjones1931}
(LJ) term to model dispersion, and the LCBOP \cite{los2003intrinsic} and
AIREBO--M \cite{o2015airebo} potentials add Morse \cite{morse1929diatomic} terms
for this purpose.  The more complex
ReaxFF \cite{srinivasan2015development} potential constructs the bond order
differently than the above potentials and includes explicit terms to account for
van der Waals (vdW), Coulombic, and under- and over-coordination energies.

These potentials have been shown to work well for a variety of applications,
but in many cases their quantitative predictions are inaccurate when compared
with first-principles and experimental results.
For example, the phonon dispersion curves of monolayer graphene at 0~K computed using
these potentials deviate largely from DFT results, especially for the
optical modes (discussed later in \sref{sec:test}).
As for interlayer interactions, the Tersoff and REBO potentials cannot be used
because they do not account for long-range dispersion interactions.
The AIREBO, AIREBO--M, LCBOP, and ReaxFF potentials do predict overall binding characteristics
between graphene layers, such as the equilibrium layer spacing and the $c$-axis
elastic modulus, but are unable to accurately distinguish energy
variations for different relative alignments of layers \cite{wen2018dihedral}.
The reason is that in addition to dispersion, the interlayer interactions include
short-range Pauli repulsion between overlapping $\pi$ orbitals of adjacent layers.
The repulsive interaction is not correctly modeled in these potentials.
The registry-dependent Kolmogorov--Crespi (KC) potential \cite{kolmogorov2005registry}
and an extension called the dihedral-angle-corrected registry-dependent interlayer potential (DRIP) \cite{wen2018dihedral}
address this by employing a term that depends on the transverse distance between
atom pairs to capture the repulsion due to orbital overlapping.
However, a major limitation of the KC potential and DRIP is that they are not
reactive, i.e.\ they require an \emph{a priori} fixed assignment of atoms into layers.
This prevents the study of many problems of interest, such as vacancy
migration between layers \cite{liu2014vacancy}.

Physics-based potentials (such as those discussed above) are devised
by selecting functional forms designed to represent the physics underlying the
material system and then fitting a handful of parameters.
In recent years, machine learning potentials \cite{behler2007generalized,
bartok2010gaussian, rupp2012fast, thompson2015spectral, shapeev2016moment, hajinazar2017stratified}
have been shown to be highly effective for a spectrum of material systems
ranging from organic molecules \cite{rupp2012fast} to alloys \cite{hajinazar2017stratified}.
Different from physics-based potentials, machine learning potentials are typically constructed by
first transforming the atomic environment information in a large dataset of first-principles results into vector representations (descriptors)
and then training general-purpose regression functions against them.
Several machine learning regression methods have been used to construct potentials, including linear regression \cite{thompson2015spectral}, kernel ridge regression \cite{rupp2012fast}, Gaussian process \cite{bartok2010gaussian}, and neural network (NN) \cite{behler2007generalized}.
Kernel ridge regression and Gaussian process are non-parametric methods, and therefore their evaluation time is proportional to the size of the training set.
This makes them computationally expensive if large datasets are used for the training (although sparsification approaches can be applied to select a representative subset of the training data for sparse model approximation).
Linear regression and NN are parametric methods, and thus their evaluation time is independent of the size of the training set.
An advantage of Gaussian process regression is that it can provide uncertainty in the predictions
(a feature the other three methods do not possess\footnote{Standard fully-connected NNs do not have the ability to provide uncertainty information,
whereas an NN trained with the dropout technique approximates a Bayesian NN, thus enabling uncertainty quantification \cite{gal2016dropout,gal2016uncertainty}.
We have explored the application of dropout NN potentials to estimate uncertainty propagation in atomistic simulations.
See \cite{wen2019dropout} for more information.}),
because it is essentially a Bayesian model.

For carbon systems, Cs{\'a}nyi \emph{et al.}\ have developed two Gaussian approximation potentials (GAPs)\footnote{GAP uses Gaussian process as the regression method.}:
one for liquid and amorphous carbon \cite{deringer2017machine} and the other for monolayer graphene \cite{rowe2018development}.
Khaliullin \emph{et al.}\ \cite{khaliullin2010graphite, khaliullin2011nucleation} have developed NN potentials to model phase transition from graphite to diamond.
Generally speaking, the \emph{transferability} (i.e.\ the ability of a potential to make accurate predictions outside its training set) of machine learning potentials is low.
Therefore, given their training sets, the GAP for liquid and amorphous carbon and the NN potentials for phase transition are not suitable for multilayer graphene structures.
The GAP for graphene is an accurate model that correctly reproduces many properties of monolayer graphene obtained from DFT \cite{rowe2018development};
however, similar to the Tersoff and REBO potentials, it lacks a description of the interlayer interactions and therefore cannot be used for multilayer
graphene structures.

In this paper, we present a new hybrid NN and physics-based
potential for multilayer graphene systems that is reactive and provides an accurate description
of both the intralayer and interlayer interactions.
The potential is referred to as ``\hnn'' (where the subscript $x$
indicates that it can be used for multiple graphene layers).
The long-range dispersion attraction
is modeled using a theoretically-motivated $r^{-6}$ term (as in the LJ potential),
and the short-range interactions are described using a general-purpose
NN. The latter include both the covalent bonds within a layer and
the repulsion due to overlapping orbitals of adjacent layers.
The inclusion of the theoretical long-range term improves the performance
of the potential since the NN does not need to learn known physics.
The parameters in the new \hnn potential are trained against a large dataset of
monolayer graphene, bilayer graphene, and graphite configurations
obtained from DFT calculations with an accurate dispersion correction.

The paper is structured as follows.
In \sref{sec:model} we introduce the new \hnn potential model
and describe the training procedure.
In \sref{sec:test}, we test the ability of the \hnn potential to reproduce
various canonical properties of interest obtained from DFT\@.
Results are compared with those of other potentials.
In \sref{sec:app}, we discuss applications of the \hnn potential to study
selected problems that are beyond the
scope of DFT: the effect of vacancies on the thermal conductivity of
monolayer graphene and interlayer friction in bilayer graphene.
The paper is summarized in \sref{sec:sum}.

\section{Definition of new model}
\label{sec:model}

\subsection{Mathematical form}

The total potential energy of a configuration consisting of $N$ atoms is
decomposed into the contributions of individual atoms
\begin{equation}
E = \sum_{\alpha=1}^N E_\alpha,
\end{equation}
where $E_\alpha$ is the energy of atom $\alpha$, composed of a
long-range interaction part and a short-range interaction part, i.e.\
$E_\alpha = E_\alpha^\text{long} + E_\alpha^\text{short}$.
The long-range dispersion attraction is modeled by a theoretically-motivated
$r^{-6}$ term as in the LJ potential,
\begin{equation}\label{eq:E:long}
  E_\alpha^\text{long} = -A\sum_{\beta\neq\alpha}^Nr_{\alpha\beta}^{-6}
  \, S_\text{up}(x_{\alpha\beta}) \, S_\text{down}(x_{\alpha\beta}),
\end{equation}
where $A$ is a fitting parameter, $r_{\alpha\beta}$ is the distance between
atoms $\alpha$ and $\beta$, and $S_\text{up}(x)$ and $S_\text{down}(x)$ are
switching functions that turn interactions on and off in certain
distance ranges.  The down switching function is defined as
\begin{equation}\label{eq:s:down}
S_\text{down}(x) =
\begin{cases}
  1, & x < 0  \\
  -6x^5 + 15x^4 -10x^3 + 1, &0 \leq x \leq 1 \\
  0, &x > 1
\end{cases}.
\end{equation}
This function monotonically decreases from one to zero over the
range $x\in[0,1]$, and has
zero first and second derivatives at both $x=0$ and $x=1$. The up
switching function is the complementary expression,
$S_\text{up}(x) = 1 - S_\text{down}(x)$.
The switches are applied within a desired distance interval
$[r^\text{min}, r^\text{max}]$ using the dimensionless
argument,
\begin{equation}\label{eq:x:r}
  x_{\alpha\beta} = \frac{r_{\alpha\beta} - r^\text{min}}{r^\text{max} - r^\text{min}}.
\end{equation}
The values of $r^\text{min}$ and $r^\text{max}$ for the up and down
switching functions are given in \tref{tab:params}.
With these values, the down switching function $S_\text{down}(x)$ causes
the potential to smoothly vanish at the cutoff $r_\text{down}^\text{max}$, and
the up switching function $S_\text{up}(x)$ turns off the long-range interactions
when the pair distance $r_{\alpha\beta}$ is smaller than $r_\text{up}^\text{min}$.

\begin{table}
\caption{Summary of parameters in the \hnn potential and hyperparameters that
define the NN structure in the short-range part of the potential.}
\label{tab:params}
\begin{ruledtabular}
\begin{tabular}{cc}
 $A$                                 &$8.3427~\text{eV}\cdot\text{\AA}^6$ \\
 $r_\text{up}^\text{min}$            &2~\ang  \\
 $r_\text{up}^\text{max}$            &4~\ang  \\
 $r_\text{down}^\text{min}$          &9~\ang  \\
 $r_\text{down}^\text{max}$          &10~\ang  \\
\hline
number of hidden layers              &3  \\
number of nodes in hidden layers     &30 \\
activation function $\act$           &$\tanh$ \\
$r_\text{cutoff}^\text{short}$       &5~\ang \\
descriptors                          &see SM \cite{supplemental} \\
weights                              &see SM \cite{supplemental} \\
biases                               &see SM \cite{supplemental} \\
\end{tabular}
\end{ruledtabular}
\end{table}

\begin{figure}
\includegraphics[width=1\columnwidth]{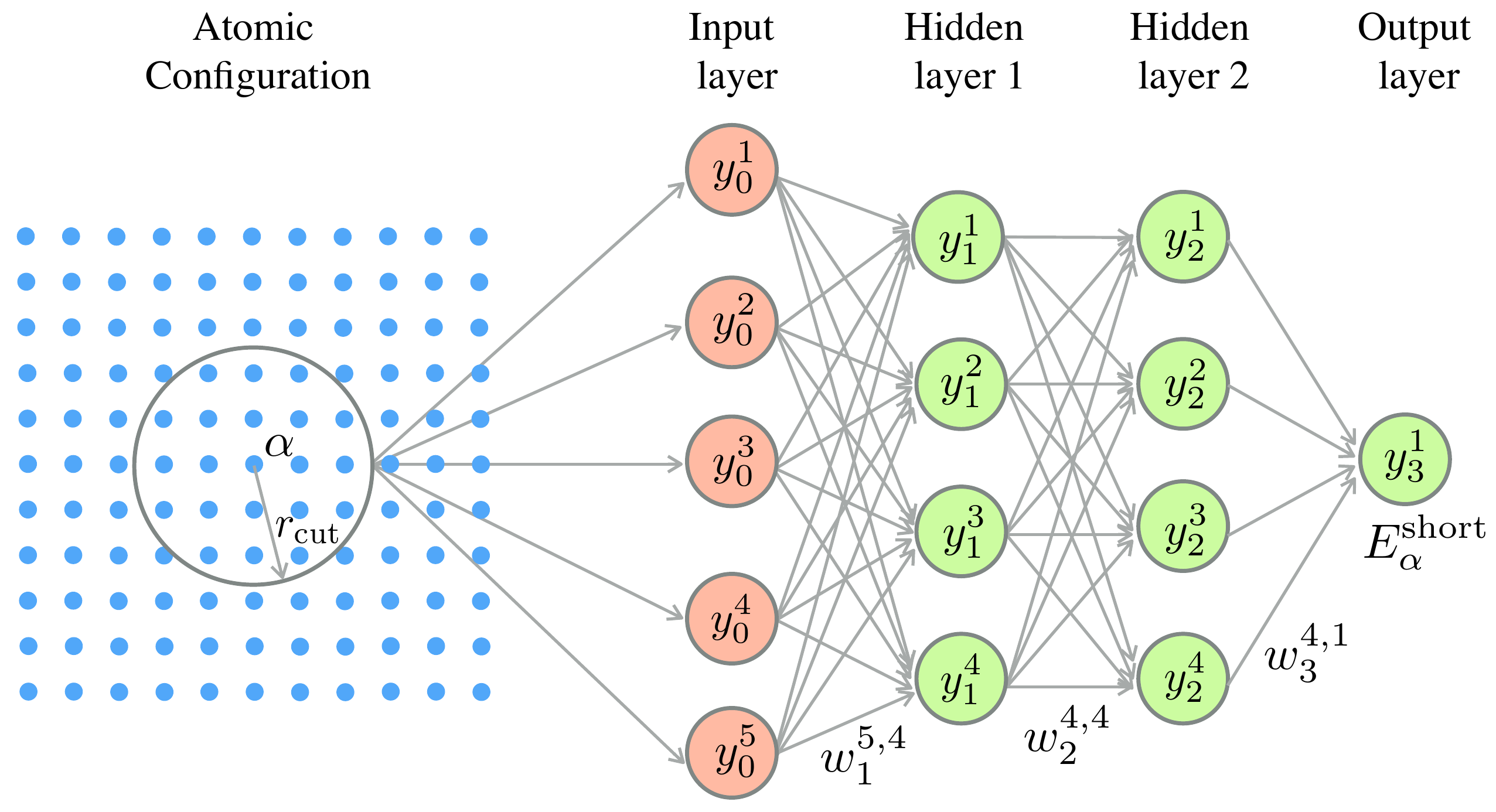}
\caption{Schematic representation of an NN potential for the
short-range energy $E_\alpha^\text{short}$ of atom $\alpha$.
This NN consists of an input layer, two hidden layers, and an output layer.
The configuration of neighbors of atom $\alpha$ within a cutoff $r_\text{cut}$
is transformed to a descriptor vector $y_0^j$ ($j=1,2,\dots,5$), which serves
as the input to the NN. The arrows connecting nodes in adjacent layers represent
weights. Biases and activation functions are not shown in
this figure. See text for explanation of the variables.
}
\label{fig:nn:struct}
\end{figure}

The short-range interactions (including both the covalent bonds within a layer
and the repulsion between overlapping orbitals of adjacent layers) are
represented by an NN as shown schematically in \fref{fig:nn:struct}.
The NN returns the short-range energy $E_\alpha^\text{short}$ of one atom
in the system (atom $\alpha$) based on the positions of itself and its neighbors
up to a cutoff distance $r_{\rm cut}$. The use of a cutoff significantly
reduces the computational cost by restricting the dependence of an atom's
energy to its local environment.

Between the input layer and the energy output layer
are so-called ``hidden'' layers that add complexity to the NN\@.
The NN in \fref{fig:nn:struct} consists of an input layer, two hidden layers,
and an output layer.
Each node in a hidden layer is connected to all nodes in the previous
layer and in the following layer. The value of node $j$ in layer $i$ is\footnote{The input layer and the output layer are indexed as the zeroth layer and third layer, respectively.}
\begin{equation}\label{eq:nn:connect}
y_i^j = \act(\sum_k y_{i-1}^{k} w_{i}^{k,j} + b_i^j ), \quad i=1,2,3,
\end{equation}
where $w_i^{k,j}$ is the weight
connecting node $k$ in layer $i-1$ and node $j$ in layer $i$, $b_i^j$ is the bias
applied to node $j$ of layer $i$, and $\act$ is an activation function (e.g.\
a hyperbolic tangent) that introduces nonlinearity into the NN\@.
More compactly, \eref{eq:nn:connect} can be written as $\bm y_i = \act(\bm
y_{i-1} \bm W_i + \bm b_i)$,\footnote{The activation function is applied
element-wise.} where $\bm y_i$ is a row vector of the node values in layer $i$,
$\bm W_i$ is a weight matrix, and $\bm b_i$ is a row vector of the biases.
For example for the NN shown in \fref{fig:nn:struct}, $\bm y_1$ and $\bm b_1$ are row vectors each with 4 elements and
$\bm W_1$ is a $5\times4$ matrix.
Consequently, the short-range atomic energy $E_\alpha^\text{short}$ can be
expressed as\footnote{Note that typically the activation function is not applied to the
output layer.}
\begin{equation}\label{eq:nn:connect:compact}
  E_\alpha^\text{short} = \act(\act(\bm y_0 \bm W_1 + \bm b_1)\bm W_2 +  \bm b_2)\bm W_3 + \bm b_3.
\end{equation}

Interatomic potentials must be invariant with respect to translation, rotation, and
inversion of space, and permutation of chemically equivalent atoms \cite{tadmor:miller:2011}.
To ensure that the NN satisfies these requirements, the environment of atom
$\alpha$, which is the input to the NN, must be transformed to a new
representation called a \emph{descriptor} that automatically satisfies
these invariances. Thus the input layer $\bm{y}_0$ is a descriptor vector
which is a function of the set of positions $\bm{r}_\alpha^\text{neigh}$
of all atoms within the neighborhood of atom $\alpha$ defined by the cutoff distance $r_\text{cut}$ (including atom $\alpha$ itself),
i.e.\footnote{The descriptor values are normalized by subtracting from each
component $y_0^j$ the mean value for this component across all atomic
environments in the training set and dividing by the standard deviation.}
\begin{equation}
y_0^j = g^j(\bm{r}_\alpha^\text{neigh}),
\end{equation}
where $j$ ranges over the components of the descriptor vector.

Various types of descriptors have been proposed in recent years including
the Coulomb matrix \cite{rupp2012fast} and bag of
bonds \cite{hansen2015machine} for molecular systems,
and the smooth overlap of atomic positions (SOAP) \cite{bartok2013representing},
symmetry functions \cite{behler2011atom},
and moment tensor \cite{shapeev2016moment} for crystalline materials.
In this work, we use symmetry functions \cite{behler2011atom, artrith2012high}, which are
discussed in detail in the Supplemental Material (SM) \cite{supplemental}.

A challenging aspect of training an NN, which is also a source of the
power and flexibility of the method, is that it is up to the developer to
select the number of descriptor terms to retain, the number of hidden
layers, the number of nodes within each hidden layer (which need
not be the same), and the activation function. It is also possible to
create different connectivity
scenarios between layers. Here we have opted for simplicity and
adopted a fully-connected network with the same number of nodes in
each hidden layer to reduce the number of hyperparameters that need to be
determined in the training process.  We chose the commonly used
hyperbolic tangent function, $\tanh(x) = (e^x - e^{-x})/(e^x + e^{-x})$,
as the nonlinear activation function $\act$.

\subsection{Dataset}

The \hnn potential parameters were determined from a dataset
of energies and forces for pristine
and defected monolayer graphene, bilayer graphene, and graphite at various states.
This includes configurations with compressed and stretched cells,
random perturbations of atoms, and configurations drawn from \emph{ab initio}
molecular dynamics (MD) trajectories at different temperatures.
The dataset consists of a total number of 14,250 configurations that are
randomly divided into a training set of 13,500 configurations (95\%) and a
test set of 750 configurations (5\%).
The dataset along with a detailed description of the configurations are
provided in the SM \cite{supplemental}.

The dataset is generated from DFT calculations using the Vienna Ab initio
Simulation Package (VASP) \cite{vasp1,vasp2}.
The exchange-correlation energy of the electrons is treated within the
generalized gradient approximated (GGA) functional of Perdew, Burke and
Ernzerhof (PBE) \cite{pbe}.
For monolayer and bilayer graphene, the supercell size in the direction
perpendicular to graphene planes is set to 30~\ang to minimize the
interaction between periodic images.
The reciprocal space is sampled using the $\Gamma$-centered Monkhorst
Pack grids \cite{monkhorst1976special} and the number of grids is chosen so
that the energy is converged to $1~\text{meV/atom}$.
The energy cutoff for the plane wave basis is set to $500~\text{eV}$.
Standard density functionals such as the local density approximation (LDA) and
GGA accurately represent Pauli repulsion in interlayer interactions, but fail to
capture vdW forces that result from dynamical correlations between fluctuating
charge distributions.\footnote{GGA predicts no binding at all at physically
meaningful spacings for graphite. LDA gives the correct interlayer spacing for
AB stacking; however, it underestimates the exfoliation energy by a factor of
two and overestimates the compressibility \cite{kolmogorov2005registry}.}
To address this limitation, various approximate corrections have been proposed
and we adopt the many-body dispersion (MBD) method \cite{tkatchenko2012accurate},
which has been shown to reproduce the more accurate adiabatic-connection
fluctuation-dissipation theory based random-phase-approximation (ACFDT--RPA) and
experimental results quite well \cite{wen2018dihedral}.

\subsection{Training}

The \hnn potential is fit in two stages: first the parameters in the
long-range part in \eref{eq:E:long} are determined, then the
parameters in the short-range NN part in \eref{eq:nn:connect:compact}.

For the long-range part, the interval bounds in the switching functions
($r_\text{up}^\text{min}$, $r_\text{up}^\text{max}$, $r_\text{down}^\text{min}$,
$r_\text{down}^\text{max}$) are listed in \tref{tab:params}.
The $r_\text{up}^\text{min}$ and $r_\text{up}^\text{max}$ values are selected
based on the graphene equilibrium lattice spacing of about 3.4~\AA,
$r_\text{down}^\text{max}$ sets the cutoff of the long-range interactions
and is based on prior experience with DRIP \cite{wen2018dihedral},
and $r_\text{down}^\text{min}$ is set a bit lower to create a smooth transition.
After fixing these, a single parameter $\bm\theta = \{A\}$ remains to
be determined.
It is optimized by minimizing a loss function $L(\bm\theta)$ that quantifies the difference
between the predictions of \eref{eq:E:long} and DFT results for
a subset of the training set
comprised of AB-stacked bilayer graphene at various layer
spacings ranging from $r_\text{up}^\text{max}$ to $r_\text{down}^\text{min}$.
The subset consists of $M=52$ configurations with concatenated coordinates
$\bm r_m$ for $m\in[1,M]$, such that $\bm r_m\in\mathbb{R}^{3N_m}$ where $N_m$
is the number of atoms in configuration $m$.
The loss function is
\begin{align} \label{eq:loss}
L(\bm\theta)
=& \sum_{m=1}^M \frac{1}{2} w^\text{e}_m \left[E(\bm r_m; \bm\theta)
  - E_m^\text{DFT} \right]^2 \nonumber \\
  +& \sum_{m=1}^M \frac{1}{2} w^\text{f}_m \left\|\bm f(\bm r_m; \bm\theta)
  - \bm f_m^\text{DFT} \right\|^2,
\end{align}
where $E(\bm r_m; \bm\theta)  $ and
$\bm{f}(\bm r_m; \bm\theta)
  = -\left.(\partial E/\partial \bm{r})\right|_{\bm{r}_m}
  \in \mathbb{R}^{3N_m}$
are the potential energy and concatenated forces for configuration $m$,
in which $E(\bm r_m; \bm\theta) = E^\text{long} = \sum_{\alpha=1}^{N_m} E_\alpha^\text{long}$.
The energy weight $w^\text{e}_m$ and force weight $w^\text{f}_m$ of
configuration $m$ have units of eV$^{-2}$ and (eV/\ang)$^{-2}$, respectively,
given energy in units of eV and forces in units of eV/\ang.
We set $w^\text{e}_m$ to $1/(N_m)^2$, and $w^\text{f}_m$ to $1/(10(N_m)^2)$.\footnote{The weights are inversely proportional to $(N_m)^2$ such that each configuration contributes more or less equally to the loss $L(\bm\theta)$. This prevents configurations with more atoms from dominating the optimization.}
The target DFT energy and forces for the long-range part
$E_m^\text{DFT}$ and $\bm f_m^\text{DFT}$ consider only interlayer interactions,
obtained in the same way as described in detail in \cite{wen2018dihedral}.
The resulting parameter $A$ is given in \tref{tab:params}.

With the long-range interactions determined, the next step is to determine
the short-range part of the potential. The same loss function in \eref{eq:loss}
is used with three differences compared with the long-range fitting:
(1) the parameters $\bm\theta$ are the weights $\bm W$ and biases $\bm b$
  in the NN;
(2) the entire training set is used;
and (3) the target energies $E_m^\text{DFT}$ and forces $\bm f_m^\text{DFT}$
are the differences between the total DFT values and the predictions from the
long-range contribution in \eref{eq:E:long}. The third item ensures that
the potential produces correct total energy and
forces when the long-range and short-range parts are used together.

The optimization was carried out using the KIM-based Learning-Integrated Fitting
Framework (KLIFF) \cite{kliff} with an L-BFGS-B minimizer \cite{zhu1997algorithm}.
KLIFF is compatible with potentials conforming to the
Knowledgebase of Interatomic Models (KIM) application programming
interface (API) \cite{tadmor2011kim}. See the SM \cite{supplemental}
(also, references\cite{plimpton1995fast,smith1996dl_poly_2,gale1997gulp,larsen2017atomic} therein)
for more details on KIM and how to use KIM potentials.
A grid search was performed to determine the optimal number of hidden layers
and nodes by fitting the potential to the training set in each case and
finding which provided the minimum loss for the test set.\footnote{The loss
of the test set is used to make the determination, rather than the
training set, to prevent overfitting.}
Using this process, it was found that 3 hidden layers with 30 nodes per layer
was the optimal choice.
The resulting energy root-mean-square error (RMSE) and forces RMSE
for the test set are 4.66~meV/atom and 41.41~meV/(\ang atom), respectively,
and 4.56~meV/atom and 41.13~meV/(\ang atom) for the training set.
See \tref{tab:params} for details of the NN parameters.

\section{Testing of the \hnn potential}
\label{sec:test}

An extensive set of calculations were performed to test the ability of the new
\hnn potential to reproduce structural, energetic, and elastic properties
of monolayer graphene, bilayer graphene, and graphite obtained from DFT\@.
A portion of the results is presented in \tref{tab:properties} together
with results from widely used potentials, \emph{ab initio} ACFDT--RPA,
and experiments.

\begingroup
\squeezetable
\begin{table*}
\caption{Summary of structural, energetic, and elastic properties computed from
the new \hnn potential and other widely used potentials.
The properties include
in-plane lattice parameter of monolayer graphene, $a$;
equilibrium layer spacing of bilayer graphene in AB stacking, $d_\text{AB}$,
bilayer graphene in AA stacking, $d_\text{AA}$, and graphite,
$d_\text{graphite}$;
interlayer binding energy of bilayer graphene, $E_\text{AB}$;
cohesive energy of monolayer graphene, $E_\text{coh}$;
single-vacancy formation energy in monolayer graphene, $E_\text{v}$;
and elastic moduli of graphite (outside parentheses) and monolayer graphene (in parentheses).
Also included are some first-principles and experimental results, as well as the computational expense relative to Tersoff.
Notes: (1) Since the Tersoff, REBO, and GAP--Gr (GAP for graphene) potentials lack the ability to model interlayer
interactions (see \sref{sec:intro}), they do not have predictions for properties related to
interlayer interactions.
(2) The KC and DRIP potentials only model interlayer interactions and therefore
cannot be used to compute the in-plane lattice parameter. Results from these
potentials used an in-plane lattice parameter of $a=2.46~\text{\AA}$.
For elastic properties, only the modulus related to stretching perpendicular
to the layers is computed in this case.}
\label{tab:properties}
\begin{ruledtabular}
\begin{tabular}{cccccccccccccc}
Method       &$a$  &$d_\text{AB}$  &$d_\text{AA}$  &$d_\text{graphite}$               &$E_\text{AB}$ &$E_\text{coh}$
 &$E_\text{v}$  &$C_{11}$  &$C_{12}$  &$C_{13}$
             &$C_{33}$  &$C_{44}$ &Time \\
             &(\ang) &(\ang)         &(\ang)          &(\ang)               &(eV/atom)
             &(eV/atom)               &(eV)    &(GPa)     &(GPa)     &(GPa)
             &(GPa)     &(GPa)  & (relative)\\
\hline
\hnn (present)  &2.467  &3.457  &3.618  &3.402  &21.63  &8.07  &8.08  &978.31~(1061.83)  &176.54~(208.77)  &$-66.74$ &40.35  &1.79  &279.4 \\
AIREBO \cite{stuart2000airebo}       &2.419  &3.392  &3.416  &3.358  &23.61  &7.43  &7.94  &1153.50~(1162.46)  &144.87~(147.64)  &0.08  &40.40  &0.28  &4.5 \\
AIREBO--M \cite{o2015airebo}  &2.420  &3.299  &3.324  &3.294  &16.18  &7.42  &7.93  &1174.25~(1157.43)  &147.66~(146.23)  &-0.02  &35.72  &0.28  &4.9 \\
LCBOP \cite{los2003intrinsic}        &2.459  &3.346  &3.365  &3.346  &12.52  &7.35  &8.13  &1049.91~(1054.32)  &157.29~(159.03)  &0.04  &29.80  &0.23  &1.6 \\
ReaxFF \cite{srinivasan2015development}       &2.462  &3.285  &3.294  &3.260  &34.59  &7.52  &7.52  &1147.67~(1119.84)  &831.84~(811.43)  &$-0.77$  &34.41  &0.15  &26.1 \\
Tersoff \cite{tersoff1989modeling}      &2.530  &       &       &       &       &7.39  &7.12  &(1274.00)  &($-240.11$) &      &       &  & 1 \\
REBO \cite{brenner2002rebo}         &2.460  &       &       &       &       &7.39  &7.82  &(1059.25)  &(148.33)  &      &       &  &1.6 \\
GAP--Gr \cite{rowe2018development}    &2.467  &       &       &       &       &7.96  &6.55  &(1108.81)  &(212.19)  &      &       &  &3814.7 \\
KC \cite{kolmogorov2005registry}           &       &3.374  &3.602  &3.337  &21.60  &      &  &  & & &34.45 & &36.6 \\
DRIP \cite{wen2018dihedral}         &       &3.439  &3.612  &3.415 &23.05  &      &  &  & & &32.00 &  &35.5 \\
\hline
DFT(PBE+MBD) &2.466  &3.426  &3.641  &3.400  &22.63  &8.06  &7.93  &1080.12~(1084.41)  &162.25~(161.25)  &$-4.63$  &33.18  &3.32  &$\sim 10^7$\\
ACFDT--RPA    &  &3.39\footnotemark[1]  &  &3.34\footnotemark[2]  & & & & & & &36\footnotemark[2]\\
\hline
\multirow{2}{*}{Experiment} &2.46\footnotemark[3] & & &3.34\footnotemark[4] & & & &1060\footnotemark[5] (1018\footnotemark[6]) &180\footnotemark[5]  &15\footnotemark[5] &36.5\footnotemark[5]  &0.27\footnotemark[5] \\
  &2.46\footnotemark[7] & & &3.356\footnotemark[8] & & & &1109\footnotemark[8] &139\footnotemark[8] &0\footnotemark[8] &38.7\footnotemark[8] &4.95\footnotemark[8]\\
\end{tabular}
\end{ruledtabular}
\footnotetext[1]{\cite{zhou2015van}.}
\footnotetext[2]{\cite{lebegue2010cohesive}.}
\footnotetext[3]{\cite{lin2012creating}.}
\footnotetext[4]{\cite{baskin1955lattice}.}
\footnotetext[5]{\cite{blakslee1970elastic}.}
\footnotetext[6]{\cite{lee2008measurement}.}
\footnotetext[7]{\cite{cooper2012experimental}.}
\footnotetext[8]{\cite{bosak2007elasticity}.}
\end{table*}
\endgroup

The in-plane lattice parameter of monolayer graphene, $a$, is obtained by
fitting the Birch--Murnaghan equation of state (EOS) \cite{birch1947finite}
(to conform to the approach used in DFT computations).
The results presented in \tref{tab:properties} show that
AIREBO and AIREBO--M underestimate the value of $a$, Tersoff overestimates it,
and the other potentials give values close to the experimental and DFT results.
\tref{tab:properties} also shows the values of the equilibrium layer spacing
for bilayer graphene in AB stacking $d_\text{AB}$, bilayer graphene in
AA stacking $d_\text{AA}$, and graphite $d_\text{graphite}$.
These values are also obtained from the Birch--Murnaghan EOS, keeping
the in-plane lattice parameter fixed to its equilibrium monolayer value.
The \hnn potential and DRIP are in good agreement with DFT(PBE+MBD) results
to which they were fit. The KC model is in better agreement
with more accurate ACDFT--RPA. The remaining potentials all underestimate
the AA separation, and have inconsistent results for AB and graphite:
AIREBO and LCBOP are accurate for both, and AIREBO--M and ReaxFF underestimate
both. Given this it is not surprising that except for \hnn,
KC, and DRIP, all of the above potentials provide inaccurate values for
$d_\text{AA} - d_\text{AB}$. The DFT value is 0.215~\AA, and the
potentials predict: 0.024~\AA\ (AIREBO),
0.025~\AA\ (AIREBO--M), 0.019~\AA\ (LCBOP), and 0.009~\AA\ (ReaxFF).
The reason for the poor accuracy is that these potentials cannot distinguish the
AA and AB stacking states. This is discussed further below.

\begin{figure}
  \includegraphics[width=1\columnwidth]{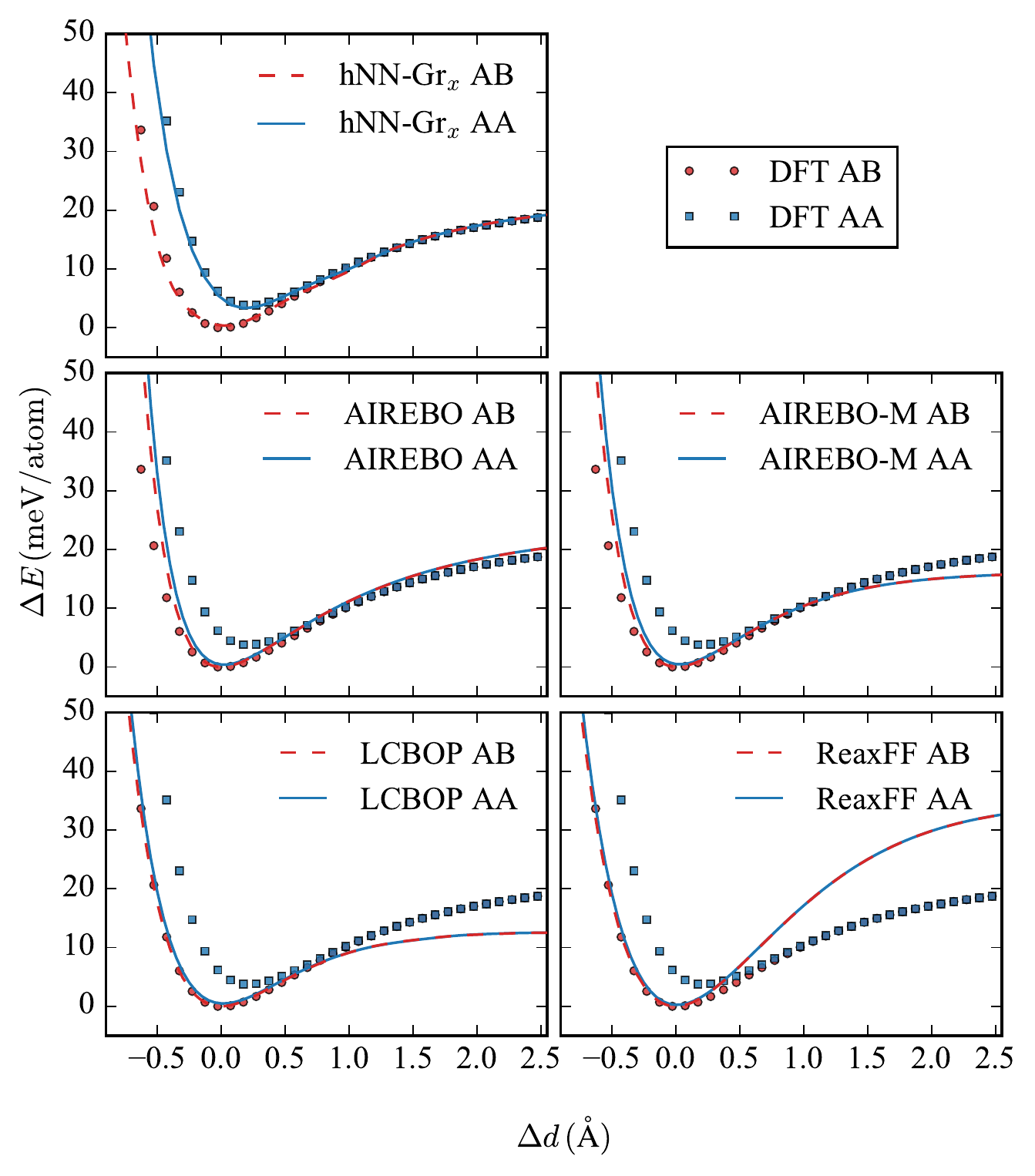}
  \caption{Interlayer binding energy of a graphene bilayer versus layer spacing
  for AB and AA stackings obtained from various potentials compared with DFT results.
   The curves are shifted such that the minimum energy in AB stacking is located
   at (0, 0). }
  \label{fig:e:vs:sep}
\end{figure}

Next, we consider energetics.
The interlayer binding energy of a graphene bilayer $E_\text{b}$ as a function of layer
spacing $d$ is shown in \fref{fig:e:vs:sep} for AB and AA stacking.
The curves are shifted such that $\Delta E = E_\text{b} - E_\text{AB}$
and $\Delta d = d - d_\text{AB}$, where $E_\text{AB}$ (listed in
\tref{tab:properties}) is the interlayer binding energy of AB-stacked
bilayer graphene at the equilibrium layer spacing $d_\text{AB}$ (i.e.\
$E_\text{AB}$ is the depth of the energy well relative to a reference state at
infinite separation).
We see that the AIREBO, AIREBO--M, LCBOP, and ReaxFF potentials give
nearly identical results for energy versus separation in the
AB and AA stacking states in contrast to DFT where a clear difference exists.
In addition, the AIREBO--M and LCBOP potentials underestimate the depth of the energy wells, whereas the ReaxFF potential overestimates it. (This can be seen
by considering the values predicted by these potentials relative to DFT
at the largest separation of $\Delta d=2.5$~\AA, which is approaching the
reference state).
The \hnn potential correctly captures the energy difference between the AB
and AA stacking states as well as the depth of the energy wells.
KC and DRIP can also capture the energy difference (see \cite{wen2018dihedral}).
Also notable is that at large separation, the curves for the two stacking states
merge since registry effects due to $\pi$-orbital overlap
become negligible and interlayer interactions are dominated by dispersion
attraction. This effect is captured correctly by the \hnn potential.

\begin{figure}
  \includegraphics[width=1\columnwidth]{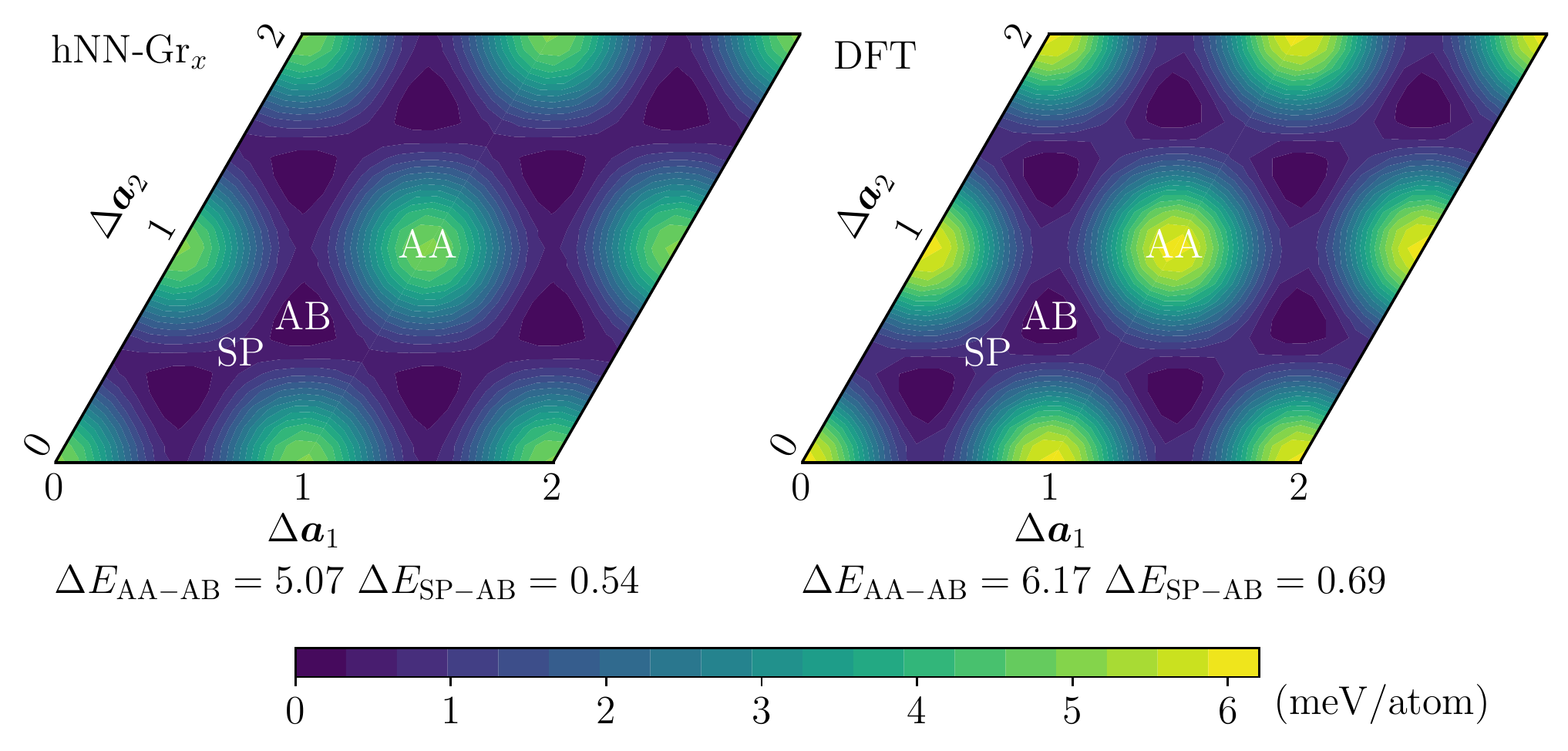}
  \caption{The GSFE of bilayer graphene obtained by sliding one layer
  relative to the other at a fixed layer spacing of $d=3.4~\text{\AA}$.
  The energy is relative to the AB state, which is $-21.53~\text{meV/atom}$
  for the new \hnn potential (on the left) and $-22.33~\text{meV/atom}$ for DFT (on the
  right).
  $\Delta E_\text{AA-AB}$ denotes the energy difference between the AA and AB states,
  and similarly $\Delta E_\text{SP-AB}$ denotes the energy difference between
  saddle point (SP) and AB states.
  The sliding parameters $\Delta \bm a_1$ and $\Delta \bm a_2$ are in
  units of lattice parameter $a=2.466~\text{\AA}$.}
  \label{fig:gsfe}
\end{figure}

A more complete view of the interlayer energetics is obtained by considering
the generalized stacking fault energy (GSFE) surface obtained by sliding one layer
relative to the other while keeping the layer spacing fixed.
\fref{fig:gsfe} shows the results for a layer spacing of $d=3.4~\text{\AA}$;
the \hnn potential is in quantitative agreement with DFT results.
The KC and DRIP GSFEs have a similar appearance (see \cite{wen2018dihedral}),
whereas the AIREBO, AIREBO--M, LCBOP, and ReaxFF GSFEs are nearly flat (not shown).

Also listed in \tref{tab:properties} are the cohesive energy
$E_\text{coh}$ and relaxed single-vacancy formation energy
$E_\text{v}$ for monolayer graphene.
The latter is computed as $E_\text{v} = E_2 -  E_1 - \mu$, where
$E_1$ and $E_2$ are the relaxed energy of monolayer graphene before and after the single
vacancy is created (by removing an atom from the simulation cell), and $\mu$ is the chemical potential of carbon, taken to be the cohesive energy $E_\text{coh}$ here.
All potentials perform reasonably well for these two properties except that the single-vacancy formation energy predicted by GAP--Gr is significantly smaller compared with the other potentials and DFT. This is likely because
GAP--Gr was only trained against configurations
drawn from MD trajectories of ideal graphene.

\begin{figure}
\begin{subfigure}{1\columnwidth}
\includegraphics[width=\columnwidth]{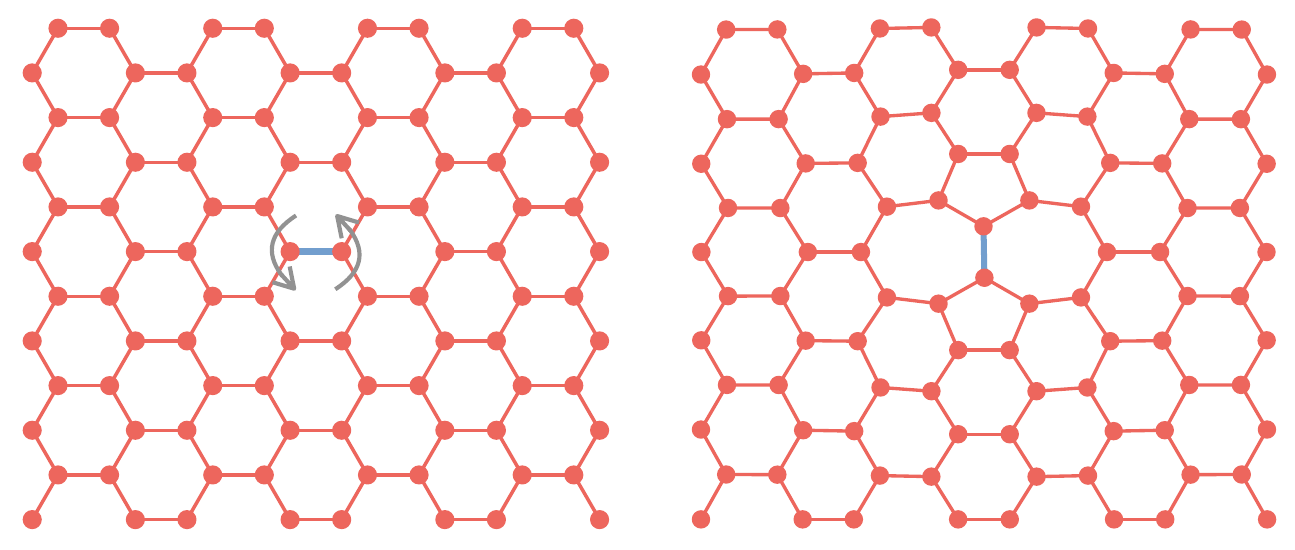}
\caption{\label{fig:rotation:setup}}
\end{subfigure}
\begin{subfigure}{1\columnwidth}
\includegraphics[width=\columnwidth]{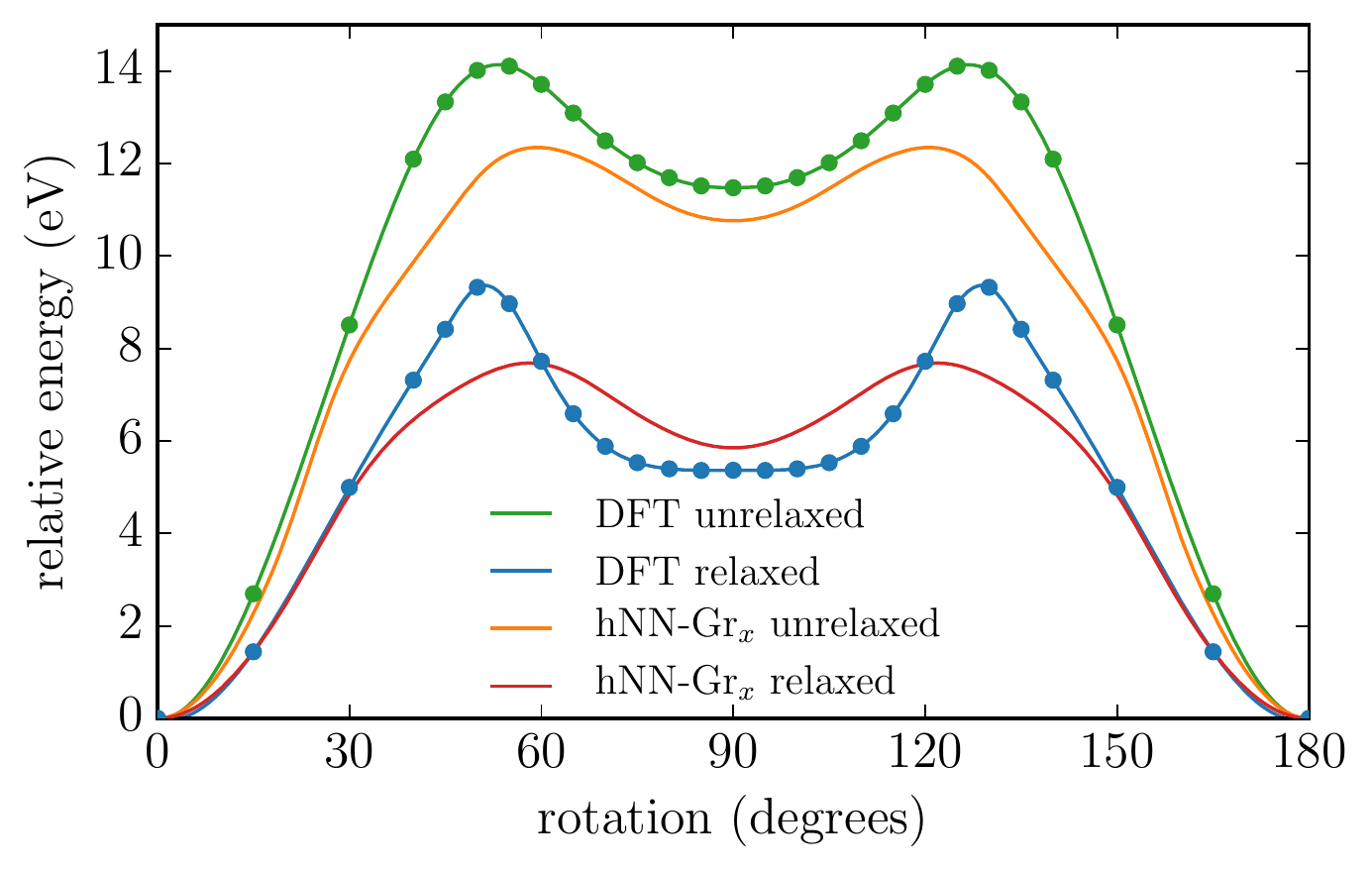}
\caption{\label{fig:energy:vs:rotation}}
\end{subfigure}
\caption{Stone--Wales defect created by rotating a pair of atoms.
(a) Monolayer graphene with a pair of atoms rotated at angles $0^\circ$ (on the
left) and $90^\circ$ (on the right),
and (b) energy versus rotation for both relaxed and unrelaxed structures
predicted by DFT and the \hnn potential.}
\end{figure}

Another interesting example, which tests the ability of the \hnn potential
to capture changes in hybridization, is the concerted exchange
mechanism first studied in graphene by
Kaxiras and Pandey\ \cite{kaxiras1988energetics}.
In this process a pair of atoms rotate by 90 degrees
converting four hexagonal rings to two pentagons and two heptagons thereby
creating a Stone--Wales (SW) defect (see \fref{fig:rotation:setup}).
To explore the energetics of the process, the total energy of a system of
96 atoms was computed as a function of the rotation angle using DFT
and the \hnn potential.
At each angle, the energy is minimized with respect to the positions
of the atoms subject to the constraint that the two rotating atoms
can only move along the line connecting them
(i.e.\ along the blue line shown in \fref{fig:rotation:setup}).

The energy versus rotation curves for DFT and \hnn are shown
in \fref{fig:energy:vs:rotation}, where the DFT results are
interpolated by a cubic spline. Overall the \hnn results follow the
DFT curve, predicting a SW defect formation energy of 5.85~eV
(red curve at rotation $90^\circ$), about 9\% higher
than the DFT prediction of 5.37~eV (blue curve at rotation $90^\circ$).
The energy barriers at the transition state are 7.68~eV at a rotation
of $58^\circ$ for \hnn, and 9.32~eV at a rotation of $51^\circ$
for DFT, a relative difference of about 17.6\%. Since \hnn is a
machine learning potential, the accuracy can be systematically
improved by augmenting the training set with configurations
along the concerted exchange path. As a comparison, we also
computed the energy versus rotation using the other potentials listed in
\tref{tab:properties} (see the results in the SM \cite{supplemental}).
None of the potentials are in very good agreement with DFT,
in particular, none capture the energy plateau in the vicinity of
the SW defect.

Finally, we consider elasticity properties.
The elastic moduli of hexagonal graphite was computed using finite differences.
The five independent components are listed in \tref{tab:properties}.
For each potential, the graphitic structure is constructed using its
corresponding in-plane lattice parameter, $a$, and equilibrium layer
spacing $d_\text{graphite}$.
In addition, the in-plane elastic moduli $C_{11}$ and $C_{12}$ of monolayer graphene were computed (values listed in parentheses).
Similar to graphite, the graphene structure is constructed using the corresponding in-plane lattice parameter of each potential, whereas the ``thickness'' of graphene (required to obtain bulk units) is assumed to be $3.34~\text{\AA}$ in all cases.
The results show that for graphite the \hnn potential is in
good agreement with DFT for $C_{11}$ (9.5\%) and $C_{12}$ (8.8\%),
reasonable agreement for $C_{33}$ (21.6\%) and $C_{44}$ (46.1\%),
and incorrect for $C_{13}$ (1340\%) (although we note that the DFT results
disagree with experiments in this case).
For graphene, the \hnn potential is in excellent agreement for $C_{11}$ (2.1\%),
but overestimates $C_{12}$ (29.5\%).
For the other potentials, notable disagreements are: (1) ReaxFF predicts significantly larger values of $C_{12}$ of both graphite and graphene;
(2) All of the potentials greatly underestimate $C_{44}$ for graphite;
(3) Tersoff overestimates $C_{11}$ and predicts negative $C_{12}$ for graphene; and
(4) GAP--Gr overestimates $C_{12}$ for graphene.

\begin{figure}
  \includegraphics[width=1\columnwidth]{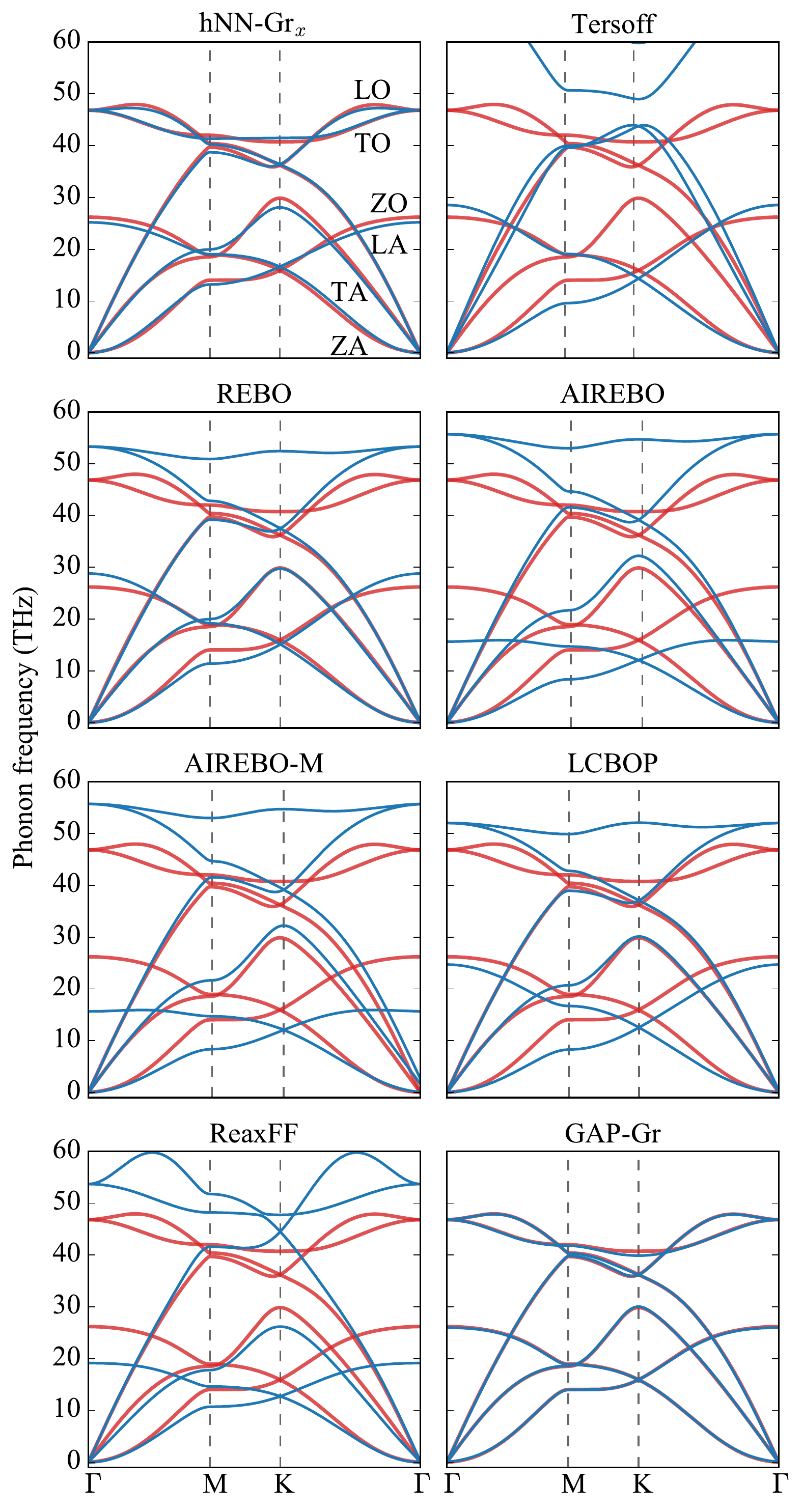}
  \caption{Phonon dispersion curves of monolayer graphene along high-symmetry points in the
  first Brillouin zone.
  The red curve is the DFT prediction, and the blue curves are results from the potentials.
  Branch labels are shown in the upper left panel, where ``L'' stands for
  longitudinal, ``T'' for transverse, ``Z'' for flexural, ``O'' for optical, and
  ``A'' for acoustic.
  Note that parts of the highest two branches by the Tersoff potential are not shown.}
  \label{fig:phonon}
\end{figure}

While the elastic moduli provide insight into the elastic behavior of the
potentials, a more complete view is gained from the phonon dispersion curves.
A number of thermodynamic properties, such as the thermal expansion coefficient
and heat capacity, can be obtained directly from dispersion relations via
calculation of the free energy.
\fref{fig:phonon} shows the phonon dispersion curves of monolayer graphene calculated
using finite differences as implemented in the phonopy package \cite{phonopy}.
The predictions of the \hnn potential and GAP--GR are in excellent agreement
with DFT\@. The other potentials provide good agreement for some phonon
branches, but not all.
REBO quantitatively predicts the shape and dispersion character of
most of the phonon branches, but fails for the high-frequency transverse optical (TO)
and longitudinal optical (LO) branches.
LCBOP, AIREBO, AIREBO--M, and ReaxFF are comparable, qualitatively
predicting the overall shapes of most curves, but are in poor
quantitative agreement with DFT. Tersoff has the worst performance
with poor qualitative agreement for most branches.
We note that a drawback common to all of the physics-based potentials
is that they fail to capture the dispersive behavior of the high-frequency
LO and TO branches, which \hnn and GAP--Gr predict with negligible error.
The phonon dispersions of bilayer graphene and graphite (not shown here) are
identical to monolayer graphene, except that the ZA branch splits into
two doubly degenerate branches near the $\Gamma$ point \cite{yan2008phonon,wirtz2004phonon}.

For the properties computed above and the potentials tested, the results
indicate that overall, machine learning potentials (both \hnn and GAP--Gr)
have higher accuracy than the physics-based potentials.
However, the accuracy comes at the price of increased computational cost.
\tref{tab:properties} shows the time (relative to Tersoff) that it takes
each potential to complete an MD trajectory of the same duration under
the canonical ensemble. The simulations were carried out using
LAMMPS \cite{plimpton1995fast, lammps} with
\hnn implemented in KIM \cite{tadmor2011kim, openkim},
GAP--Gr implemented in QUIP \cite{quip}, and the other potentials natively built
into LAMMPS.\footnote{The configuration used in the simulations is monolayer graphene
consisting of 192 atoms (bilayer graphene with 384 atoms for KC and DRIP).
Both KIM and QUIP have interfaces to LAMMPS so that their potentials can be used directly.
The simulations were performed in serial mode with one core.}
While GAP--Gr is nearly 4000 times slower than Tersoff, the \hnn
potential is much faster, only about 280 times\footnote{For the \hnn potential, the relative computational cost
of the long-range LJ part to the short-range NN part is 1:93.
Within the NN part, the ratio of the time to evaluate the descriptors
and the time associated with other computations (e.g.\ calculating energy
and forces) is 75:18. Thus it is clear that the bottleneck is the evaluation
of the descriptors.} slower than Tersoff.
As discussed in \ref{sec:intro}, this is a benefit of parametric methods; the evaluation time does not depend on the size of the training set.
Both \hnn and GAP--Gr are still significantly faster than a first-principles method like DFT, although they are significantly slower than the tested physics-based potentials.
KC and DRIP are relatively more expensive than the other physics-based potentials because to model long-range dispersion attraction,
they need to use a much larger cutoff distance. For example, DRIP uses a
cutoff of 12~\AA, whereas the other physics-based potentials considered here typically have cutoffs smaller than 5~\AA.

\section{Applications}
\label{sec:app}

The new \hnn potential is applied to two problems of interest that
are beyond the capabilities of DFT:
(1) thermal conductivity of monolayer graphene; and
(2) interlayer friction in bilayer graphene.
In both cases the effect of vacancies on the results are explored.

\subsection{Thermal conductivity}

Graphene has been reported to have extremely high thermal conductivity
 with experimentally measured values between 1500 and
2500~W/mK \cite{cai2010thermal, faugeras2010thermal, xu2014length, lee2011thermal, yousefzadi2016anisotropic} in
suspended samples at room temperature. (For comparison, copper has a
thermal conductivity of about 400~W/mK.)
Despite these efforts, accurate determination of the thermal conductivity
of graphene remains challenging because thermal transport in this material is
very sensitive to defects and experimental conditions \cite{balandin2011thermal, pereira2013divergence}.
Atomistic simulations using interatomic potentials provide an alternative
approach to study the thermal conductivity in graphene and investigate the effect of defects.
One concern is that interatomic potentials do not account for electron contributions
to thermal transport which are the dominant effect in metals. Fortunately, although
graphene is a semi-metal, at room temperature lattice vibrations account for
the majority of the thermal transport making the interatomic potential estimate
meaningful \cite{ghosh2008extremely, kim2016electronic}.
An accurate prediction of the lattice contribution depends on the ability of the potential to describe
the phonon dispersion curves, and in particular
the ZA mode associated with out-of-plane vibrations that provides the dominant
contribution to the lattice thermal conductivity in suspended
graphene \cite{lindsay2010flexural,zhang2011thermal}. As seen  in
\fref{fig:phonon}, the \hnn potential is highly accurate in predicting
all phonon dispersion branches including ZA.

The thermal conductivity is computed using the Green--Kubo method,
an equilibrium MD approach.
The Green--Kubo expression, based on linear-response theory,
is \cite{tuckerman2010statistical,schelling2002comparison}
\begin{equation}\label{eq:kappa}
\kappa_{ij}
= \frac{1}{\Omega k_\text{B} T^2} \int _0 ^\infty  \langle J_i(t) J_j(0) \rangle \,\text{d}t,
\end{equation}
where $i, j \in \{x,y,z\}$ are Cartesian components,
$k_\text{B}$ is Boltzmann's constant,
$T$ is the temperature,
$\langle J_i(t) J_j(0) \rangle$ is the heat current auto-correlation (HCA)
function expressed as a phase average,
and $\Omega$ is the volume of the system defined as the area of graphene multiplied
by the van der Waals thickness (3.457~\ang in the present case; see \tref{tab:properties}).
The upper limit of the integral in \eref{eq:kappa} can be approximated by $t_P$,
the correlation time required for the HCA to decay to zero.
In the case of an MD simulation, the phase average in the HCA is approximated
by a time average computed at discrete MD time steps.
Consequently, \eref{eq:kappa} is in fact a summation and we actually
compute \cite{schelling2002comparison}
\begin{equation}
\kappa_{ij} (t_P)
= \frac{\Delta t}{\Omega k_\text{B} T^2} \sum_{p=1}^P (Q-p)^{-1} \sum_{q=1}^{Q-p} J_i(p+q) J_j(q),
\end{equation}
where $\Delta t$ is the MD time step,
$Q$ is the total number of steps,
$P=t_P/\Delta t$ is the number of steps for integration (should be smaller than $Q$),
and $J_i(p+q)$ is the $i$th component of the heat current at step $p+q$.

A key component of the Green--Kubo method is the definition of the heat current.
We note that the heat current implemented in the
LAMMPS MD code \cite{plimpton1995fast, lammps} is intended for pair
potentials only. For many-body potentials, such as the \hnn potential, using
the LAMMPS expression can lead to incorrect results.\footnote{See
\cite{fan2015force} for a comparison of the thermal conductivity
obtained using different definitions of the heat current for the
Tersoff potential \cite{tersoff1988empirical, tersoff1989modeling}.}
In this work, we use the definition in \cite{admal2011stress}, which applies
to arbitrary many-body potentials.

\begin{figure}
\includegraphics[width=1\columnwidth]{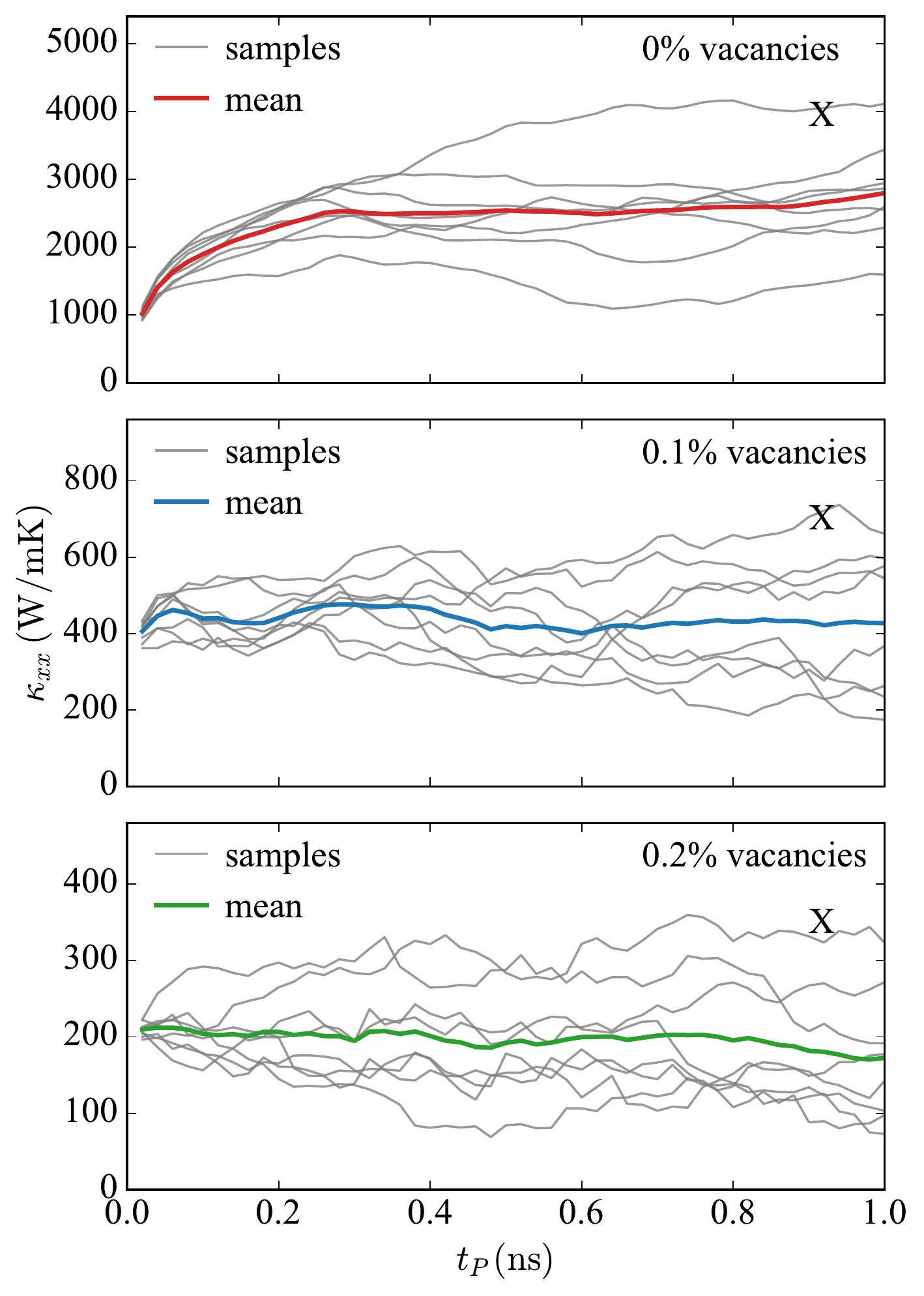}
\caption{Thermal conductivity in the $x$ direction, $\kappa_{xx}$, as a function
of $t_P$ for pristine graphene, graphene with 0.1\% vacancy
density, and graphene with 0.2\% vacancy density.
In each panel, the thin gray lines are the HCA cumulative averages obtained
from eight independent trajectories, and the thick lines (red, blue, or green)
are the means of these HCA curves.
The ``X'' denotes the sample with the largest $\kappa_{xx}$ at
$t_P=1~\text{ns}$ among the eight samples whose normalized HCA is shown in
\fref{fig:hca}.}
\label{fig:kappa}
\end{figure}

We study the thermal conductivity in pristine graphene and investigate the
impact of defects.
In practice, graphene can contain a variety of defects including single vacancies,
double vacancies, Stone--Wales defects, adatoms, dislocations, and grain
boundaries \cite{banhart2010structural, skowron2015energetics}.
Here, we focus on single vacancies, which have been experimentally shown
to be a common type of defect in graphene \cite{gass2008free}.
The base graphene system consists of a periodic rectangular supercell of size
51.25~\ang by 49.32~\ang in the $x$ (armchair) and $y$ (zigzag) directions
comprised of 960 atoms.  Separate calculations showed that
this system is sufficiently large to obtain converged thermal conductivity
for ideal graphene in agreement with previously published
results in \cite{zhang2011thermal}.
Single vacancies are generated by randomly removing atoms from
the supercell.
The equations of motion are integrated using a velocity-Verlet algorithm with
a time step of $\Delta t = 1~\text{fs}$.
The system is initially thermalized for 0.5~ns at a constant temperature of
$T=300~\text{K}$ under $NVT$ conditions (canonical ensemble)
using a Langevin thermostat. The thermostat is then switched off and
data for the Green--Kubo expression is collected under $NVE$ conditions
(microcanonical ensemble).
A time scale on the order of nanoseconds is necessary
to sufficiently converge the HCA function \cite{schelling2002comparison}.
We ran the \emph{NVE} simulation for 10~ns based on previous studies of thermal
conductivity in graphene \cite{zhang2009direct, haskins2011control}.

The thermal conductivity in the $x$ (armchair) direction, $\kappa_{xx}$, as a function
of $t_P$ for pristine graphene, graphene with a 0.1\% vacancy density
(one vacancy per supercell), graphene with a 0.2\% vacancy density
(two vacancies per supercell) is plotted in \fref{fig:kappa}.
In each case, the thermal conductivity is computed by averaging over
eight uncorrelated trajectories with different initial conditions.
We see that the majority of the samples are well converged after
$t_P=0.5~\text{ns}$, with the mean showing an even better convergence.
The thermal conductivity of pristine graphene measured at $t_P=0.5~\text{ns}$
is 2531~W/mK, in good agreement with the experimental values of 1500--2500~W/mK
for suspended graphene \cite{cai2010thermal, faugeras2010thermal, xu2014length, lee2011thermal, yousefzadi2016anisotropic}.
The thermal conductivity for the graphene with a 0.1\% vacancy density
is 415~W/mK, an 84\% reduction, and for graphene with a 0.2\% vacancy density
it is 195~W/mK, a 92\% reduction.
Similar values were obtained in the $y$ (zigzag) direction,
i.e.\ $\kappa_{yy}\approx\kappa_{xx}$ as expected due to isotropy in
the graphene plane.

\begin{figure}
\includegraphics[width=1\columnwidth]{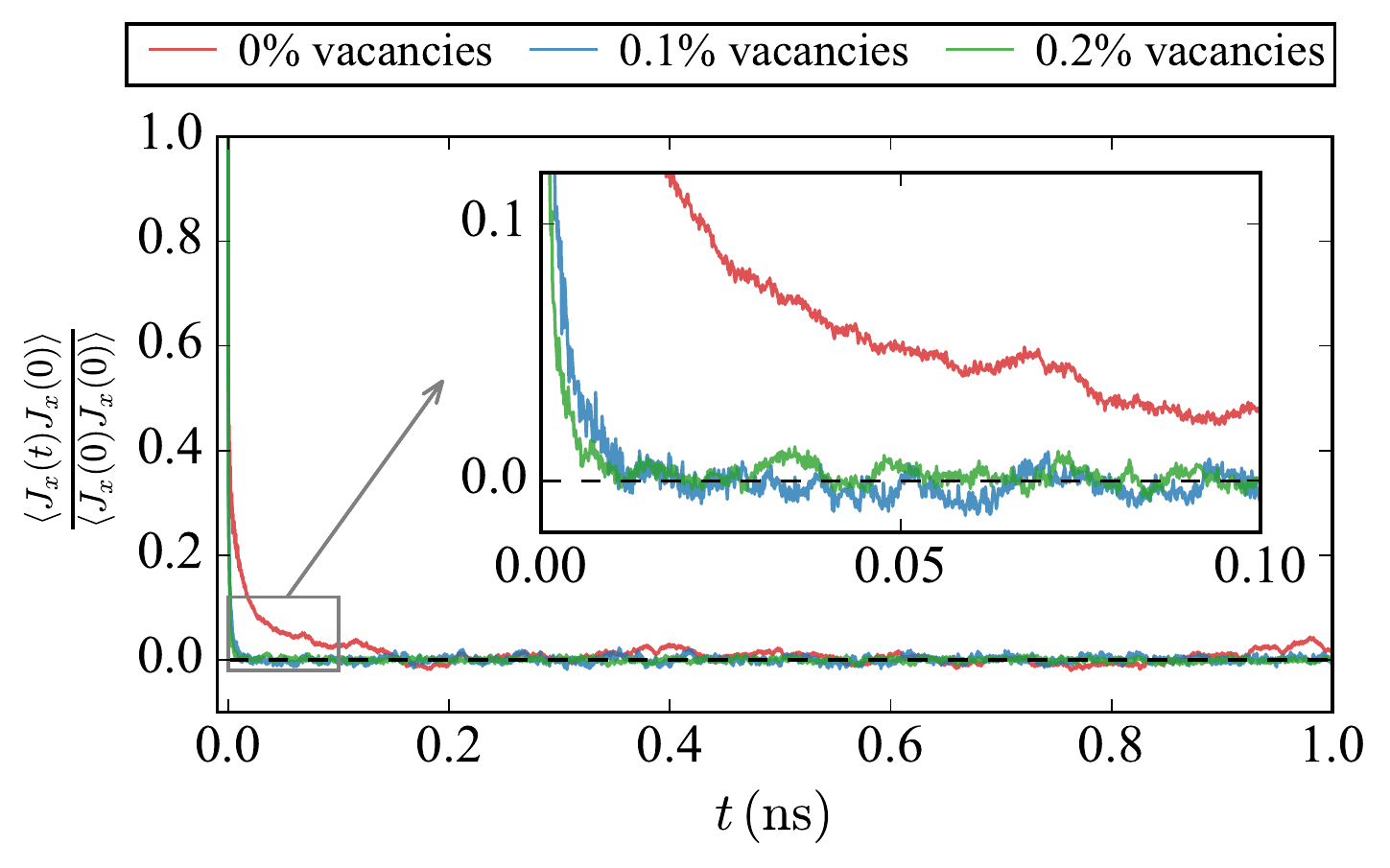}
\caption{Normalized HCA, $\left<J_x(t)J_x(0)\right> / \left<J_x(0)J_x(0)\right>$,
  as a function of time $t$ for pristine graphene, graphene with 0.1\%
  vacancy density, and graphene with 0.2\% vacancy density.
  The red, blue, and green curves are for the samples marked with an ``X''
  for graphene with 0, 0.1\% and 0.2\% vacancy density in \fref{fig:kappa}.}
\label{fig:hca}
\end{figure}

In \fref{fig:hca}, we plot the normalized HCA, $\left<J_x(t)J_x(0)\right> /
\left<J_x(0)J_x(0)\right>$, for the samples marked with an ``X'' in
\fref{fig:kappa}.
It is clear that the normalized HCA decays to zero much earlier than $t=0.5~\text{ns}$
for all
three types of graphene, indicating that $t_P=0.5~\text{ns}$ is sufficient
for calculating the thermal conductivity. Further, the decay of
the normalized HCAs for graphene containing vacancies is
much faster than that of pristine graphene, which is related to the fact
that the thermal conductivity in defective graphene is much smaller than
in pristine graphene.
(Note that $\left<J_x(0)J_x(0)\right>$ is almost the same for all three cases
and thermal conductivity is the integral of the HCA).
The underlying mechanism for the reduced thermal conductivity of graphene with vacancies is that vacancy defects are a strong
scattering source for phonons, which govern heat transport in this system.
Creation of a single vacancy leaves three carbon atoms with two-fold coordination,
effectively breaking the $sp^2$ characteristics of the local lattice.
These two-fold coordinated atoms are less likely to follow the normal pattern of
vibrations in pristine graphene and cause a significant degree of scattering \cite{haskins2011control}.

\subsection{Interlayer Friction}
Although the bonding between layers in multilayer graphene is weak,
the material still exhibits significant resistance to sliding due to
orbital overlap between layers. The friction becomes even larger when
covalent bonds are formed between adjacent layers.
Such bonds have been proposed to occur when vacancies exist in close
proximity to each other in the top and bottom layers and react to
form covalent bonds in their vicinity \cite{telling2003wigner}.
A plausible mechanism for this to happen is the creation of vacancies
through high-energy ion or electron bombardment of multilayer graphene
\cite{vicarelli:heerema:2015}.
Here, we study the effect of vacancies and interlayer covalent bonding
on friction in bilayer graphene.

\begin{figure}
  \includegraphics[width=1\columnwidth]{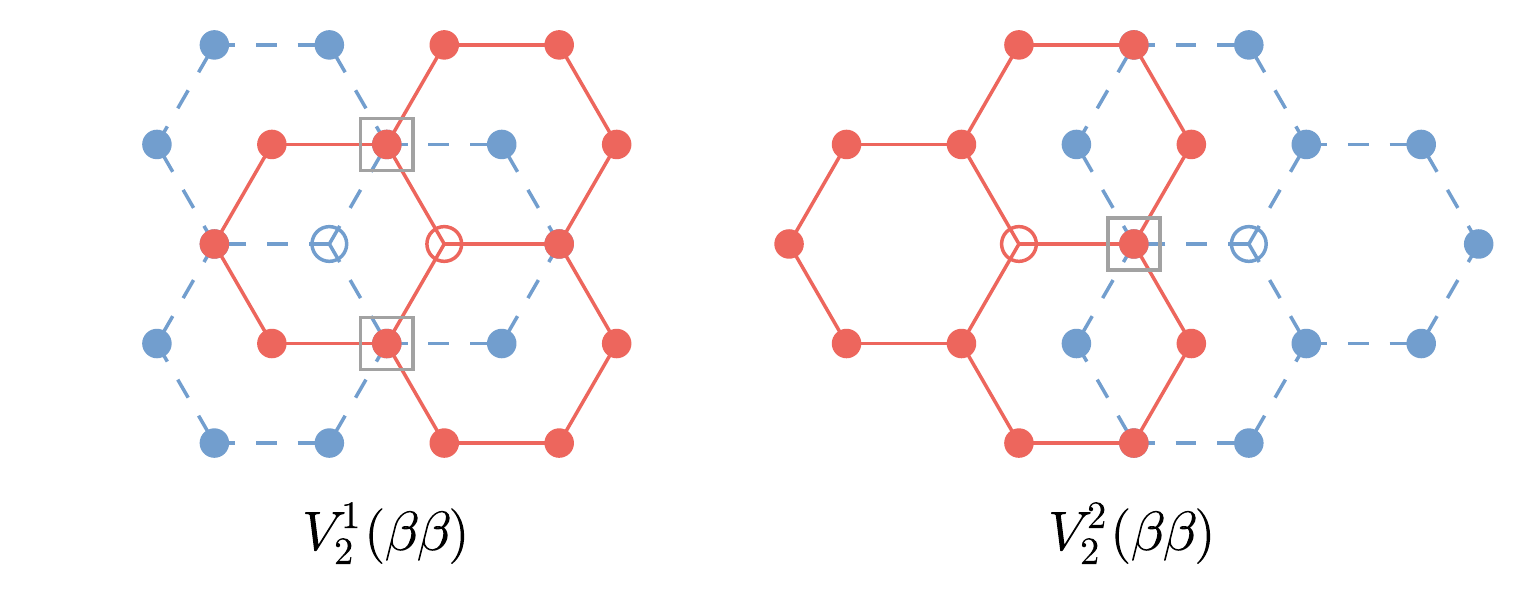}
  \caption{Close-proximity divacancies in adjacent layers of AB-stacked bilayer
  graphene that favor the formation of covalent bonds between layers.
  Hollow circles denote vacancies, and gray squares are locations where
  covalent bonds can form between atoms in adjacent layers.
  (There are two atoms in each gray square;
  the blue atom in the bottom layer is hidden by the red atom in the top layer.)
  Following the notation in \cite{telling2003wigner}, the subscript 2
  in $V_2^1(\beta\beta)$ and $V_2^2(\beta\beta)$ indicates that two single
  vacancies form a divacancy, the superscripts 1 and 2 denote first- and
  second-nearest interlayer neighbors, and $\beta$ means that a
  vacancy is located at the hexagonal ring center of the other layer.}
  \label{fig:vacancy}
\end{figure}

A number of possible interlayer divacancies can form via the coalescence of
single vacancies in adjacent layers leading to the formation of
covalent bonds \cite{telling2003wigner,teobaldi2010effect}.
We focus on the two structures shown in
\fref{fig:vacancy}, where the two vacancies are
first- and second-nearest interlayer neighbors referred to as
$V_2^1(\beta\beta)$ and $V_2^2(\beta\beta)$ (see the figure caption for an explanation of the notation).

\begin{figure}
  \includegraphics[width=1\columnwidth]{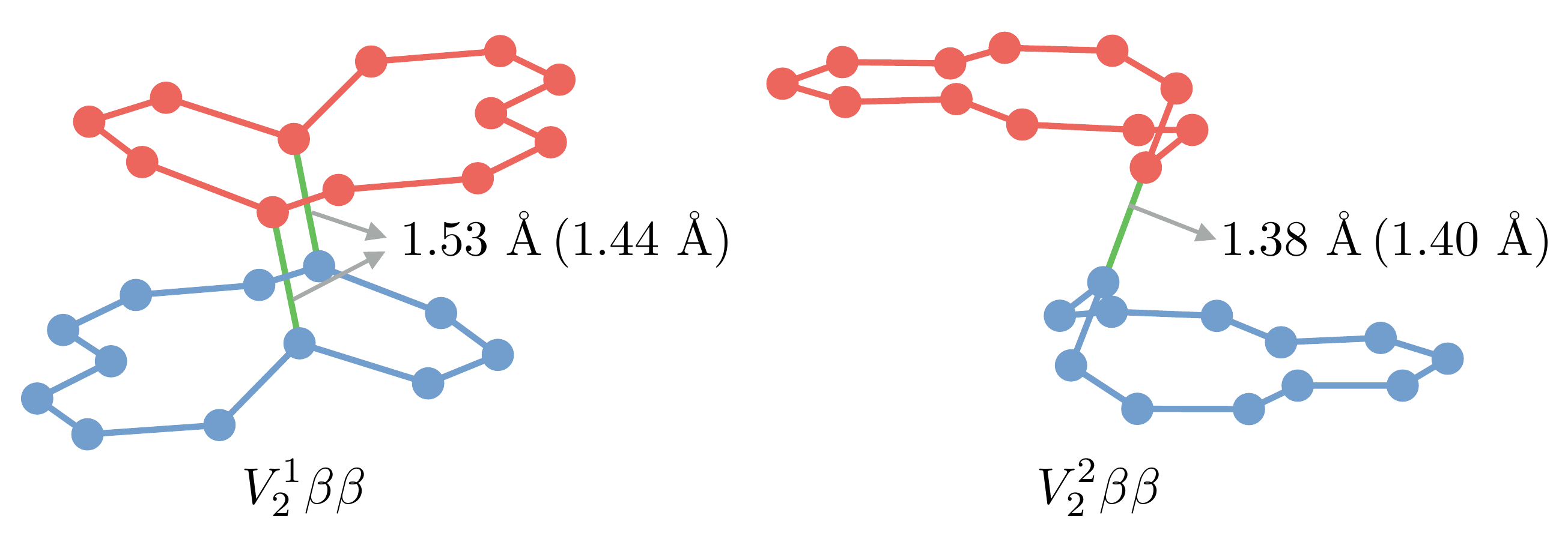}
  \caption{Core structures of the $V_2^1(\beta\beta)$ and $V_2^2(\beta\beta)$
  divacancies after relaxation.
  The interlayer covalent bond(s) formed near the divacancy are colored green.
  The bond length predicted by DFT (\hnn) is shown.}
  \label{fig:vacancy:relaxed}
\end{figure}

Graphene bilayers containing the two types of divacancies
$V_2^1(\beta\beta)$ and $V_2^2(\beta\beta)$ are fully relaxed using
DFT and the \hnn potential.
An important point is that in order for covalent bonds to form
between layers it is necessary to compress the bilayer in the direction
perpendicular to the layers, so that the layers are brought to within a
spacing of about $2.4~\text{\AA}$ prior to relaxation.
Both DFT and the \hnn potential predict the same core structure after relaxation
as shown in \fref{fig:vacancy:relaxed}.
Two interlayer covalent bonds of equal length (colored green) are
formed in the first-nearest-neighbor divacancy ($V_2^1(\beta\beta)$).
The bond length is predicted by the \hnn potential to be 1.44~\ang,
which is good agreement with the DFT value of 1.53~\ang.
The formation of the covalently-bonded divacancy leaves a two-fold coordinated
atom in each layer, which is electronically unsaturated and could be
chemically active.
For the second-nearest-neighbor divacancy ($V_2^2(\beta\beta)$)
only one bridging bond is formed with a length of 1.40~\ang according
to the \hnn potential. Again there is good agreement with DFT, which
predicts a bond length of 1.38~\ang. As expected the single bond
is stronger than the pair of bonds for the first-nearest-neighbor divacancy
as demonstrated by the shorter bond length in this case.
The $V_2^2(\beta\beta)$ divacancy leaves two two-fold coordinated
atoms in each layer, which reconstruct to form a bond (not shown) with
a bond length predicted to be 1.84~\ang by \hnn and 2.15~\ang by DFT.
(The two atoms are 2.466~\ang away from each other in pristine graphene.)

\begin{figure}[t]
  \includegraphics[width=1\columnwidth]{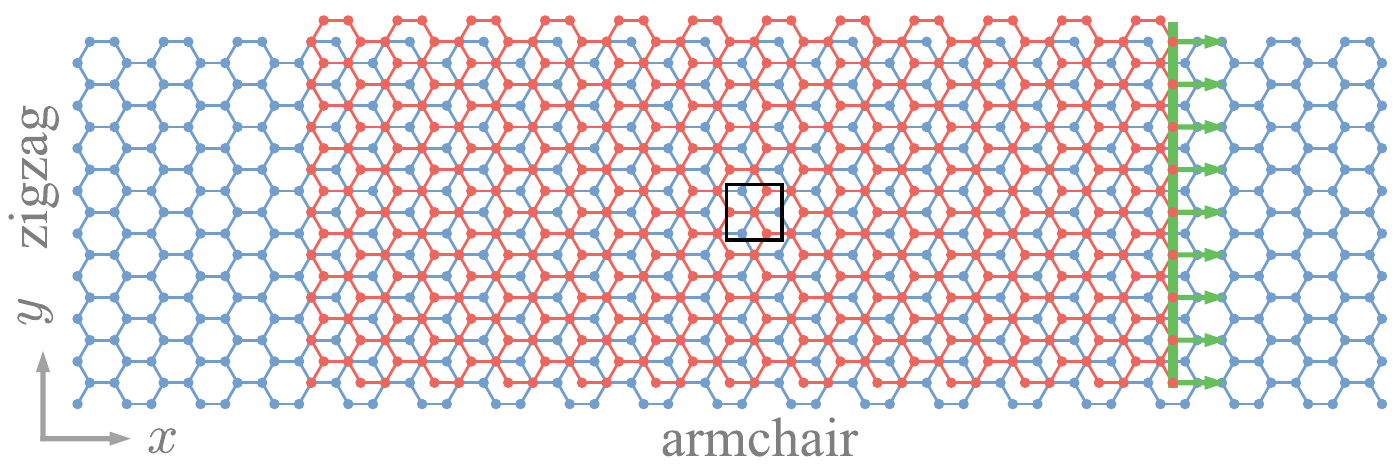}
  \caption{Representation of the simulation supercell used to compute the friction
force in bilayer graphene with and without covalently-bonded divacancies in
adjacent layers. The force required to pull the top layer to the right
along the armchair direction is measured. The black rectangle indicates the location of divacancies when included.}
  \label{fig:pull:setup}
\end{figure}

Next, we measure the interlayer friction force in bilayer graphene with
and without the two types of divacancies. The setup for this simulation
is shown in \fref{fig:pull:setup} for the armchair direction.
A graphene layer (red) is placed on top of a larger layer (blue) and pulled
to the right under displacement control conditions.
The bottom layer has a width of 76.88~\ang (in the $x$ direction) and
height 22.19~\ang (in the $y$ direction) and contains 648 atoms. The top
layer has a width of 49.83~\AA\ and 432 atoms.
When divacancies are included, they are introduced into the center of the bilayer
at the location indicated by the black rectangle in \fref{fig:pull:setup}.
Periodic boundary conditions are applied in the $x$ and $y$
directions, and the direction perpendicular to the plane is free.
Thus the system corresponds to an infinite graphene nanoribbon with
finite width in the $x$-direction (top layer) sliding on an
infinite graphene layer (bottom).
The atoms at the right end of the top layer (green shaded region)
are displaced in the $x$ direction with a step size of 0.1~\ang.
At each step, after applying the displacement to these atoms, the total
energy of the system is minimized subject to the following constraints:
(1) The atoms at the right end of the bottom layer are fixed
in all three directions;
and (2) the $x$ coordinates of the atoms at the right end of the top layer
are fixed to their displaced positions.
Following relaxation, the force $F$ required to hold the top layer in
its displaced position is computed as the total force acting on the
constrained atoms in the top layer. From this the shear stress is computed
as $\tau = F/A$, where $A$ is the area of the top layer. The shear stress
is a more useful property than the force since it can be more readily
compared across systems.

\begin{figure}[h!]
\begin{subfigure}{1\columnwidth}
\includegraphics[width=\columnwidth]{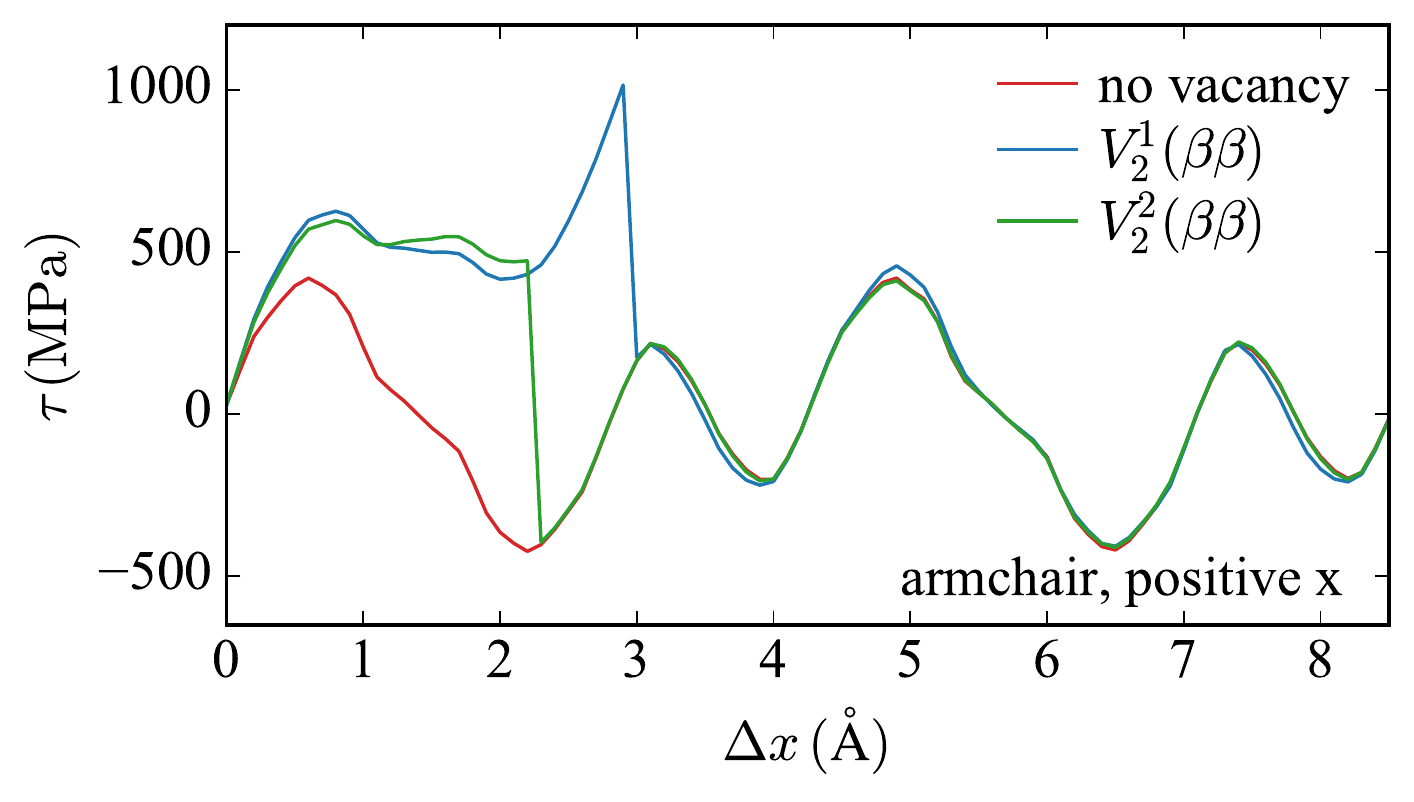}
\caption{\label{fig:fvd:armchair:pos}}
\end{subfigure}
\begin{subfigure}{1\columnwidth}
\includegraphics[width=\columnwidth]{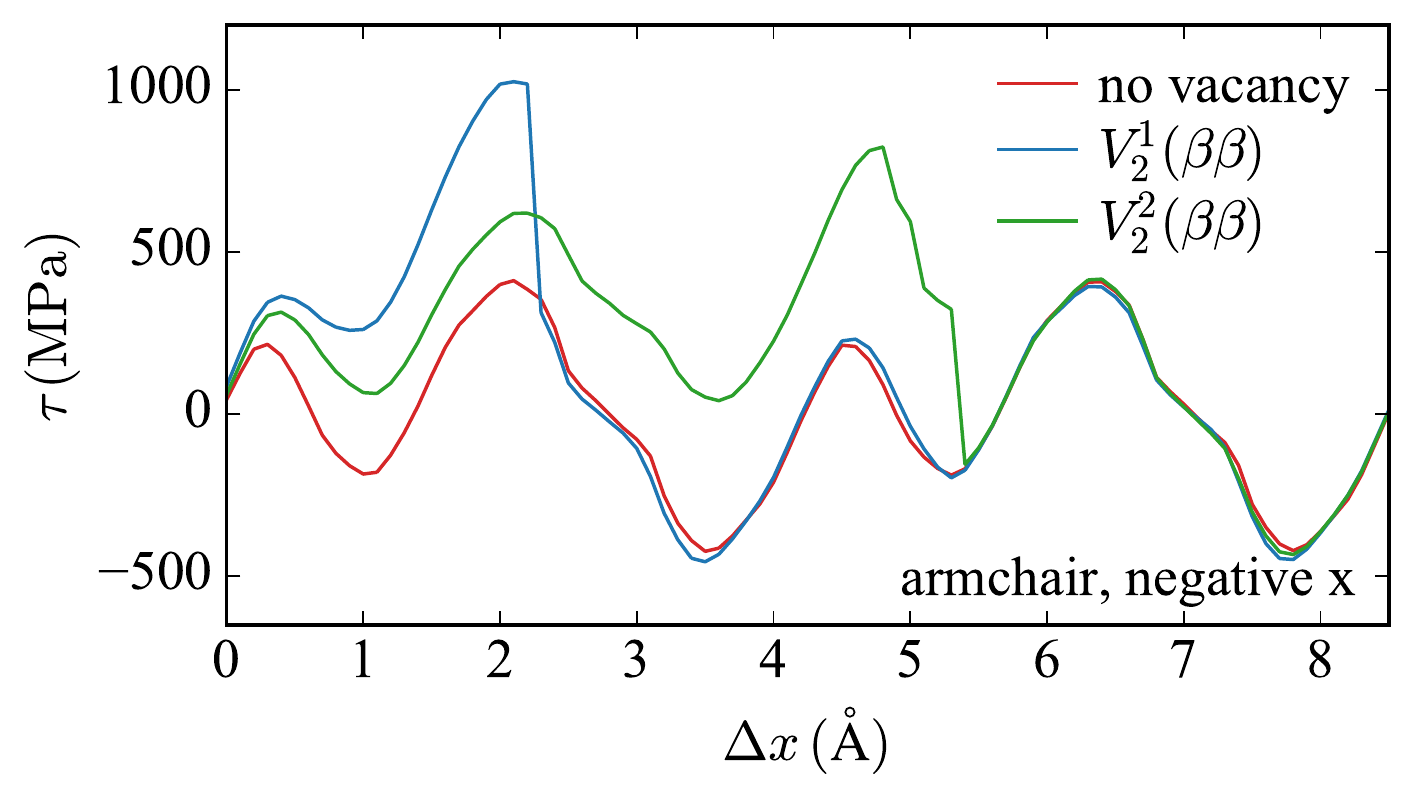}
\caption{\label{fig:fvd:armchair:neg}}
\end{subfigure}
\begin{subfigure}{1\columnwidth}
\includegraphics[width=\columnwidth]{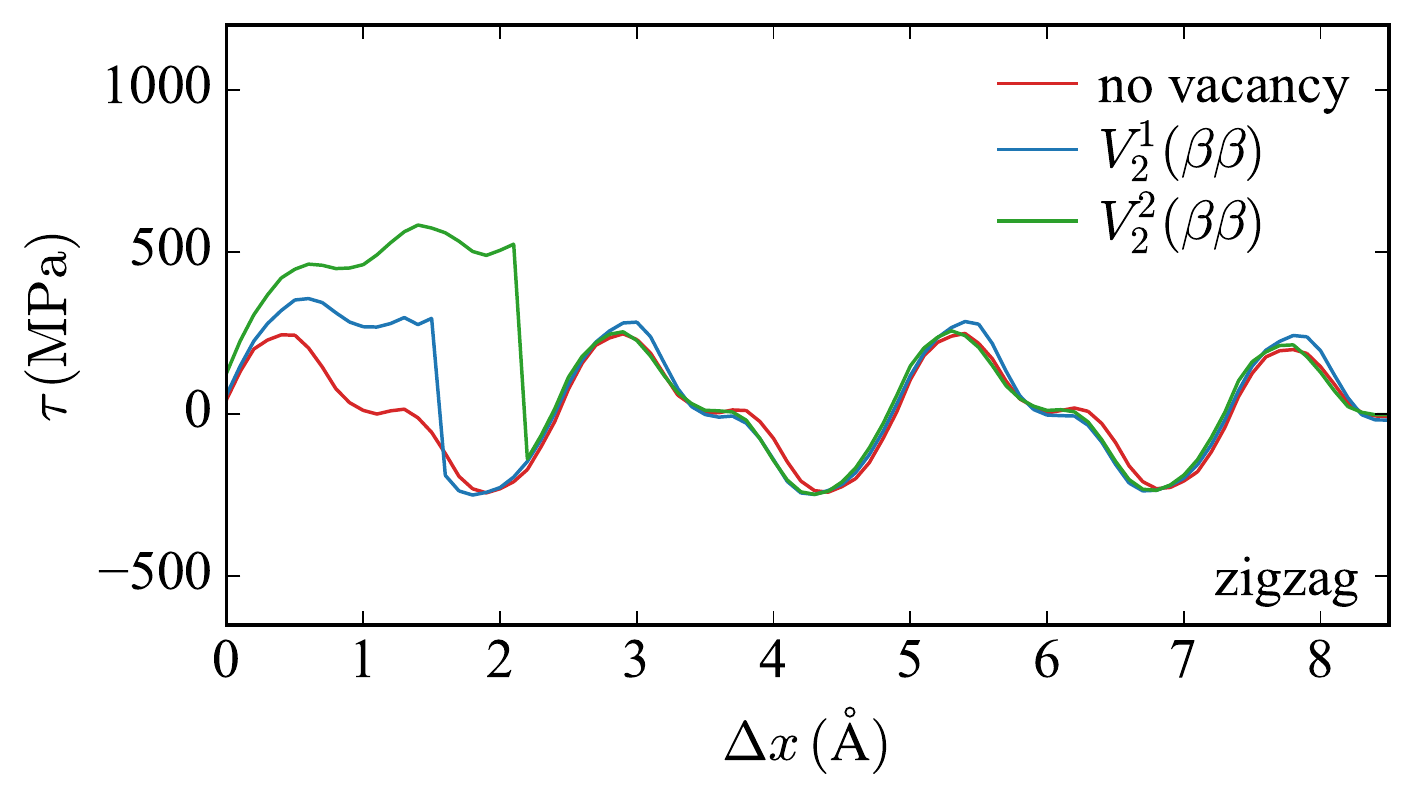}
\caption{\label{fig:fvd:zigzag}}
\end{subfigure}
\caption{Shear stress $\tau$ versus pulling distance $\Delta x$ for bilayer
graphene with and without divacancies. Three different pulling directions
are shown (see \fref{fig:pull:setup}): (a) and (b) armchair edge in the
positive and negative $x$ directions, and (c) zigzag edge in the $y$
direction (positive and negative are the same).}
\end{figure}

\fref{fig:fvd:armchair:pos} shows $\tau$ as a
function of the pulling distance $\Delta x$ along the positive
armchair direction.
For a pristine bilayer without vacancies, the maximum shear stress is
423~MPa at $\Delta x=0.6~\text{\AA}$
with a periodicity of $\sqrt{3}a=4.27~\text{\AA}$ reflecting the underlying
periodic nature of the bilayer structure.
Note that the shear stress is negative once the top layer passes the unstable
equilibrium state where it is balanced between forces pulling it forward
and backwards.
The maximum shear stress for $V_2^1(\beta\beta)$ is 1014~MPa at
$\Delta x =2.9~\text{\AA}$.
The interlayer bond breaks immediately once the
shear stress reaches this maximum, leading to an abrupt drop in the shear stress.
In contrast for the $V_2^2(\beta\beta)$ divacancy,
the interlayer bond does not break at the maximum
shear stress of 597~MPa at $\Delta x =0.8~\text{\AA}$, but instead
breaks later at a somewhat lower shear stress at $\Delta x =2.2~\text{\AA}$.
Once the interlayer bonds are broken, the $V_2^1(\beta\beta)$ and
$V_2^2(\beta\beta)$ curves follow the pristine bilayer curve
almost identically. This suggests that the presence of single vacancies
in the layers (in the absence of interlayer covalent bonding)
has a negligible effect on friction.

We expect the shear stress for pristine graphene to depend
on the pulling direction due to the changing crystallographic
orientation. The effect of the divacancies will also depend on orientation.
For example, referring to \fref{fig:vacancy},
we see that when pulling the top layer to the
right, the single vacancies in $V_2^1(\beta\beta)$ move apart, whereas
when pulling to the left they initially move closer together.
We explore friction anisotropy by considering two more directions
in \fref{fig:pull:setup}:
(1) pulling to the left along the armchair direction,
and (2) pulling upwards along the zigzag direction (downwards is the same
due to symmetry).
In the first case, the simulation setup is the same as in
\fref{fig:pull:setup}, except that the atoms on the left end of the top
layer are pulled in the negative $x$ direction.
In the second case, a bilayer is constructed with similar geometry to
\fref{fig:pull:setup}, but with the zigzag edge aligned with the $x$ direction
and the armchair edge aligned with the $y$ direction.
This system contains 370 atoms in the top layer and 560 atoms in the
bottom layer.

The shear stress versus pulling distance for these two cases are shown in
Figs.~\ref{fig:fvd:armchair:neg} and \ref{fig:fvd:zigzag}.
The results in the negative armchair direction (\fref{fig:fvd:armchair:neg})
are similar to those in the positive armchair direction
(\fref{fig:fvd:armchair:pos}), but with some differences.
The maximum shear stress for pristine graphene is the same
as in \fref{fig:fvd:armchair:pos} due to symmetry, but
for $V_2^1(\beta\beta)$ it is
1018~MPa at $\Delta x=2.2~\text{\AA}$, which is still larger
than that for $V_2^2(\beta\beta)$,
824~MPa at $\Delta x=4.8~\text{\AA}$.
However, in this orientation $V_2^1(\beta\beta)$ breaks earlier
(and immediately as before), whereas $V_2^2(\beta\beta)$ exhibits
a large amount of slip prior to bond failure.
For the zigzag direction in \fref{fig:fvd:zigzag},
the shear stress for pristine
bilayer has a periodicity of 2.466~\ang (smaller than that
in \fref{fig:fvd:armchair:pos} and \fref{fig:fvd:armchair:neg}).
The maximum shear stress for the pristine bilayer,
$V_2^1(\beta\beta)$, and $V_2^2(\beta\beta)$ are 248~MPa, 352~MPa, and
583~MPa, respectively,
all smaller than their counterparts in \fref{fig:fvd:armchair:pos} and \fref{fig:fvd:armchair:neg}.
This direction has the lowest friction resistance.

\section{Summary}
\label{sec:sum}

We have developed a hybrid NN interatomic potential for multilayer
graphene structures called ``\hnn.'' This potential employs an NN to capture the
short-range intralayer covalent bonds and interlayer orbital overlap interactions,
and a theoretically-motivated $r^{-6}$ term to model the long-range
interlayer dispersion.
The inclusion of the theoretical term improves the performance
of the potential since the NN does not need to learn known physics.
The potential parameters are determined by
training against a large dataset of energies and forces for
monolayer graphene, bilayer graphene, and graphite in various states.
The training set is computed from DFT using the PBE functional augmented with
the MBD dispersion correction to account for long-range vdW interactions.

The potential was tested against a variety of structural,
energetic, and elastic properties to which it was not directly fit.
The validation tests show that:
\begin{enumerate}
\item The \hnn potential correctly predicts the in-plane lattice parameter,
equilibrium layer spacings, interlayer binding energies, and generalized
stack fault energies for multilayer graphene structures.
An important feature is that it can distinguish the energies of bilayer
graphene in the AA and AB stacking states.
\item The \hnn potential has good agreement with DFT for the $C_{11}$ and
$C_{12}$ elastic moduli for both graphene and graphite. For the other
elastic moduli of graphite the agreement is reasonable for
$C_{33}$ and $C_{44}$, but poor for $C_{13}$. (We note however that
DFT results are inconsistent with experiments in the latter case.)
\item The phonon dispersion curves calculated from the \hnn potential are in
excellent agreement with DFT result, significantly better than any other
empirical potential, except for GAP--Gr (which is also a machine learning
potential). However GAP--Gr is limited to single-layer graphene.
\end{enumerate}

The \hnn potential was applied to several large-scale applications,
not amenable to DFT calculations.
The thermal conductivity of monolayer graphene with different
vacancy densities is computed using a Green-Kubo approach.
The thermal conductivity of pristine graphene is found to be
2531~W/mK, consistent with experimental measurements (1500--2500 W/mK).
The thermal conductivity is dramatically reduced with the addition
of vacancies due to phonon scattering: 415~W/mK for a vacancy
density of 0.1\%, and 195~W/mK for 0.2\%.

In a second application, the effect of covalent bonds between
layers in bilayer graphene on friction is explored. Such bonds
are predicted to occur when vacancies in separate layers exist
in close proximity and the bilayer is compressed.
The \hnn potential predicts the formation of interlayer
covalent bonds and a corresponding divacancy structure
in agreement with DFT. It is found that the presence
of these bonds increases the friction between layers by up to
a factor of four depending on the sliding direction.

We have shown that the new \hnn potential provides a
complete and accurate description of both the
intralayer and interlayer interactions in multilayer graphene
structures.  It can be used to study mechanical and thermal
properties of these materials, and investigate the effects of vacancy defects.
Unlike interlayer potentials like KC \cite{kolmogorov2005registry} and
DRIP \cite{wen2018dihedral} this potential does not assign
atoms membership to layers or assume a layered structure to
characterized the registry geometry. Thus, for example, \hnn
could be used to model passage of atoms between layers.

The \hnn potential is compatible with the KIM API \cite{tadmor2011kim}
and available for download from \url{https://openkim.org}
\cite{MD_435082866799_001, MO_421038499185_001}. This potential
can be used with any KIM-compliant atomistic simulation code.
(For more details on KIM, and an example of how to use the \hnn
potential in LAMMPS to compute the cohesive energy of a graphene bilayer
in AB stacking, see the SM \cite{supplemental}.)

\begin{acknowledgments}
This research was partly supported by the Army Research Office (W911NF-14-1-0247)
under the MURI program, the National Science Foundation (NSF) under grants
No.~DMR-1834251 and DMR-1834332, and through the University of Minnesota MRSEC under Award Number DMR-1420013.
The authors wish to acknowledge the Minnesota Supercomputing Institute (MSI) at the
University of Minnesota for providing resources that contributed to the results
reported in this paper.
We thank Efthimios Kaxiras and Ryan Elliott for helpful discussion.
MW thanks the University of Minnesota Doctoral Dissertation Fellowship for
supporting his research.
\end{acknowledgments}



\begin{thebibliography}{98}%
\makeatletter
\providecommand \@ifxundefined [1]{%
 \@ifx{#1\undefined}
}%
\providecommand \@ifnum [1]{%
 \ifnum #1\expandafter \@firstoftwo
 \else \expandafter \@secondoftwo
 \fi
}%
\providecommand \@ifx [1]{%
 \ifx #1\expandafter \@firstoftwo
 \else \expandafter \@secondoftwo
 \fi
}%
\providecommand \natexlab [1]{#1}%
\providecommand \enquote  [1]{``#1''}%
\providecommand \bibnamefont  [1]{#1}%
\providecommand \bibfnamefont [1]{#1}%
\providecommand \citenamefont [1]{#1}%
\providecommand \href@noop [0]{\@secondoftwo}%
\providecommand \href [0]{\begingroup \@sanitize@url \@href}%
\providecommand \@href[1]{\@@startlink{#1}\@@href}%
\providecommand \@@href[1]{\endgroup#1\@@endlink}%
\providecommand \@sanitize@url [0]{\catcode `\\12\catcode `\$12\catcode
  `\&12\catcode `\#12\catcode `\^12\catcode `\_12\catcode `\%12\relax}%
\providecommand \@@startlink[1]{}%
\providecommand \@@endlink[0]{}%
\providecommand \url  [0]{\begingroup\@sanitize@url \@url }%
\providecommand \@url [1]{\endgroup\@href {#1}{\urlprefix }}%
\providecommand \urlprefix  [0]{URL }%
\providecommand \Eprint [0]{\href }%
\providecommand \doibase [0]{https://doi.org/}%
\providecommand \selectlanguage [0]{\@gobble}%
\providecommand \bibinfo  [0]{\@secondoftwo}%
\providecommand \bibfield  [0]{\@secondoftwo}%
\providecommand \translation [1]{[#1]}%
\providecommand \BibitemOpen [0]{}%
\providecommand \bibitemStop [0]{}%
\providecommand \bibitemNoStop [0]{.\EOS\space}%
\providecommand \EOS [0]{\spacefactor3000\relax}%
\providecommand \BibitemShut  [1]{\csname bibitem#1\endcsname}%
\let\auto@bib@innerbib\@empty
\bibitem [{\citenamefont {Novoselov}\ \emph {et~al.}(2004)\citenamefont
  {Novoselov}, \citenamefont {Geim}, \citenamefont {Morozov}, \citenamefont
  {Jiang}, \citenamefont {Zhang}, \citenamefont {Dubonos}, \citenamefont
  {Grigorieva},\ and\ \citenamefont {Firsov}}]{novoselov2004electric}%
  \BibitemOpen
  \bibfield  {author} {\bibinfo {author} {\bibfnamefont {K.~S.}\ \bibnamefont
  {Novoselov}}, \bibinfo {author} {\bibfnamefont {A.~K.}\ \bibnamefont {Geim}},
  \bibinfo {author} {\bibfnamefont {S.~V.}\ \bibnamefont {Morozov}}, \bibinfo
  {author} {\bibfnamefont {D.}~\bibnamefont {Jiang}}, \bibinfo {author}
  {\bibfnamefont {Y.}~\bibnamefont {Zhang}}, \bibinfo {author} {\bibfnamefont
  {S.~V.}\ \bibnamefont {Dubonos}}, \bibinfo {author} {\bibfnamefont {I.~V.}\
  \bibnamefont {Grigorieva}},\ and\ \bibinfo {author} {\bibfnamefont {A.~A.}\
  \bibnamefont {Firsov}},\ }\bibfield  {title} {\bibinfo {title} {Electric
  field effect in atomically thin carbon films},\ }\href
  {https://doi.org/10.1126/science.1102896} {\bibfield  {journal} {\bibinfo
  {journal} {Science}\ }\textbf {\bibinfo {volume} {306}},\ \bibinfo {pages}
  {666} (\bibinfo {year} {2004})}\BibitemShut {NoStop}%
\bibitem [{\citenamefont {Neto}\ \emph {et~al.}(2009)\citenamefont {Neto},
  \citenamefont {Guinea}, \citenamefont {Peres}, \citenamefont {Novoselov},\
  and\ \citenamefont {Geim}}]{neto2009electronic}%
  \BibitemOpen
  \bibfield  {author} {\bibinfo {author} {\bibfnamefont {A.~H.~C.}\
  \bibnamefont {Neto}}, \bibinfo {author} {\bibfnamefont {F.}~\bibnamefont
  {Guinea}}, \bibinfo {author} {\bibfnamefont {N.~M.~R.}\ \bibnamefont
  {Peres}}, \bibinfo {author} {\bibfnamefont {K.~S.}\ \bibnamefont
  {Novoselov}},\ and\ \bibinfo {author} {\bibfnamefont {A.~K.}\ \bibnamefont
  {Geim}},\ }\bibfield  {title} {\bibinfo {title} {The electronic properties of
  graphene},\ }\href {https://doi.org/10.1103/revmodphys.81.109} {\bibfield
  {journal} {\bibinfo  {journal} {Rev. Mod. Phys.}\ }\textbf {\bibinfo {volume}
  {81}},\ \bibinfo {pages} {109} (\bibinfo {year} {2009})}\BibitemShut
  {NoStop}%
\bibitem [{\citenamefont {Hendry}\ \emph {et~al.}(2010)\citenamefont {Hendry},
  \citenamefont {Hale}, \citenamefont {Moger}, \citenamefont {Savchenko},\ and\
  \citenamefont {Mikhailov}}]{hendry2010coherent}%
  \BibitemOpen
  \bibfield  {author} {\bibinfo {author} {\bibfnamefont {E.}~\bibnamefont
  {Hendry}}, \bibinfo {author} {\bibfnamefont {P.~J.}\ \bibnamefont {Hale}},
  \bibinfo {author} {\bibfnamefont {J.}~\bibnamefont {Moger}}, \bibinfo
  {author} {\bibfnamefont {A.~K.}\ \bibnamefont {Savchenko}},\ and\ \bibinfo
  {author} {\bibfnamefont {S.~A.}\ \bibnamefont {Mikhailov}},\ }\bibfield
  {title} {\bibinfo {title} {Coherent nonlinear optical response of graphene},\
  }\href {https://doi.org/10.1103/physrevlett.105.097401} {\bibfield  {journal}
  {\bibinfo  {journal} {Phys. Rev. Lett.}\ }\textbf {\bibinfo {volume} {105}},\
  \bibinfo {pages} {097401} (\bibinfo {year} {2010})}\BibitemShut {NoStop}%
\bibitem [{\citenamefont {Sevik}(2014)}]{sevik2014assessment}%
  \BibitemOpen
  \bibfield  {author} {\bibinfo {author} {\bibfnamefont {C.}~\bibnamefont
  {Sevik}},\ }\bibfield  {title} {\bibinfo {title} {Assessment on lattice
  thermal properties of two-dimensional honeycomb structures:
  Graphene,h-{BN},h-{MoS}2, andh-{MoSe}2},\ }\href
  {https://doi.org/10.1103/PhysRevB.89.035422} {\bibfield  {journal} {\bibinfo
  {journal} {Phys. Rev. B}\ }\textbf {\bibinfo {volume} {89}},\ \bibinfo
  {pages} {035422} (\bibinfo {year} {2014})}\BibitemShut {NoStop}%
\bibitem [{\citenamefont {Lee}\ \emph {et~al.}(2008)\citenamefont {Lee},
  \citenamefont {Wei}, \citenamefont {Kysar},\ and\ \citenamefont
  {Hone}}]{lee2008measurement}%
  \BibitemOpen
  \bibfield  {author} {\bibinfo {author} {\bibfnamefont {C.}~\bibnamefont
  {Lee}}, \bibinfo {author} {\bibfnamefont {X.}~\bibnamefont {Wei}}, \bibinfo
  {author} {\bibfnamefont {J.~W.}\ \bibnamefont {Kysar}},\ and\ \bibinfo
  {author} {\bibfnamefont {J.}~\bibnamefont {Hone}},\ }\bibfield  {title}
  {\bibinfo {title} {Measurement of the elastic properties and intrinsic
  strength of monolayer graphene},\ }\href
  {https://doi.org/10.1126/science.1157996} {\bibfield  {journal} {\bibinfo
  {journal} {Science}\ }\textbf {\bibinfo {volume} {321}},\ \bibinfo {pages}
  {385} (\bibinfo {year} {2008})}\BibitemShut {NoStop}%
\bibitem [{\citenamefont {Geim}\ and\ \citenamefont
  {Grigorieva}(2013)}]{geim2013van}%
  \BibitemOpen
  \bibfield  {author} {\bibinfo {author} {\bibfnamefont {A.~K.}\ \bibnamefont
  {Geim}}\ and\ \bibinfo {author} {\bibfnamefont {I.~V.}\ \bibnamefont
  {Grigorieva}},\ }\bibfield  {title} {\bibinfo {title} {Van der waals
  heterostructures},\ }\href {https://doi.org/10.1038/nature12385} {\bibfield
  {journal} {\bibinfo  {journal} {Nature}\ }\textbf {\bibinfo {volume} {499}},\
  \bibinfo {pages} {419} (\bibinfo {year} {2013})}\BibitemShut {NoStop}%
\bibitem [{\citenamefont {Novoselov}\ \emph {et~al.}(2016)\citenamefont
  {Novoselov}, \citenamefont {Mishchenko}, \citenamefont {Carvalho},\ and\
  \citenamefont {Neto}}]{novoselov20162d}%
  \BibitemOpen
  \bibfield  {author} {\bibinfo {author} {\bibfnamefont {K.~S.}\ \bibnamefont
  {Novoselov}}, \bibinfo {author} {\bibfnamefont {A.}~\bibnamefont
  {Mishchenko}}, \bibinfo {author} {\bibfnamefont {A.}~\bibnamefont
  {Carvalho}},\ and\ \bibinfo {author} {\bibfnamefont {A.~H.~C.}\ \bibnamefont
  {Neto}},\ }\bibfield  {title} {\bibinfo {title} {2d materials and van der
  waals heterostructures},\ }\href {https://doi.org/10.1126/science.aac9439}
  {\bibfield  {journal} {\bibinfo  {journal} {Science}\ }\textbf {\bibinfo
  {volume} {353}},\ \bibinfo {pages} {aac9439} (\bibinfo {year}
  {2016})}\BibitemShut {NoStop}%
\bibitem [{\citenamefont {Cao}\ \emph {et~al.}(2018)\citenamefont {Cao},
  \citenamefont {Fatemi}, \citenamefont {Fang}, \citenamefont {Watanabe},
  \citenamefont {Taniguchi}, \citenamefont {Kaxiras},\ and\ \citenamefont
  {Jarillo-Herrero}}]{graphene_sc_2018}%
  \BibitemOpen
  \bibfield  {author} {\bibinfo {author} {\bibfnamefont {Y.}~\bibnamefont
  {Cao}}, \bibinfo {author} {\bibfnamefont {V.}~\bibnamefont {Fatemi}},
  \bibinfo {author} {\bibfnamefont {S.}~\bibnamefont {Fang}}, \bibinfo {author}
  {\bibfnamefont {K.}~\bibnamefont {Watanabe}}, \bibinfo {author}
  {\bibfnamefont {T.}~\bibnamefont {Taniguchi}}, \bibinfo {author}
  {\bibfnamefont {E.}~\bibnamefont {Kaxiras}},\ and\ \bibinfo {author}
  {\bibfnamefont {P.}~\bibnamefont {Jarillo-Herrero}},\ }\bibfield  {title}
  {\bibinfo {title} {Unconventional superconductivity in magic-angle graphene
  superlattices},\ }\href {https://doi.org/10.1038/nature26160} {\bibfield
  {journal} {\bibinfo  {journal} {Nature}\ }\textbf {\bibinfo {volume} {556}},\
  \bibinfo {pages} {43} (\bibinfo {year} {2018})}\BibitemShut {NoStop}%
\bibitem [{\citenamefont {Mishin}\ \emph {et~al.}(1999)\citenamefont {Mishin},
  \citenamefont {Farkas}, \citenamefont {Mehl},\ and\ \citenamefont
  {Papaconstantopoulos}}]{mishin1999interatomic}%
  \BibitemOpen
  \bibfield  {author} {\bibinfo {author} {\bibfnamefont {Y.}~\bibnamefont
  {Mishin}}, \bibinfo {author} {\bibfnamefont {D.}~\bibnamefont {Farkas}},
  \bibinfo {author} {\bibfnamefont {M.~J.}\ \bibnamefont {Mehl}},\ and\
  \bibinfo {author} {\bibfnamefont {D.~A.}\ \bibnamefont
  {Papaconstantopoulos}},\ }\bibfield  {title} {\bibinfo {title} {Interatomic
  potentials for monoatomic metals from experimental data and ab initio
  calculations},\ }\href {https://doi.org/10.1103/physrevb.59.3393} {\bibfield
  {journal} {\bibinfo  {journal} {Phys. Rev. B}\ }\textbf {\bibinfo {volume}
  {59}},\ \bibinfo {pages} {3393} (\bibinfo {year} {1999})}\BibitemShut
  {NoStop}%
\bibitem [{\citenamefont {Wen}\ \emph {et~al.}(2015)\citenamefont {Wen},
  \citenamefont {Whalen}, \citenamefont {Elliott},\ and\ \citenamefont
  {Tadmor}}]{wen2015interpolation}%
  \BibitemOpen
  \bibfield  {author} {\bibinfo {author} {\bibfnamefont {M.}~\bibnamefont
  {Wen}}, \bibinfo {author} {\bibfnamefont {S.~M.}\ \bibnamefont {Whalen}},
  \bibinfo {author} {\bibfnamefont {R.~S.}\ \bibnamefont {Elliott}},\ and\
  \bibinfo {author} {\bibfnamefont {E.~B.}\ \bibnamefont {Tadmor}},\ }\bibfield
   {title} {\bibinfo {title} {Interpolation effects in tabulated interatomic
  potentials},\ }\href {https://doi.org/10.1088/0965-0393/23/7/074008}
  {\bibfield  {journal} {\bibinfo  {journal} {Modell. Simul. Mater. Sci. Eng.}\
  }\textbf {\bibinfo {volume} {23}},\ \bibinfo {pages} {074008} (\bibinfo
  {year} {2015})}\BibitemShut {NoStop}%
\bibitem [{\citenamefont {Wen}\ \emph {et~al.}(2017)\citenamefont {Wen},
  \citenamefont {Li}, \citenamefont {Brommer}, \citenamefont {Elliott},
  \citenamefont {Sethna},\ and\ \citenamefont {Tadmor}}]{wen2017potfit}%
  \BibitemOpen
  \bibfield  {author} {\bibinfo {author} {\bibfnamefont {M.}~\bibnamefont
  {Wen}}, \bibinfo {author} {\bibfnamefont {J.}~\bibnamefont {Li}}, \bibinfo
  {author} {\bibfnamefont {P.}~\bibnamefont {Brommer}}, \bibinfo {author}
  {\bibfnamefont {R.~S.}\ \bibnamefont {Elliott}}, \bibinfo {author}
  {\bibfnamefont {J.~P.}\ \bibnamefont {Sethna}},\ and\ \bibinfo {author}
  {\bibfnamefont {E.~B.}\ \bibnamefont {Tadmor}},\ }\bibfield  {title}
  {\bibinfo {title} {A {KIM}-compliantpotfitfor fitting sloppy interatomic
  potentials: application to the {EDIP} model for silicon},\ }\href
  {https://doi.org/10.1088/0965-0393/25/1/014001} {\bibfield  {journal}
  {\bibinfo  {journal} {Modell. Simul. Mater. Sci. Eng.}\ }\textbf {\bibinfo
  {volume} {25}},\ \bibinfo {pages} {014001} (\bibinfo {year}
  {2017})}\BibitemShut {NoStop}%
\bibitem [{\citenamefont {Kolmogorov}\ and\ \citenamefont
  {Crespi}(2005)}]{kolmogorov2005registry}%
  \BibitemOpen
  \bibfield  {author} {\bibinfo {author} {\bibfnamefont {A.~N.}\ \bibnamefont
  {Kolmogorov}}\ and\ \bibinfo {author} {\bibfnamefont {V.~H.}\ \bibnamefont
  {Crespi}},\ }\bibfield  {title} {\bibinfo {title} {Registry-dependent
  interlayer potential for graphitic systems},\ }\href
  {https://doi.org/10.1103/physrevb.71.235415} {\bibfield  {journal} {\bibinfo
  {journal} {Phys. Rev. B}\ }\textbf {\bibinfo {volume} {71}},\ \bibinfo
  {pages} {235415} (\bibinfo {year} {2005})}\BibitemShut {NoStop}%
\bibitem [{\citenamefont {Zhang}\ and\ \citenamefont
  {Tadmor}(2017)}]{zhang:tadmor:2017}%
  \BibitemOpen
  \bibfield  {author} {\bibinfo {author} {\bibfnamefont {K.}~\bibnamefont
  {Zhang}}\ and\ \bibinfo {author} {\bibfnamefont {E.~B.}\ \bibnamefont
  {Tadmor}},\ }\bibfield  {title} {\bibinfo {title} {Energy and moir\'e
  patterns in 2{D} bilayers in translation and rotation: {A} study using an
  efficient discrete--continuum interlayer potential},\ }\href
  {https://doi.org/10.1016/j.eml.2016.10.010} {\bibfield  {journal} {\bibinfo
  {journal} {Extreme Mech. Lett.}\ }\textbf {\bibinfo {volume} {14}},\ \bibinfo
  {pages} {16} (\bibinfo {year} {2017})}\BibitemShut {NoStop}%
\bibitem [{\citenamefont {Zhang}\ and\ \citenamefont
  {Tadmor}(2018)}]{zhang:tadmor:2018}%
  \BibitemOpen
  \bibfield  {author} {\bibinfo {author} {\bibfnamefont {K.}~\bibnamefont
  {Zhang}}\ and\ \bibinfo {author} {\bibfnamefont {E.~B.}\ \bibnamefont
  {Tadmor}},\ }\bibfield  {title} {\bibinfo {title} {Structural and electron
  diffraction scaling of twisted graphene bilayers},\ }\href
  {https://doi.org/10.1016/j.jmps.2017.12.005} {\bibfield  {journal} {\bibinfo
  {journal} {J. Mech. Phys. Solids}\ }\textbf {\bibinfo {volume} {112}},\
  \bibinfo {pages} {225} (\bibinfo {year} {2018})}\BibitemShut {NoStop}%
\bibitem [{\citenamefont {Yoo}\ \emph {et~al.}(2019)\citenamefont {Yoo},
  \citenamefont {Engelke}, \citenamefont {Carr}, \citenamefont {Fang},
  \citenamefont {Zhang}, \citenamefont {Cazeaux}, \citenamefont {Sung},
  \citenamefont {Hovden}, \citenamefont {Tsen}, \citenamefont {Taniguchi},
  \citenamefont {Watanabe}, \citenamefont {Yi}, \citenamefont {Kim},
  \citenamefont {Luskin}, \citenamefont {Tadmor}, \citenamefont {Kaxiras},\
  and\ \citenamefont {Kim}}]{yoo:engelke:2019}%
  \BibitemOpen
  \bibfield  {author} {\bibinfo {author} {\bibfnamefont {H.}~\bibnamefont
  {Yoo}}, \bibinfo {author} {\bibfnamefont {R.}~\bibnamefont {Engelke}},
  \bibinfo {author} {\bibfnamefont {S.}~\bibnamefont {Carr}}, \bibinfo {author}
  {\bibfnamefont {S.}~\bibnamefont {Fang}}, \bibinfo {author} {\bibfnamefont
  {K.}~\bibnamefont {Zhang}}, \bibinfo {author} {\bibfnamefont
  {P.}~\bibnamefont {Cazeaux}}, \bibinfo {author} {\bibfnamefont {S.~H.}\
  \bibnamefont {Sung}}, \bibinfo {author} {\bibfnamefont {R.}~\bibnamefont
  {Hovden}}, \bibinfo {author} {\bibfnamefont {A.}~\bibnamefont {Tsen}},
  \bibinfo {author} {\bibfnamefont {T.}~\bibnamefont {Taniguchi}}, \bibinfo
  {author} {\bibfnamefont {K.}~\bibnamefont {Watanabe}}, \bibinfo {author}
  {\bibfnamefont {G.-C.}\ \bibnamefont {Yi}}, \bibinfo {author} {\bibfnamefont
  {M.}~\bibnamefont {Kim}}, \bibinfo {author} {\bibfnamefont {M.}~\bibnamefont
  {Luskin}}, \bibinfo {author} {\bibfnamefont {E.~B.}\ \bibnamefont {Tadmor}},
  \bibinfo {author} {\bibfnamefont {E.}~\bibnamefont {Kaxiras}},\ and\ \bibinfo
  {author} {\bibfnamefont {P.}~\bibnamefont {Kim}},\ }\bibfield  {title}
  {\bibinfo {title} {Atomic and electronic reconstruction at the van der waals
  interface in twisted bilayer graphene},\ }\href
  {https://doi.org/10.1038/s41563-019-0346-z} {\bibfield  {journal} {\bibinfo
  {journal} {Nat. Mater.}\ }\textbf {\bibinfo {volume} {18}},\ \bibinfo {pages}
  {448} (\bibinfo {year} {2019})}\BibitemShut {NoStop}%
\bibitem [{\citenamefont {Tersoff}(1988)}]{tersoff1988empirical}%
  \BibitemOpen
  \bibfield  {author} {\bibinfo {author} {\bibfnamefont {J.}~\bibnamefont
  {Tersoff}},\ }\bibfield  {title} {\bibinfo {title} {Empirical interatomic
  potential for carbon, with applications to amorphous carbon},\ }\href
  {https://doi.org/10.1103/physrevlett.61.2879} {\bibfield  {journal} {\bibinfo
   {journal} {Phys. Rev. Lett.}\ }\textbf {\bibinfo {volume} {61}},\ \bibinfo
  {pages} {2879} (\bibinfo {year} {1988})}\BibitemShut {NoStop}%
\bibitem [{\citenamefont {Tersoff}(1989)}]{tersoff1989modeling}%
  \BibitemOpen
  \bibfield  {author} {\bibinfo {author} {\bibfnamefont {J.}~\bibnamefont
  {Tersoff}},\ }\bibfield  {title} {\bibinfo {title} {Modeling solid-state
  chemistry: Interatomic potentials for multicomponent systems},\ }\href
  {https://doi.org/10.1103/physrevb.39.5566} {\bibfield  {journal} {\bibinfo
  {journal} {Phys. Rev. B}\ }\textbf {\bibinfo {volume} {39}},\ \bibinfo
  {pages} {5566} (\bibinfo {year} {1989})}\BibitemShut {NoStop}%
\bibitem [{\citenamefont {Brenner}(1990)}]{brenner1990physical}%
  \BibitemOpen
  \bibfield  {author} {\bibinfo {author} {\bibfnamefont {D.~W.}\ \bibnamefont
  {Brenner}},\ }\bibfield  {title} {\bibinfo {title} {Empirical potential for
  hydrocarbons for use in simulating the chemical vapor deposition of diamond
  films},\ }\href {https://doi.org/10.1103/physrevb.42.9458} {\bibfield
  {journal} {\bibinfo  {journal} {Phys. Rev. B}\ }\textbf {\bibinfo {volume}
  {42}},\ \bibinfo {pages} {9458} (\bibinfo {year} {1990})}\BibitemShut
  {NoStop}%
\bibitem [{\citenamefont {Brenner}\ \emph {et~al.}(2002)\citenamefont
  {Brenner}, \citenamefont {Shenderova}, \citenamefont {Harrison},
  \citenamefont {Stuart}, \citenamefont {Ni},\ and\ \citenamefont
  {Sinnott}}]{brenner2002rebo}%
  \BibitemOpen
  \bibfield  {author} {\bibinfo {author} {\bibfnamefont {D.~W.}\ \bibnamefont
  {Brenner}}, \bibinfo {author} {\bibfnamefont {O.~A.}\ \bibnamefont
  {Shenderova}}, \bibinfo {author} {\bibfnamefont {J.~A.}\ \bibnamefont
  {Harrison}}, \bibinfo {author} {\bibfnamefont {S.~J.}\ \bibnamefont
  {Stuart}}, \bibinfo {author} {\bibfnamefont {B.}~\bibnamefont {Ni}},\ and\
  \bibinfo {author} {\bibfnamefont {S.~B.}\ \bibnamefont {Sinnott}},\
  }\bibfield  {title} {\bibinfo {title} {A second-generation reactive empirical
  bond order ({REBO}) potential energy expression for hydrocarbons},\ }\href
  {https://doi.org/10.1088/0953-8984/14/4/312} {\bibfield  {journal} {\bibinfo
  {journal} {J. Phys.: Condens. Matter}\ }\textbf {\bibinfo {volume} {14}},\
  \bibinfo {pages} {783} (\bibinfo {year} {2002})}\BibitemShut {NoStop}%
\bibitem [{\citenamefont {Stuart}\ \emph {et~al.}(2000)\citenamefont {Stuart},
  \citenamefont {Tutein},\ and\ \citenamefont {Harrison}}]{stuart2000airebo}%
  \BibitemOpen
  \bibfield  {author} {\bibinfo {author} {\bibfnamefont {S.~J.}\ \bibnamefont
  {Stuart}}, \bibinfo {author} {\bibfnamefont {A.~B.}\ \bibnamefont {Tutein}},\
  and\ \bibinfo {author} {\bibfnamefont {J.~A.}\ \bibnamefont {Harrison}},\
  }\bibfield  {title} {\bibinfo {title} {A reactive potential for hydrocarbons
  with intermolecular interactions},\ }\href {https://doi.org/10.1063/1.481208}
  {\bibfield  {journal} {\bibinfo  {journal} {J. Chem. Phys.}\ }\textbf
  {\bibinfo {volume} {112}},\ \bibinfo {pages} {6472} (\bibinfo {year}
  {2000})}\BibitemShut {NoStop}%
\bibitem [{\citenamefont {Lennard-Jones}(1931)}]{lennardjones1931}%
  \BibitemOpen
  \bibfield  {author} {\bibinfo {author} {\bibfnamefont {J.~E.}\ \bibnamefont
  {Lennard-Jones}},\ }\bibfield  {title} {\bibinfo {title} {Cohesion},\ }\href
  {https://doi.org/10.1088/0959-5309/43/5/301} {\bibfield  {journal} {\bibinfo
  {journal} {Proc. Phys. Soc.}\ }\textbf {\bibinfo {volume} {43}},\ \bibinfo
  {pages} {461} (\bibinfo {year} {1931})}\BibitemShut {NoStop}%
\bibitem [{\citenamefont {Los}\ and\ \citenamefont
  {Fasolino}(2003)}]{los2003intrinsic}%
  \BibitemOpen
  \bibfield  {author} {\bibinfo {author} {\bibfnamefont {J.~H.}\ \bibnamefont
  {Los}}\ and\ \bibinfo {author} {\bibfnamefont {A.}~\bibnamefont {Fasolino}},\
  }\bibfield  {title} {\bibinfo {title} {Intrinsic long-range bond-order
  potential for carbon: Performance in monte carlo simulations of
  graphitization},\ }\href {https://doi.org/10.1103/physrevb.68.024107}
  {\bibfield  {journal} {\bibinfo  {journal} {Phys. Rev. B}\ }\textbf {\bibinfo
  {volume} {68}},\ \bibinfo {pages} {024107} (\bibinfo {year}
  {2003})}\BibitemShut {NoStop}%
\bibitem [{\citenamefont {O'Connor}\ \emph {et~al.}(2015)\citenamefont
  {O'Connor}, \citenamefont {Andzelm},\ and\ \citenamefont
  {Robbins}}]{o2015airebo}%
  \BibitemOpen
  \bibfield  {author} {\bibinfo {author} {\bibfnamefont {T.~C.}\ \bibnamefont
  {O'Connor}}, \bibinfo {author} {\bibfnamefont {J.}~\bibnamefont {Andzelm}},\
  and\ \bibinfo {author} {\bibfnamefont {M.~O.}\ \bibnamefont {Robbins}},\
  }\bibfield  {title} {\bibinfo {title} {{AIREBO}-m: A reactive model for
  hydrocarbons at extreme pressures},\ }\href
  {https://doi.org/10.1063/1.4905549} {\bibfield  {journal} {\bibinfo
  {journal} {J. Chem. Phys.}\ }\textbf {\bibinfo {volume} {142}},\ \bibinfo
  {pages} {024903} (\bibinfo {year} {2015})}\BibitemShut {NoStop}%
\bibitem [{\citenamefont {Morse}(1929)}]{morse1929diatomic}%
  \BibitemOpen
  \bibfield  {author} {\bibinfo {author} {\bibfnamefont {P.~M.}\ \bibnamefont
  {Morse}},\ }\bibfield  {title} {\bibinfo {title} {Diatomic molecules
  according to the wave mechanics. {II}. vibrational levels},\ }\href
  {https://doi.org/10.1103/physrev.34.57} {\bibfield  {journal} {\bibinfo
  {journal} {Phys. Rev.}\ }\textbf {\bibinfo {volume} {34}},\ \bibinfo {pages}
  {57} (\bibinfo {year} {1929})}\BibitemShut {NoStop}%
\bibitem [{\citenamefont {Srinivasan}\ \emph {et~al.}(2015)\citenamefont
  {Srinivasan}, \citenamefont {van Duin},\ and\ \citenamefont
  {Ganesh}}]{srinivasan2015development}%
  \BibitemOpen
  \bibfield  {author} {\bibinfo {author} {\bibfnamefont {S.~G.}\ \bibnamefont
  {Srinivasan}}, \bibinfo {author} {\bibfnamefont {A.~C.~T.}\ \bibnamefont {van
  Duin}},\ and\ \bibinfo {author} {\bibfnamefont {P.}~\bibnamefont {Ganesh}},\
  }\bibfield  {title} {\bibinfo {title} {Development of a {ReaxFF} potential
  for carbon condensed phases and its application to the thermal fragmentation
  of a large fullerene},\ }\href {https://doi.org/10.1021/jp510274e} {\bibfield
   {journal} {\bibinfo  {journal} {J. Phys. Chem. A}\ }\textbf {\bibinfo
  {volume} {119}},\ \bibinfo {pages} {571} (\bibinfo {year}
  {2015})}\BibitemShut {NoStop}%
\bibitem [{\citenamefont {Wen}\ \emph {et~al.}(2018)\citenamefont {Wen},
  \citenamefont {Carr}, \citenamefont {Fang}, \citenamefont {Kaxiras},\ and\
  \citenamefont {Tadmor}}]{wen2018dihedral}%
  \BibitemOpen
  \bibfield  {author} {\bibinfo {author} {\bibfnamefont {M.}~\bibnamefont
  {Wen}}, \bibinfo {author} {\bibfnamefont {S.}~\bibnamefont {Carr}}, \bibinfo
  {author} {\bibfnamefont {S.}~\bibnamefont {Fang}}, \bibinfo {author}
  {\bibfnamefont {E.}~\bibnamefont {Kaxiras}},\ and\ \bibinfo {author}
  {\bibfnamefont {E.~B.}\ \bibnamefont {Tadmor}},\ }\bibfield  {title}
  {\bibinfo {title} {Dihedral-angle-corrected registry-dependent interlayer
  potential for multilayer graphene structures},\ }\href
  {https://doi.org/10.1103/physrevb.98.235404} {\bibfield  {journal} {\bibinfo
  {journal} {Phys. Rev. B}\ }\textbf {\bibinfo {volume} {98}},\ \bibinfo
  {pages} {235404} (\bibinfo {year} {2018})}\BibitemShut {NoStop}%
\bibitem [{\citenamefont {Liu}\ \emph {et~al.}(2014)\citenamefont {Liu},
  \citenamefont {Gao}, \citenamefont {Zhang}, \citenamefont {Yan},\ and\
  \citenamefont {Ding}}]{liu2014vacancy}%
  \BibitemOpen
  \bibfield  {author} {\bibinfo {author} {\bibfnamefont {L.}~\bibnamefont
  {Liu}}, \bibinfo {author} {\bibfnamefont {J.}~\bibnamefont {Gao}}, \bibinfo
  {author} {\bibfnamefont {X.}~\bibnamefont {Zhang}}, \bibinfo {author}
  {\bibfnamefont {T.}~\bibnamefont {Yan}},\ and\ \bibinfo {author}
  {\bibfnamefont {F.}~\bibnamefont {Ding}},\ }\bibfield  {title} {\bibinfo
  {title} {Vacancy inter-layer migration in multi-layered graphene},\ }\href
  {https://doi.org/10.1039/c4nr00488d} {\bibfield  {journal} {\bibinfo
  {journal} {Nanoscale}\ }\textbf {\bibinfo {volume} {6}},\ \bibinfo {pages}
  {5729} (\bibinfo {year} {2014})}\BibitemShut {NoStop}%
\bibitem [{\citenamefont {Behler}\ and\ \citenamefont
  {Parrinello}(2007)}]{behler2007generalized}%
  \BibitemOpen
  \bibfield  {author} {\bibinfo {author} {\bibfnamefont {J.}~\bibnamefont
  {Behler}}\ and\ \bibinfo {author} {\bibfnamefont {M.}~\bibnamefont
  {Parrinello}},\ }\bibfield  {title} {\bibinfo {title} {Generalized
  neural-network representation of high-dimensional potential-energy
  surfaces},\ }\href {https://doi.org/10.1103/physrevlett.98.146401} {\bibfield
   {journal} {\bibinfo  {journal} {Phys. Rev. Lett.}\ }\textbf {\bibinfo
  {volume} {98}},\ \bibinfo {pages} {146401} (\bibinfo {year}
  {2007})}\BibitemShut {NoStop}%
\bibitem [{\citenamefont {Bart{\'{o}}k}\ \emph {et~al.}(2010)\citenamefont
  {Bart{\'{o}}k}, \citenamefont {Payne}, \citenamefont {Kondor},\ and\
  \citenamefont {Cs{\'{a}}nyi}}]{bartok2010gaussian}%
  \BibitemOpen
  \bibfield  {author} {\bibinfo {author} {\bibfnamefont {A.~P.}\ \bibnamefont
  {Bart{\'{o}}k}}, \bibinfo {author} {\bibfnamefont {M.~C.}\ \bibnamefont
  {Payne}}, \bibinfo {author} {\bibfnamefont {R.}~\bibnamefont {Kondor}},\ and\
  \bibinfo {author} {\bibfnamefont {G.}~\bibnamefont {Cs{\'{a}}nyi}},\
  }\bibfield  {title} {\bibinfo {title} {{G}aussian {A}pproximation
  {P}otentials: {T}he accuracy of quantum mechanics, without the electrons},\
  }\href {https://doi.org/10.1103/physrevlett.104.136403} {\bibfield  {journal}
  {\bibinfo  {journal} {Phys. Rev. Lett.}\ }\textbf {\bibinfo {volume} {104}},\
  \bibinfo {pages} {136403} (\bibinfo {year} {2010})}\BibitemShut {NoStop}%
\bibitem [{\citenamefont {Rupp}\ \emph {et~al.}(2012)\citenamefont {Rupp},
  \citenamefont {Tkatchenko}, \citenamefont {M{\"u}ller},\ and\ \citenamefont
  {Von~Lilienfeld}}]{rupp2012fast}%
  \BibitemOpen
  \bibfield  {author} {\bibinfo {author} {\bibfnamefont {M.}~\bibnamefont
  {Rupp}}, \bibinfo {author} {\bibfnamefont {A.}~\bibnamefont {Tkatchenko}},
  \bibinfo {author} {\bibfnamefont {K.-R.}\ \bibnamefont {M{\"u}ller}},\ and\
  \bibinfo {author} {\bibfnamefont {O.~A.}\ \bibnamefont {Von~Lilienfeld}},\
  }\bibfield  {title} {\bibinfo {title} {Fast and accurate modeling of
  molecular atomization energies with machine learning},\ }\href
  {https://doi.org/10.1103/physrevlett.108.058301} {\bibfield  {journal}
  {\bibinfo  {journal} {Phys. Rev. Lett.}\ }\textbf {\bibinfo {volume} {108}},\
  \bibinfo {pages} {058301} (\bibinfo {year} {2012})}\BibitemShut {NoStop}%
\bibitem [{\citenamefont {Thompson}\ \emph {et~al.}(2015)\citenamefont
  {Thompson}, \citenamefont {Swiler}, \citenamefont {Trott}, \citenamefont
  {Foiles},\ and\ \citenamefont {Tucker}}]{thompson2015spectral}%
  \BibitemOpen
  \bibfield  {author} {\bibinfo {author} {\bibfnamefont {A.~P.}\ \bibnamefont
  {Thompson}}, \bibinfo {author} {\bibfnamefont {L.~P.}\ \bibnamefont
  {Swiler}}, \bibinfo {author} {\bibfnamefont {C.~R.}\ \bibnamefont {Trott}},
  \bibinfo {author} {\bibfnamefont {S.~M.}\ \bibnamefont {Foiles}},\ and\
  \bibinfo {author} {\bibfnamefont {G.~J.}\ \bibnamefont {Tucker}},\ }\bibfield
   {title} {\bibinfo {title} {Spectral neighbor analysis method for automated
  generation of quantum-accurate interatomic potentials},\ }\href
  {https://doi.org/10.1016/j.jcp.2014.12.018} {\bibfield  {journal} {\bibinfo
  {journal} {J. Comput. Phys.}\ }\textbf {\bibinfo {volume} {285}},\ \bibinfo
  {pages} {316} (\bibinfo {year} {2015})}\BibitemShut {NoStop}%
\bibitem [{\citenamefont {Shapeev}(2016)}]{shapeev2016moment}%
  \BibitemOpen
  \bibfield  {author} {\bibinfo {author} {\bibfnamefont {A.~V.}\ \bibnamefont
  {Shapeev}},\ }\bibfield  {title} {\bibinfo {title} {Moment tensor potentials:
  A class of systematically improvable interatomic potentials},\ }\href
  {https://doi.org/10.1137/15m1054183} {\bibfield  {journal} {\bibinfo
  {journal} {Multiscale Model. Simul.}\ }\textbf {\bibinfo {volume} {14}},\
  \bibinfo {pages} {1153} (\bibinfo {year} {2016})}\BibitemShut {NoStop}%
\bibitem [{\citenamefont {Hajinazar}\ \emph {et~al.}(2017)\citenamefont
  {Hajinazar}, \citenamefont {Shao},\ and\ \citenamefont
  {Kolmogorov}}]{hajinazar2017stratified}%
  \BibitemOpen
  \bibfield  {author} {\bibinfo {author} {\bibfnamefont {S.}~\bibnamefont
  {Hajinazar}}, \bibinfo {author} {\bibfnamefont {J.}~\bibnamefont {Shao}},\
  and\ \bibinfo {author} {\bibfnamefont {A.~N.}\ \bibnamefont {Kolmogorov}},\
  }\bibfield  {title} {\bibinfo {title} {Stratified construction of neural
  network based interatomic models for multicomponent materials},\ }\href
  {https://doi.org/10.1103/physrevb.95.014114} {\bibfield  {journal} {\bibinfo
  {journal} {Phys. Rev. B}\ }\textbf {\bibinfo {volume} {95}},\ \bibinfo
  {pages} {014114} (\bibinfo {year} {2017})}\BibitemShut {NoStop}%
\bibitem [{\citenamefont {Gal}\ and\ \citenamefont
  {Ghahramani}(2016)}]{gal2016dropout}%
  \BibitemOpen
  \bibfield  {author} {\bibinfo {author} {\bibfnamefont {Y.}~\bibnamefont
  {Gal}}\ and\ \bibinfo {author} {\bibfnamefont {Z.}~\bibnamefont
  {Ghahramani}},\ }\bibfield  {title} {\bibinfo {title} {Dropout as a
  {B}ayesian approximation: Representing model uncertainty in deep learning},\
  }in\ \href@noop {} {\emph {\bibinfo {booktitle} {Proceedings of the 33rd
  International Conference on Machine Learning (ICML-16)}}}\ (\bibinfo {year}
  {2016})\BibitemShut {NoStop}%
\bibitem [{\citenamefont {Gal}(2016)}]{gal2016uncertainty}%
  \BibitemOpen
  \bibfield  {author} {\bibinfo {author} {\bibfnamefont {Y.}~\bibnamefont
  {Gal}},\ }\emph {\bibinfo {title} {Uncertainty in Deep Learning}},\
  \href@noop {} {Ph.D. thesis},\ \bibinfo  {school} {University of Cambridge}
  (\bibinfo {year} {2016})\BibitemShut {NoStop}%
\bibitem [{\citenamefont {Wen}\ and\ \citenamefont
  {Tadmor}(2019)}]{wen2019dropout}%
  \BibitemOpen
  \bibfield  {author} {\bibinfo {author} {\bibfnamefont {M.}~\bibnamefont
  {Wen}}\ and\ \bibinfo {author} {\bibfnamefont {E.~B.}\ \bibnamefont
  {Tadmor}},\ }\bibfield  {title} {\bibinfo {title} {Uncertainty quantification
  in molecular simulations with dropout neural network potentials},\
  }\href@noop {} {\bibfield  {journal} {\bibinfo  {journal} {submitted}\ }
  (\bibinfo {year} {2019})}\BibitemShut {NoStop}%
\bibitem [{\citenamefont {Deringer}\ and\ \citenamefont
  {Cs{\'{a}}nyi}(2017)}]{deringer2017machine}%
  \BibitemOpen
  \bibfield  {author} {\bibinfo {author} {\bibfnamefont {V.~L.}\ \bibnamefont
  {Deringer}}\ and\ \bibinfo {author} {\bibfnamefont {G.}~\bibnamefont
  {Cs{\'{a}}nyi}},\ }\bibfield  {title} {\bibinfo {title} {Machine learning
  based interatomic potential for amorphous carbon},\ }\href
  {https://doi.org/10.1103/physrevb.95.094203} {\bibfield  {journal} {\bibinfo
  {journal} {Phys. Rev. B}\ }\textbf {\bibinfo {volume} {95}},\ \bibinfo
  {pages} {094203} (\bibinfo {year} {2017})}\BibitemShut {NoStop}%
\bibitem [{\citenamefont {Rowe}\ \emph {et~al.}(2018)\citenamefont {Rowe},
  \citenamefont {Cs{\'{a}}nyi}, \citenamefont {Alf{\`{e}}},\ and\ \citenamefont
  {Michaelides}}]{rowe2018development}%
  \BibitemOpen
  \bibfield  {author} {\bibinfo {author} {\bibfnamefont {P.}~\bibnamefont
  {Rowe}}, \bibinfo {author} {\bibfnamefont {G.}~\bibnamefont {Cs{\'{a}}nyi}},
  \bibinfo {author} {\bibfnamefont {D.}~\bibnamefont {Alf{\`{e}}}},\ and\
  \bibinfo {author} {\bibfnamefont {A.}~\bibnamefont {Michaelides}},\
  }\bibfield  {title} {\bibinfo {title} {Development of a machine learning
  potential for graphene},\ }\href {https://doi.org/10.1103/physrevb.97.054303}
  {\bibfield  {journal} {\bibinfo  {journal} {Phys. Rev. B}\ }\textbf {\bibinfo
  {volume} {97}},\ \bibinfo {pages} {054303} (\bibinfo {year}
  {2018})}\BibitemShut {NoStop}%
\bibitem [{\citenamefont {Khaliullin}\ \emph {et~al.}(2010)\citenamefont
  {Khaliullin}, \citenamefont {Eshet}, \citenamefont {K{\"u}hne}, \citenamefont
  {Behler},\ and\ \citenamefont {Parrinello}}]{khaliullin2010graphite}%
  \BibitemOpen
  \bibfield  {author} {\bibinfo {author} {\bibfnamefont {R.~Z.}\ \bibnamefont
  {Khaliullin}}, \bibinfo {author} {\bibfnamefont {H.}~\bibnamefont {Eshet}},
  \bibinfo {author} {\bibfnamefont {T.~D.}\ \bibnamefont {K{\"u}hne}}, \bibinfo
  {author} {\bibfnamefont {J.}~\bibnamefont {Behler}},\ and\ \bibinfo {author}
  {\bibfnamefont {M.}~\bibnamefont {Parrinello}},\ }\bibfield  {title}
  {\bibinfo {title} {Graphite-diamond phase coexistence study employing a
  neural-network mapping of the ab initio potential energy surface},\ }\href
  {https://doi.org/10.1103/physrevb.81.100103} {\bibfield  {journal} {\bibinfo
  {journal} {Phys. Rev. B}\ }\textbf {\bibinfo {volume} {81}},\ \bibinfo
  {pages} {100103} (\bibinfo {year} {2010})}\BibitemShut {NoStop}%
\bibitem [{\citenamefont {Khaliullin}\ \emph {et~al.}(2011)\citenamefont
  {Khaliullin}, \citenamefont {Eshet}, \citenamefont {K{\"u}hne}, \citenamefont
  {Behler},\ and\ \citenamefont {Parrinello}}]{khaliullin2011nucleation}%
  \BibitemOpen
  \bibfield  {author} {\bibinfo {author} {\bibfnamefont {R.~Z.}\ \bibnamefont
  {Khaliullin}}, \bibinfo {author} {\bibfnamefont {H.}~\bibnamefont {Eshet}},
  \bibinfo {author} {\bibfnamefont {T.~D.}\ \bibnamefont {K{\"u}hne}}, \bibinfo
  {author} {\bibfnamefont {J.}~\bibnamefont {Behler}},\ and\ \bibinfo {author}
  {\bibfnamefont {M.}~\bibnamefont {Parrinello}},\ }\bibfield  {title}
  {\bibinfo {title} {Nucleation mechanism for the direct graphite-to-diamond
  phase transition},\ }\href {https://doi.org/10.1038/nmat3078} {\bibfield
  {journal} {\bibinfo  {journal} {Nat. Mater.}\ }\textbf {\bibinfo {volume}
  {10}},\ \bibinfo {pages} {693} (\bibinfo {year} {2011})}\BibitemShut
  {NoStop}%
\bibitem [{sup()}]{supplemental}%
  \BibitemOpen
  \href@noop {} {}\bibinfo {note} {See Supplemental Material at [URL will be
  inserted by publisher] for the dataset and a detailed description of it, the
  symmetry functions used as the descriptors for the neural network, the
  weights and biases parameters in the neural network, and the way to use a KIM
  potential.}\BibitemShut {Stop}%
\bibitem [{\citenamefont {Tadmor}\ and\ \citenamefont
  {Miller}(2011)}]{tadmor:miller:2011}%
  \BibitemOpen
  \bibfield  {author} {\bibinfo {author} {\bibfnamefont {E.~B.}\ \bibnamefont
  {Tadmor}}\ and\ \bibinfo {author} {\bibfnamefont {R.~E.}\ \bibnamefont
  {Miller}},\ }\href@noop {} {\emph {\bibinfo {title} {Modeling Materials:
  Continuum, Atomistic and Multiscale Techniques}}}\ (\bibinfo  {publisher}
  {Cambridge University Press},\ \bibinfo {address} {Cambridge},\ \bibinfo
  {year} {2011})\BibitemShut {NoStop}%
\bibitem [{\citenamefont {Hansen}\ \emph {et~al.}(2015)\citenamefont {Hansen},
  \citenamefont {Biegler}, \citenamefont {Ramakrishnan}, \citenamefont
  {Pronobis}, \citenamefont {Von~Lilienfeld}, \citenamefont {M{\"u}ller},\ and\
  \citenamefont {Tkatchenko}}]{hansen2015machine}%
  \BibitemOpen
  \bibfield  {author} {\bibinfo {author} {\bibfnamefont {K.}~\bibnamefont
  {Hansen}}, \bibinfo {author} {\bibfnamefont {F.}~\bibnamefont {Biegler}},
  \bibinfo {author} {\bibfnamefont {R.}~\bibnamefont {Ramakrishnan}}, \bibinfo
  {author} {\bibfnamefont {W.}~\bibnamefont {Pronobis}}, \bibinfo {author}
  {\bibfnamefont {O.~A.}\ \bibnamefont {Von~Lilienfeld}}, \bibinfo {author}
  {\bibfnamefont {K.-R.}\ \bibnamefont {M{\"u}ller}},\ and\ \bibinfo {author}
  {\bibfnamefont {A.}~\bibnamefont {Tkatchenko}},\ }\bibfield  {title}
  {\bibinfo {title} {Machine learning predictions of molecular properties:
  Accurate many-body potentials and nonlocality in chemical space},\ }\href
  {https://doi.org/10.1021/acs.jpclett.5b00831} {\bibfield  {journal} {\bibinfo
   {journal} {J. Phys. Chem. Lett.}\ }\textbf {\bibinfo {volume} {6}},\
  \bibinfo {pages} {2326} (\bibinfo {year} {2015})}\BibitemShut {NoStop}%
\bibitem [{\citenamefont {Bart{\'o}k}\ \emph {et~al.}(2013)\citenamefont
  {Bart{\'o}k}, \citenamefont {Kondor},\ and\ \citenamefont
  {Cs{\'a}nyi}}]{bartok2013representing}%
  \BibitemOpen
  \bibfield  {author} {\bibinfo {author} {\bibfnamefont {A.~P.}\ \bibnamefont
  {Bart{\'o}k}}, \bibinfo {author} {\bibfnamefont {R.}~\bibnamefont {Kondor}},\
  and\ \bibinfo {author} {\bibfnamefont {G.}~\bibnamefont {Cs{\'a}nyi}},\
  }\bibfield  {title} {\bibinfo {title} {On representing chemical
  environments},\ }\href {https://doi.org/10.1103/physrevb.87.184115}
  {\bibfield  {journal} {\bibinfo  {journal} {Phys. Rev. B}\ }\textbf {\bibinfo
  {volume} {87}},\ \bibinfo {pages} {184115} (\bibinfo {year}
  {2013})}\BibitemShut {NoStop}%
\bibitem [{\citenamefont {Behler}(2011)}]{behler2011atom}%
  \BibitemOpen
  \bibfield  {author} {\bibinfo {author} {\bibfnamefont {J.}~\bibnamefont
  {Behler}},\ }\bibfield  {title} {\bibinfo {title} {Atom-centered symmetry
  functions for constructing high-dimensional neural network potentials},\
  }\href {https://doi.org/10.1063/1.3553717} {\bibfield  {journal} {\bibinfo
  {journal} {J. Chem. Phys.}\ }\textbf {\bibinfo {volume} {134}},\ \bibinfo
  {pages} {074106} (\bibinfo {year} {2011})}\BibitemShut {NoStop}%
\bibitem [{\citenamefont {Artrith}\ and\ \citenamefont
  {Behler}(2012)}]{artrith2012high}%
  \BibitemOpen
  \bibfield  {author} {\bibinfo {author} {\bibfnamefont {N.}~\bibnamefont
  {Artrith}}\ and\ \bibinfo {author} {\bibfnamefont {J.}~\bibnamefont
  {Behler}},\ }\bibfield  {title} {\bibinfo {title} {High-dimensional neural
  network potentials for metal surfaces: A prototype study for copper},\ }\href
  {https://doi.org/10.1103/physrevb.85.045439} {\bibfield  {journal} {\bibinfo
  {journal} {Phys. Rev. B}\ }\textbf {\bibinfo {volume} {85}},\ \bibinfo
  {pages} {045439} (\bibinfo {year} {2012})}\BibitemShut {NoStop}%
\bibitem [{\citenamefont {Kresse}\ and\ \citenamefont
  {Furthm{\"u}ller}(1996{\natexlab{a}})}]{vasp1}%
  \BibitemOpen
  \bibfield  {author} {\bibinfo {author} {\bibfnamefont {G.}~\bibnamefont
  {Kresse}}\ and\ \bibinfo {author} {\bibfnamefont {J.}~\bibnamefont
  {Furthm{\"u}ller}},\ }\bibfield  {title} {\bibinfo {title} {Efficient
  iterative schemes for ab initio total-energy calculations using a plane-wave
  basis set},\ }\href {https://doi.org/10.1103/PhysRevB.54.11169} {\bibfield
  {journal} {\bibinfo  {journal} {Phys. Rev. B}\ }\textbf {\bibinfo {volume}
  {54}},\ \bibinfo {pages} {11169} (\bibinfo {year}
  {1996}{\natexlab{a}})}\BibitemShut {NoStop}%
\bibitem [{\citenamefont {Kresse}\ and\ \citenamefont
  {Furthm{\"u}ller}(1996{\natexlab{b}})}]{vasp2}%
  \BibitemOpen
  \bibfield  {author} {\bibinfo {author} {\bibfnamefont {G.}~\bibnamefont
  {Kresse}}\ and\ \bibinfo {author} {\bibfnamefont {J.}~\bibnamefont
  {Furthm{\"u}ller}},\ }\bibfield  {title} {\bibinfo {title} {Efficiency of
  ab-initio total energy calculations for metals and semiconductors using a
  plane-wave basis set},\ }\href {https://doi.org/10.1016/0927-0256(96)00008-0}
  {\bibfield  {journal} {\bibinfo  {journal} {Comput. Mater. Sci}\ }\textbf
  {\bibinfo {volume} {6}},\ \bibinfo {pages} {15} (\bibinfo {year}
  {1996}{\natexlab{b}})}\BibitemShut {NoStop}%
\bibitem [{\citenamefont {Perdew}\ \emph {et~al.}(1996)\citenamefont {Perdew},
  \citenamefont {Burke},\ and\ \citenamefont {Ernzerhof}}]{pbe}%
  \BibitemOpen
  \bibfield  {author} {\bibinfo {author} {\bibfnamefont {J.~P.}\ \bibnamefont
  {Perdew}}, \bibinfo {author} {\bibfnamefont {K.}~\bibnamefont {Burke}},\ and\
  \bibinfo {author} {\bibfnamefont {M.}~\bibnamefont {Ernzerhof}},\ }\bibfield
  {title} {\bibinfo {title} {Generalized gradient approximation made simple},\
  }\href {https://doi.org/10.1103/PhysRevLett.77.3865} {\bibfield  {journal}
  {\bibinfo  {journal} {Phys. Rev. Lett.}\ }\textbf {\bibinfo {volume} {77}},\
  \bibinfo {pages} {3865} (\bibinfo {year} {1996})}\BibitemShut {NoStop}%
\bibitem [{\citenamefont {Monkhorst}\ and\ \citenamefont
  {Pack}(1976)}]{monkhorst1976special}%
  \BibitemOpen
  \bibfield  {author} {\bibinfo {author} {\bibfnamefont {H.~J.}\ \bibnamefont
  {Monkhorst}}\ and\ \bibinfo {author} {\bibfnamefont {J.~D.}\ \bibnamefont
  {Pack}},\ }\bibfield  {title} {\bibinfo {title} {Special points for
  brillouin-zone integrations},\ }\href
  {https://doi.org/10.1103/physrevb.13.5188} {\bibfield  {journal} {\bibinfo
  {journal} {Phys. Rev. B}\ }\textbf {\bibinfo {volume} {13}},\ \bibinfo
  {pages} {5188} (\bibinfo {year} {1976})}\BibitemShut {NoStop}%
\bibitem [{\citenamefont {Tkatchenko}\ \emph {et~al.}(2012)\citenamefont
  {Tkatchenko}, \citenamefont {DiStasio}, \citenamefont {Car},\ and\
  \citenamefont {Scheffler}}]{tkatchenko2012accurate}%
  \BibitemOpen
  \bibfield  {author} {\bibinfo {author} {\bibfnamefont {A.}~\bibnamefont
  {Tkatchenko}}, \bibinfo {author} {\bibfnamefont {R.~A.}\ \bibnamefont
  {DiStasio}}, \bibinfo {author} {\bibfnamefont {R.}~\bibnamefont {Car}},\ and\
  \bibinfo {author} {\bibfnamefont {M.}~\bibnamefont {Scheffler}},\ }\bibfield
  {title} {\bibinfo {title} {Accurate and efficient method for many-body van
  der waals interactions},\ }\href
  {https://doi.org/10.1103/physrevlett.108.236402} {\bibfield  {journal}
  {\bibinfo  {journal} {Phys. Rev. Lett.}\ }\textbf {\bibinfo {volume} {108}},\
  \bibinfo {pages} {236402} (\bibinfo {year} {2012})}\BibitemShut {NoStop}%
\bibitem [{\citenamefont {Wen}\ \emph {et~al.}(2019)\citenamefont {Wen},
  \citenamefont {Elliott},\ and\ \citenamefont {Tadmor}}]{kliff}%
  \BibitemOpen
  \bibfield  {author} {\bibinfo {author} {\bibfnamefont {M.}~\bibnamefont
  {Wen}}, \bibinfo {author} {\bibfnamefont {R.~S.}\ \bibnamefont {Elliott}},\
  and\ \bibinfo {author} {\bibfnamefont {E.~B.}\ \bibnamefont {Tadmor}},\
  }\href@noop {} {\bibinfo {title} {{KLIFF}: {KIM}-based learning-integrated
  fitting framework}},\ \bibinfo {howpublished}
  {\url{https://kliff.readthedocs.io}} (\bibinfo {year} {2019})\BibitemShut
  {NoStop}%
\bibitem [{\citenamefont {Zhu}\ \emph {et~al.}(1997)\citenamefont {Zhu},
  \citenamefont {Byrd}, \citenamefont {Lu},\ and\ \citenamefont
  {Nocedal}}]{zhu1997algorithm}%
  \BibitemOpen
  \bibfield  {author} {\bibinfo {author} {\bibfnamefont {C.}~\bibnamefont
  {Zhu}}, \bibinfo {author} {\bibfnamefont {R.~H.}\ \bibnamefont {Byrd}},
  \bibinfo {author} {\bibfnamefont {P.}~\bibnamefont {Lu}},\ and\ \bibinfo
  {author} {\bibfnamefont {J.}~\bibnamefont {Nocedal}},\ }\bibfield  {title}
  {\bibinfo {title} {Algorithm 778: L-{BFGS}-b: Fortran subroutines for
  large-scale bound-constrained optimization},\ }\href
  {https://doi.org/10.1145/279232.279236} {\bibfield  {journal} {\bibinfo
  {journal} {{ACM} Transactions on Mathematical Software}\ }\textbf {\bibinfo
  {volume} {23}},\ \bibinfo {pages} {550} (\bibinfo {year} {1997})}\BibitemShut
  {NoStop}%
\bibitem [{\citenamefont {Tadmor}\ \emph {et~al.}(2011)\citenamefont {Tadmor},
  \citenamefont {Elliott}, \citenamefont {Sethna}, \citenamefont {Miller},\
  and\ \citenamefont {Becker}}]{tadmor2011kim}%
  \BibitemOpen
  \bibfield  {author} {\bibinfo {author} {\bibfnamefont {E.~B.}\ \bibnamefont
  {Tadmor}}, \bibinfo {author} {\bibfnamefont {R.~S.}\ \bibnamefont {Elliott}},
  \bibinfo {author} {\bibfnamefont {J.~P.}\ \bibnamefont {Sethna}}, \bibinfo
  {author} {\bibfnamefont {R.~E.}\ \bibnamefont {Miller}},\ and\ \bibinfo
  {author} {\bibfnamefont {C.~A.}\ \bibnamefont {Becker}},\ }\bibfield  {title}
  {\bibinfo {title} {The potential of atomistic simulations and the
  knowledgebase of interatomic models},\ }\href
  {https://doi.org/10.1007/s11837-011-0102-6} {\bibfield  {journal} {\bibinfo
  {journal} {{JOM}}\ }\textbf {\bibinfo {volume} {63}},\ \bibinfo {pages} {17}
  (\bibinfo {year} {2011})}\BibitemShut {NoStop}%
\bibitem [{\citenamefont {Plimpton}(1995)}]{plimpton1995fast}%
  \BibitemOpen
  \bibfield  {author} {\bibinfo {author} {\bibfnamefont {S.}~\bibnamefont
  {Plimpton}},\ }\bibfield  {title} {\bibinfo {title} {Fast parallel algorithms
  for short-range molecular dynamics},\ }\href
  {https://doi.org/10.1006/jcph.1995.1039} {\bibfield  {journal} {\bibinfo
  {journal} {J. Comput. Phys.}\ }\textbf {\bibinfo {volume} {117}},\ \bibinfo
  {pages} {1} (\bibinfo {year} {1995})}\BibitemShut {NoStop}%
\bibitem [{\citenamefont {Smith}\ and\ \citenamefont
  {Forester}(1996)}]{smith1996dl_poly_2}%
  \BibitemOpen
  \bibfield  {author} {\bibinfo {author} {\bibfnamefont {W.}~\bibnamefont
  {Smith}}\ and\ \bibinfo {author} {\bibfnamefont {T.}~\bibnamefont
  {Forester}},\ }\bibfield  {title} {\bibinfo {title} {{DL}{\_}{POLY}{\_}2.0: A
  general-purpose parallel molecular dynamics simulation package},\ }\href
  {https://doi.org/10.1016/S0263-7855(96)00043-4} {\bibfield  {journal}
  {\bibinfo  {journal} {J. Mol. Graphics}\ }\textbf {\bibinfo {volume} {14}},\
  \bibinfo {pages} {136} (\bibinfo {year} {1996})}\BibitemShut {NoStop}%
\bibitem [{\citenamefont {Gale}(1997)}]{gale1997gulp}%
  \BibitemOpen
  \bibfield  {author} {\bibinfo {author} {\bibfnamefont {J.~D.}\ \bibnamefont
  {Gale}},\ }\bibfield  {title} {\bibinfo {title} {{GULP}: A computer program
  for the symmetry-adapted simulation of solids},\ }\href
  {https://doi.org/10.1039/a606455h} {\bibfield  {journal} {\bibinfo  {journal}
  {J. Chem. Soc.-Farad. Trans.}\ }\textbf {\bibinfo {volume} {93}},\ \bibinfo
  {pages} {629} (\bibinfo {year} {1997})}\BibitemShut {NoStop}%
\bibitem [{\citenamefont {Larsen}\ and\ \citenamefont
  {Others}(2017)}]{larsen2017atomic}%
  \BibitemOpen
  \bibfield  {author} {\bibinfo {author} {\bibfnamefont {A.~H.}\ \bibnamefont
  {Larsen}}\ and\ \bibinfo {author} {\bibnamefont {Others}},\ }\bibfield
  {title} {\bibinfo {title} {The atomic simulation environment{\textemdash}a
  {P}ython library for working with atoms},\ }\href
  {https://doi.org/10.1088/1361-648x/aa680e} {\bibfield  {journal} {\bibinfo
  {journal} {J. Phys.: Condens. Matter}\ }\textbf {\bibinfo {volume} {29}},\
  \bibinfo {pages} {273002} (\bibinfo {year} {2017})}\BibitemShut {NoStop}%
\bibitem [{\citenamefont {Zhou}\ \emph {et~al.}(2015)\citenamefont {Zhou},
  \citenamefont {Han}, \citenamefont {Dai}, \citenamefont {Sun},\ and\
  \citenamefont {Srolovitz}}]{zhou2015van}%
  \BibitemOpen
  \bibfield  {author} {\bibinfo {author} {\bibfnamefont {S.}~\bibnamefont
  {Zhou}}, \bibinfo {author} {\bibfnamefont {J.}~\bibnamefont {Han}}, \bibinfo
  {author} {\bibfnamefont {S.}~\bibnamefont {Dai}}, \bibinfo {author}
  {\bibfnamefont {J.}~\bibnamefont {Sun}},\ and\ \bibinfo {author}
  {\bibfnamefont {D.~J.}\ \bibnamefont {Srolovitz}},\ }\bibfield  {title}
  {\bibinfo {title} {van der waals bilayer energetics: Generalized
  stacking-fault energy of graphene, boron nitride, and graphene/boron nitride
  bilayers},\ }\href {https://doi.org/10.1103/physrevb.92.155438} {\bibfield
  {journal} {\bibinfo  {journal} {Phys. Rev. B}\ }\textbf {\bibinfo {volume}
  {92}},\ \bibinfo {pages} {155438} (\bibinfo {year} {2015})}\BibitemShut
  {NoStop}%
\bibitem [{\citenamefont {Leb{\`{e}}gue}\ \emph {et~al.}(2010)\citenamefont
  {Leb{\`{e}}gue}, \citenamefont {Harl}, \citenamefont {Gould}, \citenamefont
  {{\'{A}}ngy{\'{a}}n}, \citenamefont {Kresse},\ and\ \citenamefont
  {Dobson}}]{lebegue2010cohesive}%
  \BibitemOpen
  \bibfield  {author} {\bibinfo {author} {\bibfnamefont {S.}~\bibnamefont
  {Leb{\`{e}}gue}}, \bibinfo {author} {\bibfnamefont {J.}~\bibnamefont {Harl}},
  \bibinfo {author} {\bibfnamefont {T.}~\bibnamefont {Gould}}, \bibinfo
  {author} {\bibfnamefont {J.~G.}\ \bibnamefont {{\'{A}}ngy{\'{a}}n}}, \bibinfo
  {author} {\bibfnamefont {G.}~\bibnamefont {Kresse}},\ and\ \bibinfo {author}
  {\bibfnamefont {J.~F.}\ \bibnamefont {Dobson}},\ }\bibfield  {title}
  {\bibinfo {title} {Cohesive properties and asymptotics of the dispersion
  interaction in graphite by the random phase approximation},\ }\href
  {https://doi.org/10.1103/physrevlett.105.196401} {\bibfield  {journal}
  {\bibinfo  {journal} {Phys. Rev. Lett.}\ }\textbf {\bibinfo {volume} {105}},\
  \bibinfo {pages} {196401} (\bibinfo {year} {2010})}\BibitemShut {NoStop}%
\bibitem [{\citenamefont {Lin}\ and\ \citenamefont
  {Zhang}(2012)}]{lin2012creating}%
  \BibitemOpen
  \bibfield  {author} {\bibinfo {author} {\bibfnamefont {L.}~\bibnamefont
  {Lin}}\ and\ \bibinfo {author} {\bibfnamefont {S.}~\bibnamefont {Zhang}},\
  }\bibfield  {title} {\bibinfo {title} {Creating high yield water soluble
  luminescent graphene quantum dots via exfoliating and disintegrating carbon
  nanotubes and graphite flakes},\ }\href {https://doi.org/10.1039/c2cc35559k}
  {\bibfield  {journal} {\bibinfo  {journal} {Chem. Commun.}\ }\textbf
  {\bibinfo {volume} {48}},\ \bibinfo {pages} {10177} (\bibinfo {year}
  {2012})}\BibitemShut {NoStop}%
\bibitem [{\citenamefont {Baskin}\ and\ \citenamefont
  {Meyer}(1955)}]{baskin1955lattice}%
  \BibitemOpen
  \bibfield  {author} {\bibinfo {author} {\bibfnamefont {Y.}~\bibnamefont
  {Baskin}}\ and\ \bibinfo {author} {\bibfnamefont {L.}~\bibnamefont {Meyer}},\
  }\bibfield  {title} {\bibinfo {title} {Lattice constants of graphite at low
  temperatures},\ }\href {https://doi.org/10.1103/physrev.100.544} {\bibfield
  {journal} {\bibinfo  {journal} {Phys. Rev.}\ }\textbf {\bibinfo {volume}
  {100}},\ \bibinfo {pages} {544} (\bibinfo {year} {1955})}\BibitemShut
  {NoStop}%
\bibitem [{\citenamefont {Blakslee}\ \emph {et~al.}(1970)\citenamefont
  {Blakslee}, \citenamefont {Proctor}, \citenamefont {Seldin}, \citenamefont
  {Spence},\ and\ \citenamefont {Weng}}]{blakslee1970elastic}%
  \BibitemOpen
  \bibfield  {author} {\bibinfo {author} {\bibfnamefont {O.~L.}\ \bibnamefont
  {Blakslee}}, \bibinfo {author} {\bibfnamefont {D.~G.}\ \bibnamefont
  {Proctor}}, \bibinfo {author} {\bibfnamefont {E.~J.}\ \bibnamefont {Seldin}},
  \bibinfo {author} {\bibfnamefont {G.~B.}\ \bibnamefont {Spence}},\ and\
  \bibinfo {author} {\bibfnamefont {T.}~\bibnamefont {Weng}},\ }\bibfield
  {title} {\bibinfo {title} {Elastic constants of compression-annealed
  pyrolytic graphite},\ }\href {https://doi.org/10.1063/1.1659428} {\bibfield
  {journal} {\bibinfo  {journal} {J. Appl. Phys.}\ }\textbf {\bibinfo {volume}
  {41}},\ \bibinfo {pages} {3373} (\bibinfo {year} {1970})}\BibitemShut
  {NoStop}%
\bibitem [{\citenamefont {Cooper}\ \emph {et~al.}(2012)\citenamefont {Cooper},
  \citenamefont {D'Anjou}, \citenamefont {Ghattamaneni}, \citenamefont
  {Harack}, \citenamefont {Hilke}, \citenamefont {Horth}, \citenamefont
  {Majlis}, \citenamefont {Massicotte}, \citenamefont {Vandsburger},
  \citenamefont {Whiteway},\ and\ \citenamefont {Yu}}]{cooper2012experimental}%
  \BibitemOpen
  \bibfield  {author} {\bibinfo {author} {\bibfnamefont {D.~R.}\ \bibnamefont
  {Cooper}}, \bibinfo {author} {\bibfnamefont {B.}~\bibnamefont {D'Anjou}},
  \bibinfo {author} {\bibfnamefont {N.}~\bibnamefont {Ghattamaneni}}, \bibinfo
  {author} {\bibfnamefont {B.}~\bibnamefont {Harack}}, \bibinfo {author}
  {\bibfnamefont {M.}~\bibnamefont {Hilke}}, \bibinfo {author} {\bibfnamefont
  {A.}~\bibnamefont {Horth}}, \bibinfo {author} {\bibfnamefont
  {N.}~\bibnamefont {Majlis}}, \bibinfo {author} {\bibfnamefont
  {M.}~\bibnamefont {Massicotte}}, \bibinfo {author} {\bibfnamefont
  {L.}~\bibnamefont {Vandsburger}}, \bibinfo {author} {\bibfnamefont
  {E.}~\bibnamefont {Whiteway}},\ and\ \bibinfo {author} {\bibfnamefont
  {V.}~\bibnamefont {Yu}},\ }\bibfield  {title} {\bibinfo {title} {Experimental
  review of graphene},\ }\href {https://doi.org/10.5402/2012/501686} {\bibfield
   {journal} {\bibinfo  {journal} {{ISRN} Condens. Matter Phys.}\ }\textbf
  {\bibinfo {volume} {2012}},\ \bibinfo {pages} {1} (\bibinfo {year}
  {2012})}\BibitemShut {NoStop}%
\bibitem [{\citenamefont {Bosak}\ \emph {et~al.}(2007)\citenamefont {Bosak},
  \citenamefont {Krisch}, \citenamefont {Mohr}, \citenamefont {Maultzsch},\
  and\ \citenamefont {Thomsen}}]{bosak2007elasticity}%
  \BibitemOpen
  \bibfield  {author} {\bibinfo {author} {\bibfnamefont {A.}~\bibnamefont
  {Bosak}}, \bibinfo {author} {\bibfnamefont {M.}~\bibnamefont {Krisch}},
  \bibinfo {author} {\bibfnamefont {M.}~\bibnamefont {Mohr}}, \bibinfo {author}
  {\bibfnamefont {J.}~\bibnamefont {Maultzsch}},\ and\ \bibinfo {author}
  {\bibfnamefont {C.}~\bibnamefont {Thomsen}},\ }\bibfield  {title} {\bibinfo
  {title} {Elasticity of single-crystalline graphite: Inelastic x-ray
  scattering study},\ }\href {https://doi.org/10.1103/physrevb.75.153408}
  {\bibfield  {journal} {\bibinfo  {journal} {Phys. Rev. B}\ }\textbf {\bibinfo
  {volume} {75}},\ \bibinfo {pages} {153408} (\bibinfo {year}
  {2007})}\BibitemShut {NoStop}%
\bibitem [{\citenamefont {Birch}(1947)}]{birch1947finite}%
  \BibitemOpen
  \bibfield  {author} {\bibinfo {author} {\bibfnamefont {F.}~\bibnamefont
  {Birch}},\ }\bibfield  {title} {\bibinfo {title} {Finite elastic strain of
  cubic crystals},\ }\href {https://doi.org/10.1103/physrev.71.809} {\bibfield
  {journal} {\bibinfo  {journal} {Phys. Rev.}\ }\textbf {\bibinfo {volume}
  {71}},\ \bibinfo {pages} {809} (\bibinfo {year} {1947})}\BibitemShut
  {NoStop}%
\bibitem [{\citenamefont {Kaxiras}\ and\ \citenamefont
  {Pandey}(1988)}]{kaxiras1988energetics}%
  \BibitemOpen
  \bibfield  {author} {\bibinfo {author} {\bibfnamefont {E.}~\bibnamefont
  {Kaxiras}}\ and\ \bibinfo {author} {\bibfnamefont {K.~C.}\ \bibnamefont
  {Pandey}},\ }\bibfield  {title} {\bibinfo {title} {Energetics of defects and
  diffusion mechanisms in graphite},\ }\href
  {https://doi.org/10.1103/physrevlett.61.2693} {\bibfield  {journal} {\bibinfo
   {journal} {Phys. Rev. Lett.}\ }\textbf {\bibinfo {volume} {61}},\ \bibinfo
  {pages} {2693} (\bibinfo {year} {1988})}\BibitemShut {NoStop}%
\bibitem [{\citenamefont {Togo}\ and\ \citenamefont {Tanaka}(2015)}]{phonopy}%
  \BibitemOpen
  \bibfield  {author} {\bibinfo {author} {\bibfnamefont {A.}~\bibnamefont
  {Togo}}\ and\ \bibinfo {author} {\bibfnamefont {I.}~\bibnamefont {Tanaka}},\
  }\bibfield  {title} {\bibinfo {title} {First principles phonon calculations
  in materials science},\ }\href
  {https://doi.org/10.1016/j.scriptamat.2015.07.021} {\bibfield  {journal}
  {\bibinfo  {journal} {Scr. Mater.}\ }\textbf {\bibinfo {volume} {108}},\
  \bibinfo {pages} {1} (\bibinfo {year} {2015})}\BibitemShut {NoStop}%
\bibitem [{\citenamefont {Yan}\ \emph {et~al.}(2008)\citenamefont {Yan},
  \citenamefont {Ruan},\ and\ \citenamefont {Chou}}]{yan2008phonon}%
  \BibitemOpen
  \bibfield  {author} {\bibinfo {author} {\bibfnamefont {J.-A.}\ \bibnamefont
  {Yan}}, \bibinfo {author} {\bibfnamefont {W.~Y.}\ \bibnamefont {Ruan}},\ and\
  \bibinfo {author} {\bibfnamefont {M.~Y.}\ \bibnamefont {Chou}},\ }\bibfield
  {title} {\bibinfo {title} {Phonon dispersions and vibrational properties of
  monolayer, bilayer, and trilayer graphene: Density-functional perturbation
  theory},\ }\href {https://doi.org/10.1103/physrevb.77.125401} {\bibfield
  {journal} {\bibinfo  {journal} {Phys. Rev. B}\ }\textbf {\bibinfo {volume}
  {77}},\ \bibinfo {pages} {125401} (\bibinfo {year} {2008})}\BibitemShut
  {NoStop}%
\bibitem [{\citenamefont {Wirtz}\ and\ \citenamefont
  {Rubio}(2004)}]{wirtz2004phonon}%
  \BibitemOpen
  \bibfield  {author} {\bibinfo {author} {\bibfnamefont {L.}~\bibnamefont
  {Wirtz}}\ and\ \bibinfo {author} {\bibfnamefont {A.}~\bibnamefont {Rubio}},\
  }\bibfield  {title} {\bibinfo {title} {The phonon dispersion of graphite
  revisited},\ }\href {https://doi.org/10.1016/j.ssc.2004.04.042} {\bibfield
  {journal} {\bibinfo  {journal} {Solid State Commun.}\ }\textbf {\bibinfo
  {volume} {131}},\ \bibinfo {pages} {141} (\bibinfo {year}
  {2004})}\BibitemShut {NoStop}%
\bibitem [{lam(2019)}]{lammps}%
  \BibitemOpen
  \href@noop {} {\bibinfo {title} {Large-scale atomic/molecular massively
  parallel simulator ({LAMMPS})}},\ \bibinfo {howpublished}
  {\url{http://lammps.sandia.gov}} (\bibinfo {year} {2019})\BibitemShut
  {NoStop}%
\bibitem [{ope(2019)}]{openkim}%
  \BibitemOpen
  \href@noop {} {\bibinfo {title} {Open knowledgebase of interatomic models
  ({OpenKIM})}},\ \bibinfo {howpublished} {\url{https://openkim.org}} (\bibinfo
  {year} {2019})\BibitemShut {NoStop}%
\bibitem [{qui(2019)}]{quip}%
  \BibitemOpen
  \href@noop {} {\bibinfo {title} {{QUIP}: a collection of software tools to
  carry out molecular dynamics simulations}},\ \bibinfo {howpublished}
  {\url{http://www.libatoms.org/Home/Software}} (\bibinfo {year}
  {2019})\BibitemShut {NoStop}%
\bibitem [{\citenamefont {Cai}\ \emph {et~al.}(2010)\citenamefont {Cai},
  \citenamefont {Moore}, \citenamefont {Zhu}, \citenamefont {Li}, \citenamefont
  {Chen}, \citenamefont {Shi},\ and\ \citenamefont {Ruoff}}]{cai2010thermal}%
  \BibitemOpen
  \bibfield  {author} {\bibinfo {author} {\bibfnamefont {W.}~\bibnamefont
  {Cai}}, \bibinfo {author} {\bibfnamefont {A.~L.}\ \bibnamefont {Moore}},
  \bibinfo {author} {\bibfnamefont {Y.}~\bibnamefont {Zhu}}, \bibinfo {author}
  {\bibfnamefont {X.}~\bibnamefont {Li}}, \bibinfo {author} {\bibfnamefont
  {S.}~\bibnamefont {Chen}}, \bibinfo {author} {\bibfnamefont {L.}~\bibnamefont
  {Shi}},\ and\ \bibinfo {author} {\bibfnamefont {R.~S.}\ \bibnamefont
  {Ruoff}},\ }\bibfield  {title} {\bibinfo {title} {Thermal transport in
  suspended and supported monolayer graphene grown by chemical vapor
  deposition},\ }\href {https://doi.org/10.1021/nl9041966} {\bibfield
  {journal} {\bibinfo  {journal} {Nano Lett.}\ }\textbf {\bibinfo {volume}
  {10}},\ \bibinfo {pages} {1645} (\bibinfo {year} {2010})}\BibitemShut
  {NoStop}%
\bibitem [{\citenamefont {Faugeras}\ \emph {et~al.}(2010)\citenamefont
  {Faugeras}, \citenamefont {Faugeras}, \citenamefont {Orlita}, \citenamefont
  {Potemski}, \citenamefont {Nair},\ and\ \citenamefont
  {Geim}}]{faugeras2010thermal}%
  \BibitemOpen
  \bibfield  {author} {\bibinfo {author} {\bibfnamefont {C.}~\bibnamefont
  {Faugeras}}, \bibinfo {author} {\bibfnamefont {B.}~\bibnamefont {Faugeras}},
  \bibinfo {author} {\bibfnamefont {M.}~\bibnamefont {Orlita}}, \bibinfo
  {author} {\bibfnamefont {M.}~\bibnamefont {Potemski}}, \bibinfo {author}
  {\bibfnamefont {R.~R.}\ \bibnamefont {Nair}},\ and\ \bibinfo {author}
  {\bibfnamefont {A.~K.}\ \bibnamefont {Geim}},\ }\bibfield  {title} {\bibinfo
  {title} {Thermal conductivity of graphene in corbino membrane geometry},\
  }\href {https://doi.org/10.1021/nn9016229} {\bibfield  {journal} {\bibinfo
  {journal} {{ACS} Nano}\ }\textbf {\bibinfo {volume} {4}},\ \bibinfo {pages}
  {1889} (\bibinfo {year} {2010})}\BibitemShut {NoStop}%
\bibitem [{\citenamefont {Xu}\ \emph {et~al.}(2014)\citenamefont {Xu},
  \citenamefont {Pereira}, \citenamefont {Wang}, \citenamefont {Wu},
  \citenamefont {Zhang}, \citenamefont {Zhao}, \citenamefont {Bae},
  \citenamefont {Bui}, \citenamefont {Xie}, \citenamefont {Thong},
  \citenamefont {Hong}, \citenamefont {Loh}, \citenamefont {Donadio},
  \citenamefont {Li},\ and\ \citenamefont {Özyilmaz}}]{xu2014length}%
  \BibitemOpen
  \bibfield  {author} {\bibinfo {author} {\bibfnamefont {X.}~\bibnamefont
  {Xu}}, \bibinfo {author} {\bibfnamefont {L.~F.~C.}\ \bibnamefont {Pereira}},
  \bibinfo {author} {\bibfnamefont {Y.}~\bibnamefont {Wang}}, \bibinfo {author}
  {\bibfnamefont {J.}~\bibnamefont {Wu}}, \bibinfo {author} {\bibfnamefont
  {K.}~\bibnamefont {Zhang}}, \bibinfo {author} {\bibfnamefont
  {X.}~\bibnamefont {Zhao}}, \bibinfo {author} {\bibfnamefont {S.}~\bibnamefont
  {Bae}}, \bibinfo {author} {\bibfnamefont {C.~T.}\ \bibnamefont {Bui}},
  \bibinfo {author} {\bibfnamefont {R.}~\bibnamefont {Xie}}, \bibinfo {author}
  {\bibfnamefont {J.~T.~L.}\ \bibnamefont {Thong}}, \bibinfo {author}
  {\bibfnamefont {B.~H.}\ \bibnamefont {Hong}}, \bibinfo {author}
  {\bibfnamefont {K.~P.}\ \bibnamefont {Loh}}, \bibinfo {author} {\bibfnamefont
  {D.}~\bibnamefont {Donadio}}, \bibinfo {author} {\bibfnamefont
  {B.}~\bibnamefont {Li}},\ and\ \bibinfo {author} {\bibfnamefont
  {B.}~\bibnamefont {Özyilmaz}},\ }\bibfield  {title} {\bibinfo {title}
  {Length-dependent thermal conductivity in suspended single-layer graphene},\
  }\href {https://doi.org/10.1038/ncomms4689} {\bibfield  {journal} {\bibinfo
  {journal} {Nat. Commun.}\ }\textbf {\bibinfo {volume} {5}},\ \bibinfo {pages}
  {3689} (\bibinfo {year} {2014})}\BibitemShut {NoStop}%
\bibitem [{\citenamefont {Lee}\ \emph {et~al.}(2011)\citenamefont {Lee},
  \citenamefont {Yoon}, \citenamefont {Kim}, \citenamefont {Lee},\ and\
  \citenamefont {Cheong}}]{lee2011thermal}%
  \BibitemOpen
  \bibfield  {author} {\bibinfo {author} {\bibfnamefont {J.-U.}\ \bibnamefont
  {Lee}}, \bibinfo {author} {\bibfnamefont {D.}~\bibnamefont {Yoon}}, \bibinfo
  {author} {\bibfnamefont {H.}~\bibnamefont {Kim}}, \bibinfo {author}
  {\bibfnamefont {S.~W.}\ \bibnamefont {Lee}},\ and\ \bibinfo {author}
  {\bibfnamefont {H.}~\bibnamefont {Cheong}},\ }\bibfield  {title} {\bibinfo
  {title} {Thermal conductivity of suspended pristine graphene measured by
  raman spectroscopy},\ }\href {https://doi.org/10.1103/physrevb.83.081419}
  {\bibfield  {journal} {\bibinfo  {journal} {Phys. Rev. B}\ }\textbf {\bibinfo
  {volume} {83}},\ \bibinfo {pages} {081419} (\bibinfo {year}
  {2011})}\BibitemShut {NoStop}%
\bibitem [{\citenamefont {Nobakht}\ and\ \citenamefont
  {Shin}(2016)}]{yousefzadi2016anisotropic}%
  \BibitemOpen
  \bibfield  {author} {\bibinfo {author} {\bibfnamefont {A.~Y.}\ \bibnamefont
  {Nobakht}}\ and\ \bibinfo {author} {\bibfnamefont {S.}~\bibnamefont {Shin}},\
  }\bibfield  {title} {\bibinfo {title} {Anisotropic control of thermal
  transport in graphene/si heterostructures},\ }\href
  {https://doi.org/10.1063/1.4971873} {\bibfield  {journal} {\bibinfo
  {journal} {J. Appl. Phys.}\ }\textbf {\bibinfo {volume} {120}},\ \bibinfo
  {pages} {225111} (\bibinfo {year} {2016})}\BibitemShut {NoStop}%
\bibitem [{\citenamefont {Balandin}(2011)}]{balandin2011thermal}%
  \BibitemOpen
  \bibfield  {author} {\bibinfo {author} {\bibfnamefont {A.~A.}\ \bibnamefont
  {Balandin}},\ }\bibfield  {title} {\bibinfo {title} {Thermal properties of
  graphene and nanostructured carbon materials},\ }\href
  {https://doi.org/10.1038/nmat3064} {\bibfield  {journal} {\bibinfo  {journal}
  {Nat. Mater.}\ }\textbf {\bibinfo {volume} {10}},\ \bibinfo {pages} {569}
  (\bibinfo {year} {2011})}\BibitemShut {NoStop}%
\bibitem [{\citenamefont {Pereira}\ and\ \citenamefont
  {Donadio}(2013)}]{pereira2013divergence}%
  \BibitemOpen
  \bibfield  {author} {\bibinfo {author} {\bibfnamefont {L.~F.~C.}\
  \bibnamefont {Pereira}}\ and\ \bibinfo {author} {\bibfnamefont
  {D.}~\bibnamefont {Donadio}},\ }\bibfield  {title} {\bibinfo {title}
  {Divergence of the thermal conductivity in uniaxially strained graphene},\
  }\href {https://doi.org/10.1103/physrevb.87.125424} {\bibfield  {journal}
  {\bibinfo  {journal} {Phys. Rev. B}\ }\textbf {\bibinfo {volume} {87}},\
  \bibinfo {pages} {125424} (\bibinfo {year} {2013})}\BibitemShut {NoStop}%
\bibitem [{\citenamefont {Ghosh}\ \emph {et~al.}(2008)\citenamefont {Ghosh},
  \citenamefont {Calizo}, \citenamefont {Teweldebrhan}, \citenamefont
  {Pokatilov}, \citenamefont {Nika}, \citenamefont {Balandin}, \citenamefont
  {Bao}, \citenamefont {Miao},\ and\ \citenamefont {Lau}}]{ghosh2008extremely}%
  \BibitemOpen
  \bibfield  {author} {\bibinfo {author} {\bibfnamefont {S.}~\bibnamefont
  {Ghosh}}, \bibinfo {author} {\bibfnamefont {I.}~\bibnamefont {Calizo}},
  \bibinfo {author} {\bibfnamefont {D.}~\bibnamefont {Teweldebrhan}}, \bibinfo
  {author} {\bibfnamefont {E.~P.}\ \bibnamefont {Pokatilov}}, \bibinfo {author}
  {\bibfnamefont {D.~L.}\ \bibnamefont {Nika}}, \bibinfo {author}
  {\bibfnamefont {A.~A.}\ \bibnamefont {Balandin}}, \bibinfo {author}
  {\bibfnamefont {W.}~\bibnamefont {Bao}}, \bibinfo {author} {\bibfnamefont
  {F.}~\bibnamefont {Miao}},\ and\ \bibinfo {author} {\bibfnamefont {C.~N.}\
  \bibnamefont {Lau}},\ }\bibfield  {title} {\bibinfo {title} {Extremely high
  thermal conductivity of graphene: Prospects for thermal management
  applications in nanoelectronic circuits},\ }\href
  {https://doi.org/10.1063/1.2907977} {\bibfield  {journal} {\bibinfo
  {journal} {Appl. Phys. Lett.}\ }\textbf {\bibinfo {volume} {92}},\ \bibinfo
  {pages} {151911} (\bibinfo {year} {2008})}\BibitemShut {NoStop}%
\bibitem [{\citenamefont {Kim}\ \emph {et~al.}(2016)\citenamefont {Kim},
  \citenamefont {Park},\ and\ \citenamefont {Marzari}}]{kim2016electronic}%
  \BibitemOpen
  \bibfield  {author} {\bibinfo {author} {\bibfnamefont {T.~Y.}\ \bibnamefont
  {Kim}}, \bibinfo {author} {\bibfnamefont {C.-H.}\ \bibnamefont {Park}},\ and\
  \bibinfo {author} {\bibfnamefont {N.}~\bibnamefont {Marzari}},\ }\bibfield
  {title} {\bibinfo {title} {The electronic thermal conductivity of graphene},\
  }\href {https://doi.org/10.1021/acs.nanolett.5b05288} {\bibfield  {journal}
  {\bibinfo  {journal} {Nano Lett.}\ }\textbf {\bibinfo {volume} {16}},\
  \bibinfo {pages} {2439} (\bibinfo {year} {2016})}\BibitemShut {NoStop}%
\bibitem [{\citenamefont {Lindsay}\ \emph {et~al.}(2010)\citenamefont
  {Lindsay}, \citenamefont {Broido},\ and\ \citenamefont
  {Mingo}}]{lindsay2010flexural}%
  \BibitemOpen
  \bibfield  {author} {\bibinfo {author} {\bibfnamefont {L.}~\bibnamefont
  {Lindsay}}, \bibinfo {author} {\bibfnamefont {D.~A.}\ \bibnamefont
  {Broido}},\ and\ \bibinfo {author} {\bibfnamefont {N.}~\bibnamefont
  {Mingo}},\ }\bibfield  {title} {\bibinfo {title} {Flexural phonons and
  thermal transport in graphene},\ }\href
  {https://doi.org/10.1103/physrevb.82.115427} {\bibfield  {journal} {\bibinfo
  {journal} {Phys. Rev. B}\ }\textbf {\bibinfo {volume} {82}},\ \bibinfo
  {pages} {115427} (\bibinfo {year} {2010})}\BibitemShut {NoStop}%
\bibitem [{\citenamefont {Zhang}\ \emph {et~al.}(2011)\citenamefont {Zhang},
  \citenamefont {Lee},\ and\ \citenamefont {Cho}}]{zhang2011thermal}%
  \BibitemOpen
  \bibfield  {author} {\bibinfo {author} {\bibfnamefont {H.}~\bibnamefont
  {Zhang}}, \bibinfo {author} {\bibfnamefont {G.}~\bibnamefont {Lee}},\ and\
  \bibinfo {author} {\bibfnamefont {K.}~\bibnamefont {Cho}},\ }\bibfield
  {title} {\bibinfo {title} {Thermal transport in graphene and effects of
  vacancy defects},\ }\href {https://doi.org/10.1103/physrevb.84.115460}
  {\bibfield  {journal} {\bibinfo  {journal} {Phys. Rev. B}\ }\textbf {\bibinfo
  {volume} {84}},\ \bibinfo {pages} {115460} (\bibinfo {year}
  {2011})}\BibitemShut {NoStop}%
\bibitem [{\citenamefont {Tuckerman}(2010)}]{tuckerman2010statistical}%
  \BibitemOpen
  \bibfield  {author} {\bibinfo {author} {\bibfnamefont {M.}~\bibnamefont
  {Tuckerman}},\ }\href@noop {} {\emph {\bibinfo {title} {Statistical
  mechanics: theory and molecular simulation}}}\ (\bibinfo  {publisher} {Oxford
  university press},\ \bibinfo {year} {2010})\BibitemShut {NoStop}%
\bibitem [{\citenamefont {Schelling}\ \emph {et~al.}(2002)\citenamefont
  {Schelling}, \citenamefont {Phillpot},\ and\ \citenamefont
  {Keblinski}}]{schelling2002comparison}%
  \BibitemOpen
  \bibfield  {author} {\bibinfo {author} {\bibfnamefont {P.~K.}\ \bibnamefont
  {Schelling}}, \bibinfo {author} {\bibfnamefont {S.~R.}\ \bibnamefont
  {Phillpot}},\ and\ \bibinfo {author} {\bibfnamefont {P.}~\bibnamefont
  {Keblinski}},\ }\bibfield  {title} {\bibinfo {title} {Comparison of
  atomic-level simulation methods for computing thermal conductivity},\ }\href
  {https://doi.org/10.1103/physrevb.65.144306} {\bibfield  {journal} {\bibinfo
  {journal} {Phys. Rev. B}\ }\textbf {\bibinfo {volume} {65}},\ \bibinfo
  {pages} {144306} (\bibinfo {year} {2002})}\BibitemShut {NoStop}%
\bibitem [{\citenamefont {Fan}\ \emph {et~al.}(2015)\citenamefont {Fan},
  \citenamefont {Pereira}, \citenamefont {Wang}, \citenamefont {Zheng},
  \citenamefont {Donadio},\ and\ \citenamefont {Harju}}]{fan2015force}%
  \BibitemOpen
  \bibfield  {author} {\bibinfo {author} {\bibfnamefont {Z.}~\bibnamefont
  {Fan}}, \bibinfo {author} {\bibfnamefont {L.~F.~C.}\ \bibnamefont {Pereira}},
  \bibinfo {author} {\bibfnamefont {H.-Q.}\ \bibnamefont {Wang}}, \bibinfo
  {author} {\bibfnamefont {J.-C.}\ \bibnamefont {Zheng}}, \bibinfo {author}
  {\bibfnamefont {D.}~\bibnamefont {Donadio}},\ and\ \bibinfo {author}
  {\bibfnamefont {A.}~\bibnamefont {Harju}},\ }\bibfield  {title} {\bibinfo
  {title} {Force and heat current formulas for many-body potentials in
  molecular dynamics simulations with applications to thermal conductivity
  calculations},\ }\href {https://doi.org/10.1103/physrevb.92.094301}
  {\bibfield  {journal} {\bibinfo  {journal} {Phys. Rev. B}\ }\textbf {\bibinfo
  {volume} {92}},\ \bibinfo {pages} {094301} (\bibinfo {year}
  {2015})}\BibitemShut {NoStop}%
\bibitem [{\citenamefont {Admal}\ and\ \citenamefont
  {Tadmor}(2011)}]{admal2011stress}%
  \BibitemOpen
  \bibfield  {author} {\bibinfo {author} {\bibfnamefont {N.~C.}\ \bibnamefont
  {Admal}}\ and\ \bibinfo {author} {\bibfnamefont {E.~B.}\ \bibnamefont
  {Tadmor}},\ }\bibfield  {title} {\bibinfo {title} {Stress and heat flux for
  arbitrary multibody potentials: A unified framework},\ }\href
  {https://doi.org/10.1063/1.3582905} {\bibfield  {journal} {\bibinfo
  {journal} {J. Chem. Phys.}\ }\textbf {\bibinfo {volume} {134}},\ \bibinfo
  {pages} {184106} (\bibinfo {year} {2011})}\BibitemShut {NoStop}%
\bibitem [{\citenamefont {Banhart}\ \emph {et~al.}(2010)\citenamefont
  {Banhart}, \citenamefont {Kotakoski},\ and\ \citenamefont
  {Krasheninnikov}}]{banhart2010structural}%
  \BibitemOpen
  \bibfield  {author} {\bibinfo {author} {\bibfnamefont {F.}~\bibnamefont
  {Banhart}}, \bibinfo {author} {\bibfnamefont {J.}~\bibnamefont {Kotakoski}},\
  and\ \bibinfo {author} {\bibfnamefont {A.~V.}\ \bibnamefont
  {Krasheninnikov}},\ }\bibfield  {title} {\bibinfo {title} {Structural defects
  in graphene},\ }\href {https://doi.org/10.1021/nn102598m} {\bibfield
  {journal} {\bibinfo  {journal} {{ACS} Nano}\ }\textbf {\bibinfo {volume}
  {5}},\ \bibinfo {pages} {26} (\bibinfo {year} {2010})}\BibitemShut {NoStop}%
\bibitem [{\citenamefont {Skowron}\ \emph {et~al.}(2015)\citenamefont
  {Skowron}, \citenamefont {Lebedeva}, \citenamefont {Popov},\ and\
  \citenamefont {Bichoutskaia}}]{skowron2015energetics}%
  \BibitemOpen
  \bibfield  {author} {\bibinfo {author} {\bibfnamefont {S.~T.}\ \bibnamefont
  {Skowron}}, \bibinfo {author} {\bibfnamefont {I.~V.}\ \bibnamefont
  {Lebedeva}}, \bibinfo {author} {\bibfnamefont {A.~M.}\ \bibnamefont
  {Popov}},\ and\ \bibinfo {author} {\bibfnamefont {E.}~\bibnamefont
  {Bichoutskaia}},\ }\bibfield  {title} {\bibinfo {title} {Energetics of atomic
  scale structure changes in graphene},\ }\href
  {https://doi.org/10.1039/c4cs00499j} {\bibfield  {journal} {\bibinfo
  {journal} {Chem. Soc. Rev.}\ }\textbf {\bibinfo {volume} {44}},\ \bibinfo
  {pages} {3143} (\bibinfo {year} {2015})}\BibitemShut {NoStop}%
\bibitem [{\citenamefont {Gass}\ \emph {et~al.}(2008)\citenamefont {Gass},
  \citenamefont {Bangert}, \citenamefont {Bleloch}, \citenamefont {Wang},
  \citenamefont {Nair},\ and\ \citenamefont {Geim}}]{gass2008free}%
  \BibitemOpen
  \bibfield  {author} {\bibinfo {author} {\bibfnamefont {M.~H.}\ \bibnamefont
  {Gass}}, \bibinfo {author} {\bibfnamefont {U.}~\bibnamefont {Bangert}},
  \bibinfo {author} {\bibfnamefont {A.~L.}\ \bibnamefont {Bleloch}}, \bibinfo
  {author} {\bibfnamefont {P.}~\bibnamefont {Wang}}, \bibinfo {author}
  {\bibfnamefont {R.~R.}\ \bibnamefont {Nair}},\ and\ \bibinfo {author}
  {\bibfnamefont {A.~K.}\ \bibnamefont {Geim}},\ }\bibfield  {title} {\bibinfo
  {title} {Free-standing graphene at atomic resolution},\ }\href
  {https://doi.org/10.1038/nnano.2008.280} {\bibfield  {journal} {\bibinfo
  {journal} {Nat. Nanotechnol.}\ }\textbf {\bibinfo {volume} {3}},\ \bibinfo
  {pages} {676} (\bibinfo {year} {2008})}\BibitemShut {NoStop}%
\bibitem [{\citenamefont {Zhang}\ \emph {et~al.}(2009)\citenamefont {Zhang},
  \citenamefont {Tang}, \citenamefont {Girit}, \citenamefont {Hao},
  \citenamefont {Martin}, \citenamefont {Zettl}, \citenamefont {Crommie},
  \citenamefont {Shen},\ and\ \citenamefont {Wang}}]{zhang2009direct}%
  \BibitemOpen
  \bibfield  {author} {\bibinfo {author} {\bibfnamefont {Y.}~\bibnamefont
  {Zhang}}, \bibinfo {author} {\bibfnamefont {T.-T.}\ \bibnamefont {Tang}},
  \bibinfo {author} {\bibfnamefont {C.}~\bibnamefont {Girit}}, \bibinfo
  {author} {\bibfnamefont {Z.}~\bibnamefont {Hao}}, \bibinfo {author}
  {\bibfnamefont {M.~C.}\ \bibnamefont {Martin}}, \bibinfo {author}
  {\bibfnamefont {A.}~\bibnamefont {Zettl}}, \bibinfo {author} {\bibfnamefont
  {M.~F.}\ \bibnamefont {Crommie}}, \bibinfo {author} {\bibfnamefont {Y.~R.}\
  \bibnamefont {Shen}},\ and\ \bibinfo {author} {\bibfnamefont
  {F.}~\bibnamefont {Wang}},\ }\bibfield  {title} {\bibinfo {title} {Direct
  observation of a widely tunable bandgap in bilayer graphene},\ }\href
  {https://doi.org/10.1038/nature08105} {\bibfield  {journal} {\bibinfo
  {journal} {Nature}\ }\textbf {\bibinfo {volume} {459}},\ \bibinfo {pages}
  {820} (\bibinfo {year} {2009})}\BibitemShut {NoStop}%
\bibitem [{\citenamefont {Haskins}\ \emph {et~al.}(2011)\citenamefont
  {Haskins}, \citenamefont {K{\i}nac{\i}}, \citenamefont {Sevik}, \citenamefont
  {Sevin{\c{c}}li}, \citenamefont {Cuniberti},\ and\ \citenamefont
  {{\c{C}}a{\u{g}}{\i}n}}]{haskins2011control}%
  \BibitemOpen
  \bibfield  {author} {\bibinfo {author} {\bibfnamefont {J.}~\bibnamefont
  {Haskins}}, \bibinfo {author} {\bibfnamefont {A.}~\bibnamefont
  {K{\i}nac{\i}}}, \bibinfo {author} {\bibfnamefont {C.}~\bibnamefont {Sevik}},
  \bibinfo {author} {\bibfnamefont {H.}~\bibnamefont {Sevin{\c{c}}li}},
  \bibinfo {author} {\bibfnamefont {G.}~\bibnamefont {Cuniberti}},\ and\
  \bibinfo {author} {\bibfnamefont {T.}~\bibnamefont {{\c{C}}a{\u{g}}{\i}n}},\
  }\bibfield  {title} {\bibinfo {title} {Control of thermal and electronic
  transport in defect-engineered graphene nanoribbons},\ }\href
  {https://doi.org/10.1021/nn200114p} {\bibfield  {journal} {\bibinfo
  {journal} {{ACS} Nano}\ }\textbf {\bibinfo {volume} {5}},\ \bibinfo {pages}
  {3779} (\bibinfo {year} {2011})}\BibitemShut {NoStop}%
\bibitem [{\citenamefont {Telling}\ \emph {et~al.}(2003)\citenamefont
  {Telling}, \citenamefont {Ewels}, \citenamefont {El-Barbary},\ and\
  \citenamefont {Heggie}}]{telling2003wigner}%
  \BibitemOpen
  \bibfield  {author} {\bibinfo {author} {\bibfnamefont {R.~H.}\ \bibnamefont
  {Telling}}, \bibinfo {author} {\bibfnamefont {C.~P.}\ \bibnamefont {Ewels}},
  \bibinfo {author} {\bibfnamefont {A.~A.}\ \bibnamefont {El-Barbary}},\ and\
  \bibinfo {author} {\bibfnamefont {M.~I.}\ \bibnamefont {Heggie}},\ }\bibfield
   {title} {\bibinfo {title} {Wigner defects bridge the graphite gap},\ }\href
  {https://doi.org/10.1038/nmat876} {\bibfield  {journal} {\bibinfo  {journal}
  {Nat. Mater.}\ }\textbf {\bibinfo {volume} {2}},\ \bibinfo {pages} {333}
  (\bibinfo {year} {2003})}\BibitemShut {NoStop}%
\bibitem [{\citenamefont {Vicarelli}\ \emph {et~al.}(2015)\citenamefont
  {Vicarelli}, \citenamefont {Heerema}, \citenamefont {Dekker},\ and\
  \citenamefont {Zandbergen}}]{vicarelli:heerema:2015}%
  \BibitemOpen
  \bibfield  {author} {\bibinfo {author} {\bibfnamefont {L.}~\bibnamefont
  {Vicarelli}}, \bibinfo {author} {\bibfnamefont {S.~J.}\ \bibnamefont
  {Heerema}}, \bibinfo {author} {\bibfnamefont {C.}~\bibnamefont {Dekker}},\
  and\ \bibinfo {author} {\bibfnamefont {H.~W.}\ \bibnamefont {Zandbergen}},\
  }\bibfield  {title} {\bibinfo {title} {Controlling defects in graphene for
  optimizing the electrical properties of graphene nanodevices},\ }\href
  {https://doi.org/10.1021/acsnano.5b01762} {\bibfield  {journal} {\bibinfo
  {journal} {ACS Nano}\ }\textbf {\bibinfo {volume} {9}},\ \bibinfo {pages}
  {3428} (\bibinfo {year} {2015})}\BibitemShut {NoStop}%
\bibitem [{\citenamefont {Teobaldi}\ \emph {et~al.}(2010)\citenamefont
  {Teobaldi}, \citenamefont {Ohnishi}, \citenamefont {Tanimura},\ and\
  \citenamefont {Shluger}}]{teobaldi2010effect}%
  \BibitemOpen
  \bibfield  {author} {\bibinfo {author} {\bibfnamefont {G.}~\bibnamefont
  {Teobaldi}}, \bibinfo {author} {\bibfnamefont {H.}~\bibnamefont {Ohnishi}},
  \bibinfo {author} {\bibfnamefont {K.}~\bibnamefont {Tanimura}},\ and\
  \bibinfo {author} {\bibfnamefont {A.~L.}\ \bibnamefont {Shluger}},\
  }\bibfield  {title} {\bibinfo {title} {The effect of van der waals
  interactions on the properties of intrinsic defects in graphite},\ }\href
  {https://doi.org/10.1016/j.carbon.2010.07.029} {\bibfield  {journal}
  {\bibinfo  {journal} {Carbon}\ }\textbf {\bibinfo {volume} {48}},\ \bibinfo
  {pages} {4145} (\bibinfo {year} {2010})}\BibitemShut {NoStop}%
\bibitem [{\citenamefont {Wen}(2019{\natexlab{a}})}]{MD_435082866799_001}%
  \BibitemOpen
  \bibfield  {author} {\bibinfo {author} {\bibfnamefont {M.}~\bibnamefont
  {Wen}},\ }\href {https://doi.org/10.25950/9fa4935a} {\bibinfo {title} {{A}
  hybrid neural network model driver for multilayer two-dimensional materials
  developed by {W}en and {T}admor (2019) v001}},\ \bibinfo {howpublished}
  {OpenKIM, \url{https://doi:10.25950/ff8f563a}} (\bibinfo {year}
  {2019}{\natexlab{a}})\BibitemShut {NoStop}%
\bibitem [{\citenamefont {Wen}(2019{\natexlab{b}})}]{MO_421038499185_001}%
  \BibitemOpen
  \bibfield  {author} {\bibinfo {author} {\bibfnamefont {M.}~\bibnamefont
  {Wen}},\ }\href {https://doi.org/10.25950/a74cc44e} {\bibinfo {title} {{A}
  hybrid neural network potential for multilayer graphene systems developed by
  {W}en and {T}admor (2019) v001}},\ \bibinfo {howpublished} {OpenKIM,
  \url{https://doi.org/10.25950/a74cc44e}} (\bibinfo {year}
  {2019}{\natexlab{b}})\BibitemShut {NoStop}%
\end{thebibliography}
%

\end{document}


\title{Supplemental Material for: A hybrid neural network potential for multilayer graphene}

\author{Mingjian Wen}
\affiliation{Department of Aerospace Engineering and Mechanics, University of Minnesota, Minneapolis, MN 55455, USA}

\author{Ellad B. Tadmor}
\email[Author to whom correspondence should be addressed: ]{tadmor@umn.edu}
\affiliation{Department of Aerospace Engineering and Mechanics, University of Minnesota, Minneapolis, MN 55455, USA}

\maketitle

\section{Dataset}
\label{sec:data:set}

The dataset consists of energies and forces for pristine and defected monolayer graphene,
bilayer graphene, and graphite in various states.
The configurations in the dataset are generated in two ways: (1)
crystals with distortions (compression and stretching of the simulation cell
together with random perturbations of atoms),
and (2) configurations drawn from \emph{ab initio} molecular dynamics (AIMD)
trajectories at 300, 900, and 1500~K.

For monolayer graphene, the configurations include:
\begin{itemize}
  \item pristine
  \begin{itemize}
    \item In-plane compressed and stretched monolayers
    \item AIMD trajectories
  \end{itemize}
  \item defected
  \begin{itemize}
    \item Configurations from the minimization of a monolayer with a single vacancy
    \item AIMD trajectories of monolayers with a single vacancy
  \end{itemize}
\end{itemize}

For bilayer graphene, the configurations include:
\begin{itemize}
  \item pristine
  \begin{itemize}
    \item AB-stacked bilayers with compression and stretching in the basal plane
    \item Bilayers with different translational registry (e.g.\ AA, AB, and
    SP) at various layer separations
    \item Twisted bilayers with different twisting angles at various layer
    separations
    \item AIMD trajectories of twisted bilayers and bilayers in AB and AA stackings
  \end{itemize}
  \item defected
  \begin{itemize}
    \item Configurations from the minimization of a bilayer with a single vacancy in each layer
    \item AIMD trajectories of a bilayer with a single vacancy in one layer and
    the other layer pristine
    \item AIMD trajectories of a bilayer with a single vacancy in each layer;
    Initial configuration without interlayer bonds
    \item AIMD trajectories of a bilayer with a single vacancy in each layer;
    Initial configuration with interlayer bonds formed
  \end{itemize}
\end{itemize}

For graphite, the configurations include:
\begin{itemize}
  \item pristine
  \begin{itemize}
    \item Graphite with compression and stretching in the basal plane
    \item Graphite with compression and stretching along the $c$-axis
    \item AIMD trajectories
  \end{itemize}
\end{itemize}

\section{Atomic neighborhood descriptors}
\label{sec:descriptors}

The descriptors used to characterize atomic neighborhoods in this work are the
symmetry functions proposed by Behler and coworkers \cite{behler2011atom, artrith2012high}.
The function used to describe the radial environment of atom $\alpha$ is
\begin{equation}\label{eq:g2}
    G_\alpha^2 = \sum_{\beta\neq\alpha}^N e^{-\eta (r_{\alpha\beta} - R_\text{s})^2} f_\text{c}(r_{\alpha\beta}),
\end{equation}
and the function used for the angular environment of atom $\alpha$ is
\begin{equation}\label{eq:g3}
    G_\alpha^4 = 2^{1-\zeta} \sum_{\beta \neq \alpha}^N\sum_{\substack{\gamma>\beta\\ \gamma\neq \alpha}}^N
    (1+\lambda \cos\theta_{\beta\alpha\gamma})^\zeta
    e^{-\eta(r_{\alpha\beta}^2 + r_{\alpha\gamma}^2 + r_{\beta\gamma}^2)}
    f_\text{c}(r_{\alpha\beta}) f_\text{c}(r_{\alpha\gamma}) f_\text{c}(r_{\beta\gamma}),
\end{equation}
where $N$ is the total number of atoms,
$r_{\alpha\beta}$ is the distance between atoms $\alpha$ and $\beta$, and
$\theta_{\beta\alpha\gamma}$ is the angle between bonds $\beta$--$\alpha$ and $\gamma$--$\alpha$ with the vertex at
atom $\alpha$.
The cutoff function takes the form
\begin{equation}
f_\text{c}(r) =
\begin{cases}
  \dfrac{1}{2} \left(\cos\dfrac{\pi r}{R_\text{c}} + 1\right),  & r\leq R_\text{c} \\
  0, &r > R_\text{c}
\end{cases},
\end{equation}
in which the cutoff distance $R_\text{c}$ is set to $r_\text{cutoff}^\text{short} =
5~\text{\AA}$ defined in the main text.
The hyperparameters $\eta$ and $R_\text{s}$ in \eref{eq:g2} are listed in
\tref{tab:params:g2} and the hyperparameters $\zeta, \lambda$, and $\eta$ in
\eref{eq:g3} are listed in in \tref{tab:params:g4}.
The hyperparameters are taken from \cite{artrith2012high}.

\renewcommand{\arraystretch}{0.65}
\newcolumntype{C}[1]{>{\centering\let\newline\\\arraybackslash\hspace{0pt}}m{#1}}
\begin{table}[H]
\caption{Hyperparameters used in the radial descriptor $G_\alpha^2$.}
\label{tab:params:g2}
\centering
\begin{tabular}{C{10mm}C{20mm}C{20mm}}
\toprule
No.  & $\eta$ (Bohr$^{-2}$)  &$R_\text{s}$ (Bohr) \\
\hline
1    &0.001  &0\\
2    &0.01   &0\\
3    &0.02   &0\\
4    &0.035  &0\\
5    &0.06   &0\\
6    &0.1    &0\\
7    &0.2    &0\\
8    &0.4    &0\\
\botrule
\end{tabular}
\end{table}

\begin{table}[H]
\caption{Hyperparameters used in the angular descriptor $G_\alpha^4$.}
\label{tab:params:g4}
\centering
\begin{tabular}{C{10mm}C{10mm}C{10mm}C{20mm}|C{10mm}C{10mm}C{10mm}C{20mm}}
\toprule
No.  & $\zeta$  &$\lambda$  &$\eta$ (Bohr$^{-2}$)  &No.  & $\zeta$  &$\lambda$  &$\eta$ (Bohr$^{-2}$) \\
\hline
1  & 1  &$-1$  &0.0001  &23  & 2  &$-1$  &0.025 \\
2  & 1  &$ 1$  &0.0001  &24  & 2  &$ 1$  &0.025 \\
3  & 2  &$-1$  &0.0001  &25  & 4  &$-1$  &0.025 \\
4  & 2  &$ 1$  &0.0001  &26  & 4  &$ 1$  &0.025 \\
5  & 1  &$-1$  &0.003  &27  &16  &$-1$  &0.025 \\
6  & 1  &$ 1$  &0.003  &28  &16  &$ 1$  &0.025 \\
7  & 2  &$-1$  &0.003  &29  & 1  &$-1$  &0.045 \\
8  & 2  &$ 1$  &0.003  &30  & 1  &$ 1$  &0.045 \\
9  & 1  &$-1$  &0.008  &31  & 2  &$-1$  &0.045 \\
10  & 1  &$ 1$  &0.008  &32  & 2  &$ 1$  &0.045 \\
11  & 2  &$-1$  &0.008  &33  & 4  &$-1$  &0.045 \\
12  & 2  &$ 1$  &0.008  &34  & 4  &$ 1$  &0.045 \\
13  & 1  &$-1$  &0.015  &35  &16  &$-1$  &0.045 \\
14  & 1  &$ 1$  &0.015  &36  &16  &$ 1$  &0.045 \\
15  & 2  &$-1$  &0.015  &37  & 1  &$-1$  &0.08 \\
16  & 2  &$ 1$  &0.015  &38  & 1  &$ 1$  &0.08 \\
17  & 4  &$-1$  &0.015  &39  & 2  &$-1$  &0.08 \\
18  & 4  &$ 1$  &0.015  &40  & 2  &$ 1$  &0.08 \\
19  &16  &$-1$  &0.015  &41  & 4  &$-1$  &0.08 \\
20  &16  &$ 1$  &0.015  &42  & 4  &$ 1$  &0.08 \\
21  & 1  &$-1$  &0.025  &43  &16  &$ 1$  &0.08 \\
22  & 1  &$ 1$  &0.025  & & & & \\
\botrule
\end{tabular}
\end{table}

\section{Concerted exchange in monolayer graphene}

\begin{figure}
\includegraphics[width=0.7\columnwidth]{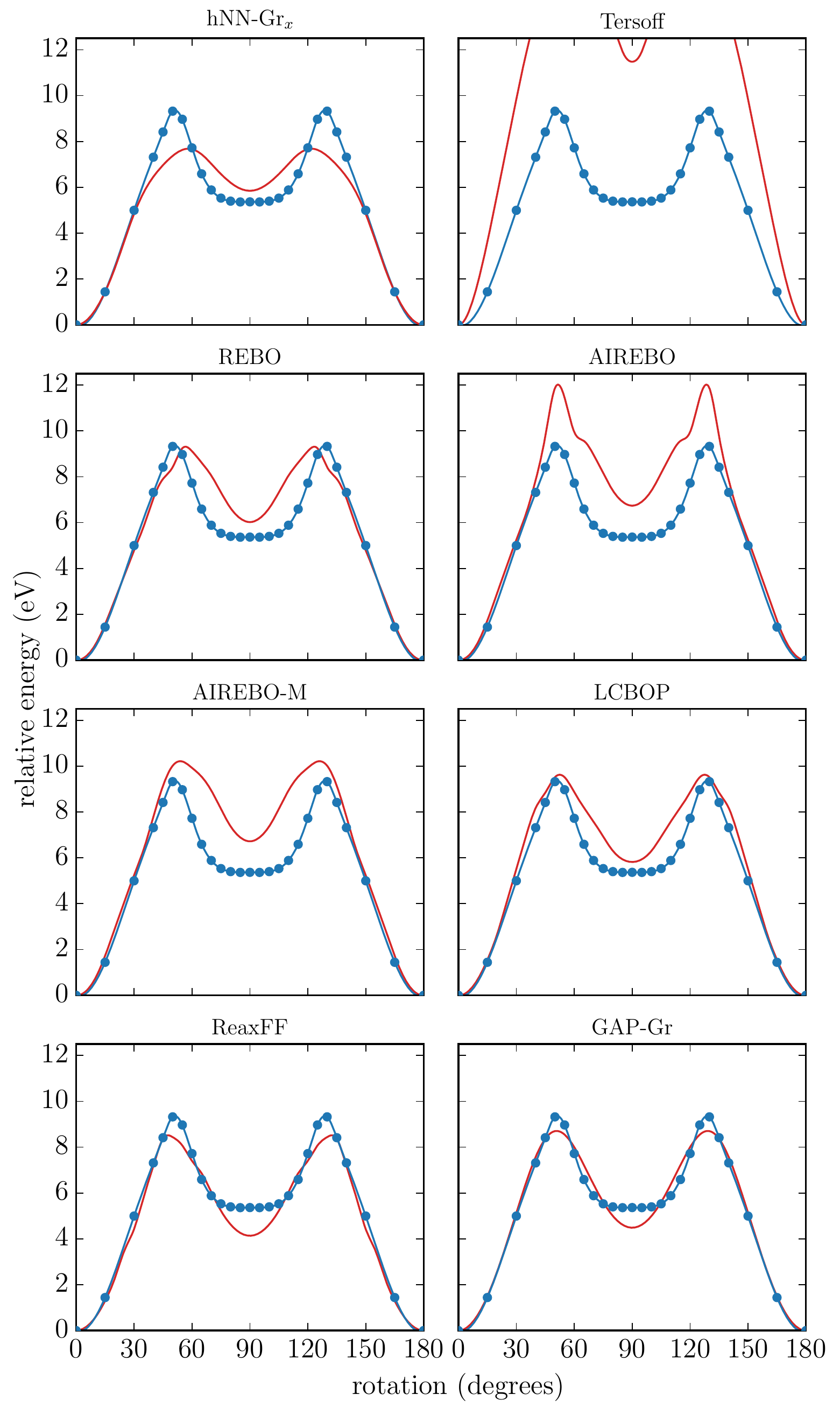}
\caption{\label{fig:e:v:rotation}Energy versus angle in creating the
Stone--Wales defect by rotating a pair of atoms in a monolayer graphene
predicted by DFT (blue dots) and potentials (red curve).
Note that parts of the prediction from the Tersoff potential are not shown, but
the energy barrier and its location are listed in \tref{tab:formation:energy}.}
\end{figure}

\begin{table}
\caption{\label{tab:formation:energy}Formation energy and the energy barrier to
form the Stone--Wales defect predicted by various potentials compared to the DFT
result.}
\begin{ruledtabular}
\begin{tabular}{cccc}
 Method   &Formation energy (eV)  &Energy barrier (eV)  & Barrier location ($^\circ$) \\
\hline
DFT       &5.37     &9.32     &51     \\
\hnn      &5.85     &7.68     &58     \\
AIREBO    &6.74     &12.02    &52     \\
AIREBO--M &6.72     &10.21    &54     \\
LCBOP     &5.82     &9.62     &52     \\
ReaxFF    &4.14     &8.52     &48     \\
Tersoff   &11.48   &19.01     &71     \\
REBO      &6.02     &9.31     &57     \\
GAP--Gr   &4.49     &8.71     &51     \\
\end{tabular}
\end{ruledtabular}
\end{table}

In the concerted exchange mechanism a pair of atoms is rotated
by 90 degrees to form a Stone--Wales defect. The energy verses
angle curves for a number of empirical potentials along with DFT
results are presented in \fref{fig:e:v:rotation}.
For each potential and for DFT, the initial monolayer graphene
was constructed using its corresponding equilibrium in-plane lattice parameter.
At each angle, the two rotating atoms are constrained to relax along the line
connecting them, whereas the other atoms are free to relax within the graphene
plane as discussed in the main text.
The DFT prediction is interpolated using a cubic spline.
We see from \fref{fig:e:v:rotation} that in general the ReaxFF and GAP--Gr
potentials follow the shape of the DFT prediction better than the other
potentials.
The Stone--Wales formation energy, and the maximum energy barrier and its
location are given in \tref{tab:formation:energy}.
The \hnn and LCBOP potentials give the closest formation energy to DFT, the
REBO potential predicts the best energy barrier, while the GAP--Gr potential
predicts the correct barrier location.
None of the potentials are able to predict the plateau around the
Stone--Wales defect structure at $90^\circ$.

\section{Using the Open Knowledgebase of Interatomic Models (OpenKIM)}
\label{sec:kim}

The Open Knowledgebase of Interatomic Models (OpenKIM)
(\url{https://openkim.org}) is an open-source, publicly accessible repository of
classical interatomic potentials and their predictions for material
properties, which can be visualized and compared with first-principles data.
Interatomic potentials archived in OpenKIM, called ``KIM Models,''
can be used with simulation codes
that are compatible with the KIM Application Programming Interface (API),
such as LAMMPS \cite{plimpton1995fast, lammps}, ASE \cite{larsen2017atomic,ase},
DL\_POLY \cite{dlpoly}, and GULP \cite{gale1997gulp,gulp}.
There are two types of OpenKIM models: Portable Models (PMs) that can be used
in multiple simulation codes, and Simulator Models (SMs) that only work with
a specific simulator. In either case, using a KIM Model is a simple matter of
providing the model's identifier (i.e.\ it's KIM ID), to a simulation
code as specified in its documentation.

As an example, we describe how a KIM Model is used with LAMMPS (August 2019
release of LAMMPS with KIM API version 2.1.).
To use KIM Models with LAMMPS, you will need to have the KIM API, your desired
potential, and a version of LAMMPS with KIM-support enabled,
installed on your machine. For instructions on how to do so,
see the KIM API website (\url{https://openkim.org/kim-api/}) and
the LAMMPS documentation (\url{https://lammps.sandia.gov/doc/Manual.html}).
In a LAMMPS input script, a KIM Model is then selected and activated
using the {\sf kim\_init} and {\sf kim\_interactions} commands. The following
example script shows how to compute the cohesive energy of an AB
graphene bilayer using the potential developed in this paper and implemented as
a KIM Portable Model \cite{MD_435082866799_001,MO_421038499185_001}.
\begin{verbatim}
################################################################################
# Compute the energy of a unit cell of bilayer graphene in AB stacking
# (consisting of 4 atoms) with a lattice parameter of a=2.467 Angstrom and a
# layer spacing of c = 3.457 Angstrom.
# Periodic boundary conditions are applied in the two in-plane directions,
# while the out-of-plane direction is free.
################################################################################

# Initialize interatomic potential (KIM model) and units
atom_style    atomic
kim_init      hNN_WenTadmor_2019Grx_C__MO_421038499185_001 metal

# boundary conditions
boundary      p p f

variable      a equal 2.467
variable      c equal 3.457
variable      coa2 equal 2.0*$c/$a

# create a honeycomb lattice (two layers)
lattice       custom  $a  &
                    a1      1.0         0.0               0.0      &
                    a2      0.5         $(sqrt(3.0)/2.0)  0.0      &
                    a3      0.0         0.0               ${coa2}  &
                    basis   0.0         0.0               0.0      &
                    basis   $(1.0/3.0)  $(1.0/3.0)        0.0      &
                    basis   $(1.0/3.0)  $(1.0/3.0)        0.5      &
                    basis   $(2.0/3.0)  $(2.0/3.0)        0.5      &
                    spacing 1.0         $(sqrt(3.0)/2.0)  ${coa2}

# create simulation box and atoms
region        reg prism 0 1  0 1  0 1  0.5 0 0  units lattice
create_box    1 reg
create_atoms  1 box

# set mass of carbon
mass          1 12

# specify atom type to chemical species mapping for the KIM model
kim_interactions C

# compute energy
run 0

# print energy
variable      E equal pe
variable      Eperatom equal $E/4.0
print """
================================================================================
The potential energy of a unit cell of bilayer graphene in AB stacking is:
$E eV, (i.e., ${Eperatom} eV/atom).

The used in-plane lattice parameter is a = 2.467 Angstrom and the layer spacing
is c = 3.457 Angstrom.
================================================================================
"""
\end{verbatim}

As noted above, the advantage of releasing a potential as a KIM Portable Model
(instead of implementing it within a single code like LAMMPS) is that it will
work with all simulation codes compatible with the KIM API.
In addition, all interatomic potentials archived in OpenKIM are issued a
``KIM ID'' (with a three-digit version number to track changes) and a
digital object identifier (DOI) that can be cited in publications.
Citing the model DOI makes it possible for the reader to download the
exact interatomic potential (code and parameters) used in the paper
so that they can replicate the results if desired or use it for their
own purposes.

\newpage
%